\documentclass[preprint]{aastex62}
\usepackage{subfigure}
\pdfoutput=1

\graphicspath{{./}{figures/}}

\received{2019 October 26}
\revised{2020 January 19}
\accepted{2020 March 13}
\submitjournal{The Astrophysical Journal}

\shorttitle{Gemini NIFS Observations of Markarian~3}
\shortauthors{Gnilka et al.}
\begin{document}

\title{GEMINI NEAR INFRARED FIELD SPECTROGRAPH OBSERVATIONS OF THE SEYFERT 2 GALAXY MRK~3: FEEDING AND FEEDBACK ON GALACTIC AND NUCLEAR SCALES}

\correspondingauthor{D. Michael Crenshaw}
\email{crenshaw@astro.gsu.edu}

\author{C. L. Gnilka}
\affil{Department of Physics and Astronomy,
Georgia State University,
25 Park Place, Suite 605,
Atlanta, GA 30303, USA}

\author[0000-0002-6465-3639]{D. M. Crenshaw}
\affil{Department of Physics and Astronomy,
Georgia State University,
25 Park Place, Suite 605,
Atlanta, GA 30303, USA}

\author[0000-0002-3365-8875]{Travis C. Fischer}
\affil{Space Telescope Science Institute, 3700 San Martin Drive, Baltimore, MD 21218, USA}

\author[0000-0002-4917-7873]{M. Revalski}
\affil{Space Telescope Science Institute, 3700 San Martin Drive, Baltimore, MD 21218, USA}

\author[0000-0001-8658-2723]{B. Meena}
\affil{Department of Physics and Astronomy,
Georgia State University,
25 Park Place, Suite 605,
Atlanta, GA 30303, USA}

\author[0000-0001-5099-8700]{F. Martinez}
\affil{Department of Physics and Astronomy,
Georgia State University,
25 Park Place, Suite 605,
Atlanta, GA 30303, USA}

\author[0000-0001-5862-2150]{G. E. Polack}
\affil{Department of Physics and Astronomy,
Georgia State University,
25 Park Place, Suite 605,
Atlanta, GA 30303, USA}

\author{C. Machuca}
\affil{Department of Astronomy,
University of Wisconsin,
Madison, WI 53706, USA}

\author{D. Dashtamirova}
\affil{Space Telescope Science Institute, 3700 San Martin Drive, Baltimore, MD 21218, USA}

\author[0000-0002-6928-9848]{Steven B. Kraemer}
\affil{Institute for Astrophysics and Computational Sciences,
Department of Physics,
The Catholic University of America,
Washington, DC 20064, USA}

\author[0000-0003-2450-3246]{H. R. Schmitt}
\affil{Naval Research Laboratory,
Washington, DC 20375, USA}

\author[0000-0003-0483-3723]{R. A. Riffel}
\affil{Departamento de F\'isica, Centro de Ci\^encias Naturais e Exatas, Universidade Federal de Santa Maria, 97105-900 Santa Maria, RS, Brazil}
\affil{Department of Physics \& Astronomy, Johns Hopkins University, Bloomberg Center, 3400 N. Charles St, Baltimore, MD 21218, USA}

\author[0000-0003-1772-0023]{T. Storchi-Bergmann}
\affil{Departamento de Astronomia, Universidade Federal do Rio Grande do Sul, IF, CP 15051, 91501-970 Porto Alegre, RS, Brazil}



\begin{abstract}

We explore the kinematics of the stars, ionized gas, and warm molecular gas
in the Seyfert 2 galaxy Mrk~3 (UGC~3426) on nuclear and galactic scales with {\it Gemini} Near-Infrared Field Spectrograph (NIFS) observations, previous {\it Hubble Space Telescope} data, and new long-slit spectra from the {\it Apache Point Observatory} ({\it APO}) 3.5 m telescope. The {\it APO} spectra are consistent with our previous suggestion that a galactic-scale gas/dust disk at PA $=$ 129\arcdeg, offset from the major axis of the host S0 galaxy at PA $=$ 28\arcdeg, is responsible for the orientation of the extended narrow-line region (ENLR). The disk is fed by an H~I tidal stream from a gas-rich spiral galaxy (UGC~3422) $\sim$100 kpc to the NW of Mrk 3, and is ionized by the AGN to a distance of at least $\sim$20\arcsec\ ($\sim$5.4 kpc) from the central supermassive black hole (SMBH). The kinematics within at least 320 pc of the SMBH are dominated by outflows with radial (line of sight) velocities up to 1500 km s$^{-1}$ in the ionized gas and 500 km s$^{-1}$ in the warm molecular gas, consistent with in situ heating, ionization, and acceleration of ambient gas to produce the narrow-line region (NLR) outflows. There is a disk of ionized and warm molecular gas within $\sim$400 pc of the SMBH that has re-oriented close to the stellar major axis but is counter-rotating, consistent with claims of external fueling of AGN in S0 galaxies.

\end{abstract}



\section{Introduction} \label{sec:intro}
Mrk~3 (UGC~3426) is a nearby S0 galaxy \citep{Windhorst et al.(2002)} with a bright active galactic nucleus (AGN) that is classified as a Seyfert 2 based on its strong, high-ionization narrow emission lines with full-width at half-maximum (FWHM) $\leq$ 2000 km s$^{-1}$ and apparent lack of broader lines in the optical, although spectropolarimetry reveals broad emission lines from a hidden Seyfert 1 nucleus \citep{Schmidt and Miller(1985), Miller and Goodrich(1990), Tran(1995)}. At a redshift of $z = 0.01351$ based on H~I 21-cm observations \citep{Tifft and Cocke(1988)}, Mrk~3 is at Hubble flow distance of $\sim$55 Mpc for H$_0$ $= 73$ km s$^{-1}$ Mpc$^{-1}$, so that 1$''$ corresponds to a transverse size of 270 pc.

\citet{Collins et al.(2009)} estimate that the AGN in Mrk 3 has a bolometric luminosity of $L_{bol} = 2 \times$ 10$^{45}$ erg s$^{-1}$ based on the unabsorbed X-ray 2 -- 10 keV continuum flux from a fit to {\it ASCA} data \citep{Turner et al.(1997)} and their adopted spectral energy distribution (SED). Assuming the mass of the AGN's supermassive black hole (SMBH) is $M_{BH} \approx$ 4.5 $\times$ 10$^{8}$ M$_\sun$, based on the M$_{BH}$ -- stellar velocity dispersion relationship \citep{Woo and Urry(2002)}, the AGN's Eddington ratio is $\sim$0.035 (with considerable uncertainty given that both of the above measurements are indirect).

Due to its proximity and relative brightness, Mrk~3 has been extensively studied in nearly all wave bands \citep{Collins et al.(2005), Crenshaw et al.(2010), Bogdan et al.(2017)}. {\it Hubble Space Telescope} ({\it HST}) [O~III] images with the Faint Object Camera (FOC) show bright emission-line knots in its narrow-line region (NLR) near the AGN in a backwards ``S'' configuration and fainter emission-line arcs within a projected bicone geometry defining the extended NLR (ENLR).  High-resolution radio observations show radio knots in the NLR that are roughly coincident with the [O~III] knots but more collimated and at a slightly different position angle (PA) \citep{Kukula et al.(1999)}. X-ray observations with the {\it Chandra X-ray Observatory} show highly ionized gas that is also roughly coincident with, but possibly more extended than, the [O~III] emitting clouds in the NLR \citep{Bogdan et al.(2017)}.

Our studies of Mrk~3 beginning with {\it HST} STIS spectra found that the NLR gas is outflowing in a roughly biconical geometry, with increasing radial (line of sight) velocity to a peak value of $\sim$600 km s$^{-1}$ at a projected distance of $\sim$0.\arcsec2 from the SMBH along PA $=$ 71\arcdeg, followed by declining radial velocity to the systemic velocity at a projected distance of $\sim$1.\arcsec2 ($\sim$320 pc) (Ruiz et al. 2001). In a subsequent study using the STIS long slit spectra at PA $= 71\arcdeg$ over its full UV and optical range, we determined the reddening and physical conditions in the NLR from the emission-line ratios and photoionization modeling, finding evidence for screening of the NLR gas by absorbers closer to the nucleus and even more shielding of gas outside the nominal bicone, which we suggested is responsible for most of the low-ionization gas (Collins et al. 2005, 2009).

In \citet{Crenshaw et al.(2010)}, we re-examined the geometry of Mrk~3 to address the different PAs and opening angles of the NLR and ENLR, the nature of the emission-line arcs in ENLR, and the backwards ``S'' shape of the NLR. We found that the structures and orientations of the NLR and ENLR could be explained by the intersection of a broader ionizing bicone with a gas and dust disk that is not coplanar with the stellar disk. Based on the locations of the dust lanes and the relative orientations of the host galaxy and ENLR, we suggested that the gas and dust disk in Mrk~3 lies at a PA of $\sim$129\arcdeg, which is offset from the stellar disk by $\sim$100\arcdeg\ in an eastward direction. \citep{Crenshaw et al.(2010)}. We suggested that this gas/dust disk is the result of external feeding due to an encounter with a gas-rich spiral galaxy (UGC 3422) that is $\sim$100 kpc to the NW of Mrk~3, which can be seen as a bridge of H~I gas between the two galaxies in 21 cm observations \citep{Noordermeer et al.(2005)}.

Given the above findings, Mrk~3 provides an interesting nearby case of AGN fueling from an external galaxy through an ongoing tidal interaction and strong outflows that are likely a result of this interaction. Thus, it provides an opportunity to explore the feeding and feedback processes and their connections on both nuclear and galactic scales, particularly through kinematic studies. To pursue this goal, we obtained new observations with {\it Gemini North's} Near-Infrared Field Spectrograph (NIFS) as well as the Astrophysical Research Consortium's 3.5 m telescope and Dual Imaging Spectrograph (DIS) at {\it Apache Point Observatory} ({\it APO}). We also took another look at our previous {\it HST} STIS long-slit spectra of the NLR for comparison.

\section{OBSERVATIONS AND DATA REDUCTION} \label{sec:obs}
\begin{figure}[ht!]
%\plotone{mrk3_judy_schmidt.pdf}
\includegraphics[width=\textwidth]{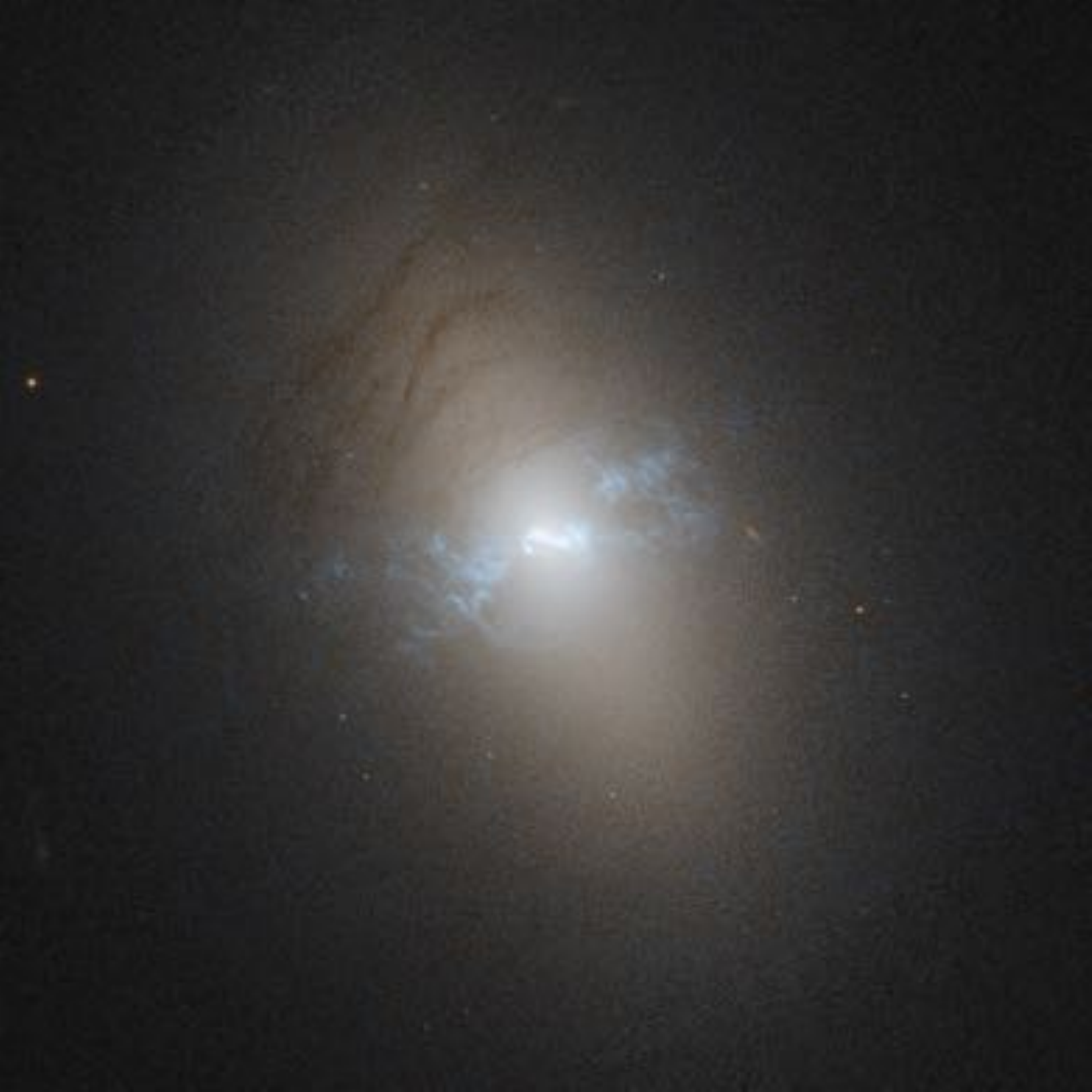}
\caption{Composite 50$''$ $\times$ 50$''$ image of Mrk~3 from {\it HST} WFPC2 F814W (red) and F660W (blue) archive observations, showing the orientation of the host galaxy, dust lanes, and the ionized gas in the NLR and ENLR. North is up and east is to the left. Image credit: NASA / ESA / Judy Schmidt. Used with the permission of Judy Schmidt. \label{fig:mrk3_galaxy}}
\end{figure}

We show a composite color image of Mrk~3 in Figure \ref{fig:mrk3_galaxy}. The stellar continuum emission highlighted in red shows the S0 host galaxy at a position angle (PA) of 28\arcdeg\ \citep{Schmitt and Kinney(2000)}, and an unusual set of dust lanes that do {\it not} lie along the major axis as pointed out in \citet{Crenshaw et al.(2010)}. The extended narrow-line region (ENLR) seen in blue consists of ionized gas at a PA $\approx$ 110\arcdeg, nearly perpendicular to the large-scale stellar axis, which appears to be the continuation of dust lanes into the ionizing AGN bicone as seen in several other AGN, including Mrk~573 \citep{Fischer et al.(2010), Fischer et al.(2017)}. The narrow-line region (NLR) seen in white consists of very luminous ionized gas in a backwards ``S'' shape surrounding the nucleus and with a full extent of $\sim$2.\arcsec4 ($\sim$650 pc) \citep{Capetti et al.(1995), Capetti et al.(1996), Capetti et al.(1999), Schmitt and Kinney(1996), Schmitt et al.(2003)}. We describe our spectroscopic observations below.

\subsection{Gemini Near Infrared Spectrograph (NIFS)} \label{subsec:nifs}

We observed Mrk~3 with {\it Gemini North's} Near-Infrared Field Spectrograph (NIFS) \citep{McGregor et al.(2003)} and Altitude Conjugate Adaptive Optics for the Infrared (ALTAIR) system using the bright central nucleus as a natural guide star ({\it Gemini} program GN-2010A-Q-56). We obtained observations in the Z, J, and K-long bands in single pointings on separate occasions in the NIFS 3$'' \times 3''$ field of view (FOV) as described in Table \ref{tab:nifs}. We followed a standard object-sky-object dithering sequence to obtain six individual 600 s on-source integrations for a total integration time of 3600 s in each band. We were able to constrain the observations to occur during excellent weather and image quality (within the best 20\% of nights during the observing semester), resulting in an angular resolution of $\sim$0.\arcsec1 (FWHM) in all bands with the AO system as determined from the point-spread functions of the telluric standard stars. The spectral resolving power ($=$ $\lambda$/$\Delta\lambda$, where $\Delta\lambda$ is a resolution element) shown in Table \ref{tab:nifs} for each grating gives a velocity resolution of 50 -- 60 km s$^{-1}$ (FWHM).

\begin{deluxetable*}{ccccc}[ht!]
\tablecaption{Gemini NIFS Observations of Mrk~3 \label{tab:nifs}}
\tablecolumns{5}
%\tablenum{1}
\tablewidth{0pt}
\tablehead{
\colhead{UT date} & 
\colhead{Grating} & \colhead{Integration} &
\colhead{Wavelength Range} & \colhead{Spectral Res.}  \\
\colhead{} & \colhead{} &
\colhead{(sec)} & \colhead{(\micron)} & \colhead{($\lambda$/$\Delta\lambda$)} 
}
\startdata
2010-10-11 & Z & 3600 &0.94 -- 1.15 &4990 \\
2010-11-05 & J & 3600 &1.15 -- 1.33 &6040 \\
2010-11-04 & K-long & 3600 &2.10 -- 2.54 &5290 \\
\enddata
\end{deluxetable*}

We reduced the data using tasks contained in the NIFS subpackage within the GEMINI IRAF package as well as standard IRAF tasks. The reduction process includes image trimming, flat fielding, sky subtraction, s-distortion correction, and wavelength calibration. We corrected the frames for telluric absorption features and flux calibrated the spectra by interpolating blackbody functions to the spectra of telluric standard (early A-type) stars to generate the sensitivity functions. We median combined the individual integrations into a single data cube for each band using the gemcombine task. We aligned the data cubes from the three different bands so that the continuum flux centroids are located at the center of each FOV, as justified below. The NIFS instrument samples the observations in $0.\arcsec043 \times 0.\arcsec103$ bins \citep{McGregor et al.(2003)}, which are then resampled during data reduction to $0.\arcsec05 \times 0.\arcsec05$ in their image planes. Together with the spectral dimension, the reduction provides calibrated data cubes with dimensions of $60 \times 60 \times 2040$ spaxels. Due to the alignment, the bottom (most southern) two rows in the NIFS image plane spanning 0.\arcsec1 are blank.

We followed the procedures of \citet{Menezes et al.(2014)} for additional corrections to the NIFS calibrated data cubes that first require resampling them to a finer (0.\arcsec025 $\times$ 0.\arcsec025) spatial grid. We did not correct for differential atmospheric refraction (DAR), which is a function of wavelength and zenith distance \citep{Filippenko(1982)}, across any of the bands because the calculated shift of the continuum centroid is $\leq$ 0.\arcsec025 for our near-IR observations. However, we did effectively correct for DAR between bands by aligning the broad-band continuum centroids as described above. The offset between bands has the same magnitude ($\sim$0.\arcsec05) and direction predicted from the DAR equation \citep{Menezes et al.(2014)}, indicating no significant physical offset in the continuum centroids between the near-IR bands. We also used the Butterworth spatial filtering suggested by \citet{Menezes et al.(2014)} to remove periodic, high-frequency noise from the data cubes, with a cutoff frequency of 0.30 Nyquists to remove most of the periodic noise without significantly affecting the fluxes. We applied their suggested Richardson-Lucy deconvolution to correct for the broad wings of the point-spread function (PSF) typical for ground-based AO using a Gaussian PSF constructed from telluric stars observed before or after the AGN and interpolated as a function of wavelength. This procedure results in more distinct and higher-contrast emission-line clouds in images extracted from the corrected NIFS data as shown in \citet{Menezes et al.(2014)} and specifically in Mrk~3 by \citet{Pope et al.(2016)}. Finally, we resampled the data cubes back to the original spatial grid of 0.\arcsec05 $\times$ 0.\arcsec05.

\subsection{Apache Point Observatory Dual Imaging Spectrograph (APO DIS)} \label{subsec:apo}
We obtained long-slit spectra of Mrk~3 across the entire galaxy, which extends over $\sim$80\arcsec\ along its major axis in the I band \citep{Schmitt and Kinney(2000)}, using the Astrophysical Research Consortium's 3.5 m telescope and Dual Imaging Spectrograph (DIS) at the {\it Apache Point Observatory} ({\it APO}) . DIS provides simultaneous blue and red spectra, and we chose the B1200 and R1200 medium-dispersion gratings and a 2.\arcsec0 wide slit to obtain spectral resolving powers of $R = 4000 - 5500$, similar to that of NIFS, at the expense of limited wavelength coverage and non-overlapping spectra. The angular scale in the cross-dispersion direction is 0.\arcsec42\ pixel$^{-1}$ in the blue and 0.\arcsec40\ pixel$^{-1}$ in the red.

As shown in Table \ref{tab:apo}, we obtained two sets of blue and red spectra at position angles of 30\arcdeg, 90\arcdeg, and 129\arcdeg. The first set covers the regions around the [O~III] $\lambda$5007 and H$\alpha$ emission lines and the second set covers a multitude of emission lines in the blue and the Ca~II triplet stellar absorption lines in the red. We obtained these spectra on clear or mostly clear nights at low airmass with reasonable seeing, as determined from cross-dispersion profiles of standard stars and shown in Table \ref{tab:apo}.

\begin{deluxetable*}{ccccccc}[ht!]
\tablecaption{APO DIS Observations of Mrk~3 \label{tab:apo}}
\tablecolumns{7}
%\tablenum{2}
\tablewidth{0pt}
\tablehead{
\colhead{UT date} & \colhead{Integration} &
\colhead{Wavelength Range} & \colhead{Wavelength Range} &
\colhead{Pos. Angle} & \colhead{Airmass} & \colhead{Seeing} \\ % 2nd number from logs
\colhead{} & \colhead{(sec)} &
\colhead{Blue (\AA)} & \colhead{(Red (\AA)} &
\colhead{(\degr)} & \colhead{} & \colhead{($''$)}
}
\startdata
2014-10-25 &2400 &4743 -- 5554 &5987 -- 7175 &30  &1.27 &1.4 \\
2014-12-24 &2400 &4743 -- 5543 &5982 -- 7170 &129 &1.33 &3.0 \\
2015-02-19 &2400 &4475 -- 5575 &5982 -- 7171 &90  &1.28 &1.5 \\
2016-12-01 &3600 &3332 -- 4597 &8020 -- 9170 &30  &1.28 &1.5 \\
2016-12-01 &3600 &3332 -- 4597 &8020 -- 9170 &90  &1.42 &1.5 \\
2017-09-19 &3600 &4264 -- 5537\tablenotemark{a} &7995 -- 9149 &129 &1.42 &1.8 \\ 
%2017-10-21 &3600 &4240 -- 5514 &6022 -- 7211 &60  &1.36 &2.6\tablenotemark{a} \\
%2017-10-21 &3600 &4240 -- 5514 &6021 -- 7210 &150 &1.29 &2.6\tablenotemark{a} \\ %2017-11-23 &3600 &4277 -- 5538 &7978 -- 9131 &330 &1.31 &1.4\tablenotemark{a} \\
%2017-11-23 &3600 &4277 -- 5543 &7978 -- 9131 &240 &1.38 &1.4\tablenotemark{a} \\
\enddata
\tablenotetext{a}{Affected by instrument-scattered light.}
%% \tablecomments{}
\end{deluxetable*}

During the latter part of our observations of Mrk~3, DIS began to show signs of scattered light in the blue and red sides of the instrument due to contamination of the field-correcting optics from an apparent dewar leak. Despite efforts to clean the optics periodically, the contamination was extensive enough in observations beginning on 2017 September 19 to produce noticeable wings in the cross-dispersion profiles of standard stars and ``halos'' around the emission lines of Mrk 3, significantly affecting kinematic measurements. Our measurements of the cross-dispersion wings indicate that the blue side on the above date was affected but the red side was not. Observations on later dates at position angles of 60\arcdeg\ and 150\arcdeg\ all show enhanced broad wings and, along with the blue spectrum from 2017 September 19, were not included in our analysis.

We reduced the DIS long-slit spectra using standard IRAF routines and the {\it APO} atmospheric extinction curve to produce two-dimensional flux-calibrated spectral images with constant wavelength along each column \citep{Revalski et al.(2018a)}. We performed additional corrections in IDL by resampling the spectral images to produce constant spatial location along each row (correcting for a slight tilt of the slit with respect to the cross-dispersion direction) and subtracting the night sky lines using regions outside of the galaxy. The blue and red spectra do not overlap and we are interested in accurate relative fluxes of the emission lines, so we scaled the DIS spectra separately in flux to match a spectrum of Mrk~3 spanning 4000 -- 7700 \AA, which we obtained with the {\it Lowell Observatory} 1.8 m telescope and deVeny Spectrograph through a 4.\arcsec0 wide slit in photometric conditions on 2010 March 18.

\subsection{Hubble Space Telescope Data} \label{subsec:hst}
We use some of the {\it HST} images and spectra from our previous studies \citep{Ruiz et al.(2001), Collins et al.(2005)} to supplement our analysis in this paper. Table \ref{tab:hst} lists these data. For the Space Telescope Imaging Spectrograph (STIS) spectra, we used a 52\arcsec $\times$ 0.\arcsec1 slit along PA $=$ 71\arcdeg. Their spectral resolving powers are lower than those of the other spectra that we obtained, yielding a velocity resolution of 300 - 400 km s$^{-1}$, but they provide the best angular contrast between emission-line clouds along the slit due to {\it HST's} high Strehl ratio at a resolution of 0.\arcsec1 (FWHM).

\begin{deluxetable*}{ccccccc}[ht!]
\tablecaption{Selected {\it HST} Observations of Mrk~3 \label{tab:hst}}
\tablecolumns{7}
%\tablenum{3}
\tablewidth{0pt}
\tablehead{
\colhead{UT date} & \colhead{Detector} & \colhead{Filter/Grating} & \colhead{Integration} & \colhead{Wavelength Range} &
\colhead{Spectral Res.}  & \colhead{Pos. Angle}  \\
\colhead{} & \colhead{} & \colhead{} &
\colhead{(sec)} & \colhead{(\AA)} &
\colhead{($\lambda$/$\Delta\lambda$)}  & \colhead{(\arcdeg)}
}
\startdata
%1994-03-20 & FOC   & F502M & 750  & 4973 -- 5047   &     & \\
%1994-03-20 & FOC   & F550M & 1196 & 4973 -- 5047   &     & \\
1997-10-20 & WFPC2 & F606W & 500  & 4835 -- 7035   &     & \\ 
2000-08-19 & WFPC2 & F814W & 160  & 7324 -- 9082   &     &  \\
2000-01-16 & STIS  & G430L & 1080 & 2652 -- 5947   & 900 & 71 \\
2000-01-16 & STIS  & G750L & 1080 & 4821 -- 10,680 & 700 & 71 \\
\enddata
\end{deluxetable*}

%Our processing of the FOC F502M image containing primarly [O~III] $\lambda\lambda$4959, 5007 emission is described in \citet{Crenshaw et al.(2010)} and includes the subtraction of continuum emission using a scaled version of the FOC F550M image.
We used the WFPC2 F606W image as described in \citet{Crenshaw et al.(2010)} to produce a structure map (see \citet{Pogge and Martini(2002)} that highlights emission-line features due primarily to [O~III] and H$\alpha$ as well as dust features in the host galaxy. Our analysis of the F814W image using GALFIT is described in Section \ref{subsec:photometry}. Our reduction of the STIS spectra to produce wavelength and flux-calibrated spectral images is described in \citet{Ruiz et al.(2001)} and we show STIS spectral images of various emission lines in \citet{Collins et al.(2005)}.

\begin{figure}[ht!]
\plotone{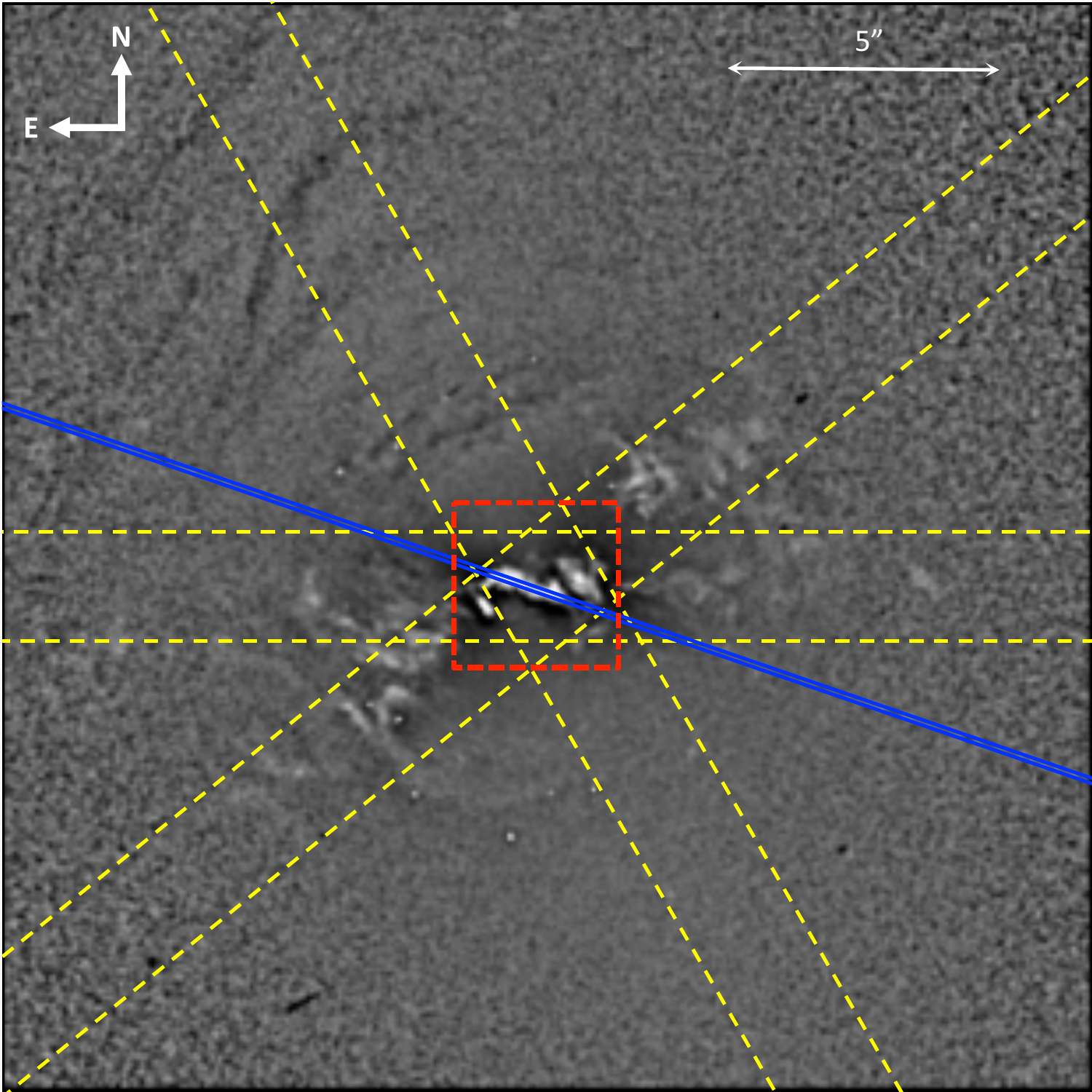}
\caption{{\it HST} F606W stucture map of the inner regions of Mrk~3 (1$''$ $=$ 270 pc). Bright regions correspond to [O~III] and continuum emission and darker regions are due to dust lanes. The bright arcs in the inner regions are gas/dust spirals that cross into the AGN radiation bicone and become at least partially ionized. The figure shows the locations of the {\it Gemini} NIFS field of view (red) encompassing the backward S-shape of the NLR, {\it HST} STIS slit (blue), and {\it APO} DIS slits (yellow).\label{fig:mrk3_slits}}
\end{figure}

\bigskip\bigskip

\subsection{Observational Footprints} \label{subsec:footprints}
Figure \ref{fig:mrk3_slits} shows a structure map of the {\it HST} F606W image, expanded around the inner regions of Mrk~3, with the locations of our observations superimposed. The NIFS 3\arcsec $\times$ 3\arcsec\ FOV encompasses the bright backwards ``S''-shaped NLR. The STIS long slit lies along the inner, linear portion of the NLR. The {\it APO} DIS long slits are close to the stellar major axis at PA $=$ 30\arcdeg, the proposed gas disk major axis at PA $=$ 129\arcdeg,  and an intermediate location at PA $=$90\arcdeg. All of the {\it APO} slits intersect at the nucleus and contain most of the NLR over just 2 -- 3 angular resolution elements; their primary function is to trace the kinematics of the ENLR and host galaxy.

\section{ANALYSIS AND RESULTS} \label{sec:analysis}

\subsection{Host Galaxy Photometry} \label{subsec:photometry}
\begin{figure}[ht!]
\plotone{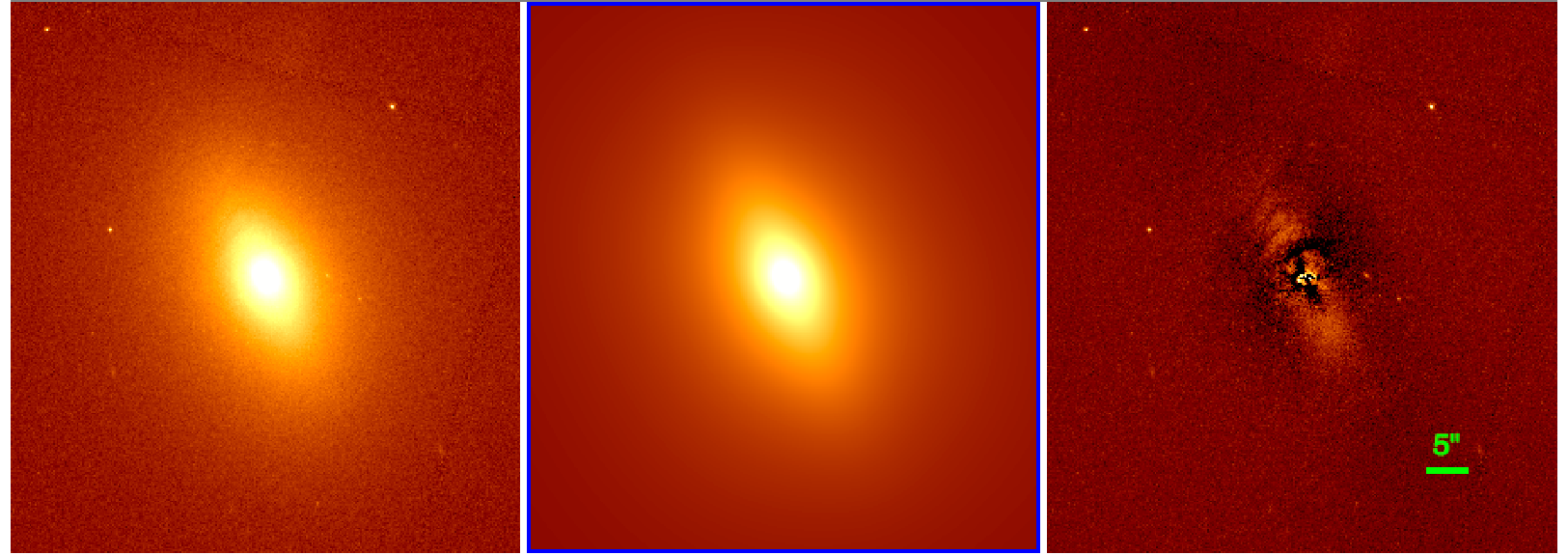}
\caption{{Left: {\it HST} WFPC2 F814W continuum image of Mrk 3. Center: Best fit GALFIT model (3 components). Right: Residual image obtained by subtracting the model from the image, showing the dust lanes to the NE and the NLR emission near the nucleus.
\label{fig:mrk3_galfit}}}
\end{figure}

To characterize the morphological components of the host galaxy of Mrk~3, we decomposed the {\it HST}/WFPC2 F814W image using GALFIT version 3.0.5 \citep{Peng et al.(2002), Peng et al.(2010)}. We show the original image, GALFIT model, and resulting residual map in Figure \ref{fig:mrk3_galfit}. We find the best fitting model is composed of three S\'{e}rsic components with parameters given in Table \ref{tab:galfit}. 

Components 1 and 3 have very similar PAs, axial ratios, and S\'{e}rsic indices, and the first two values match with those from ellipse fitting of ground-based imaging in the I band (PA $=$ 28\arcdeg, b/a $=$ 0.84) \citep{Schmitt and Kinney(2000)}. Given their S\'{e}rsic indices, these likely represent a bulge or pseudo-bulge. Component 2 has a similar PA as well, and a S\'{e}rsic index that indicates it is an exponential disk, in which case its axial ratio gives an inclination of 65\arcdeg. We find no evidence of a large stellar bar or oval in the GALFIT results or in the PA and ellipticity plots of \citet{Schmitt and Kinney(2000)}. The lack of a bar or stellar spiral arms in the GALFIT residual image in Figure \ref{fig:mrk3_galfit} supports the S0 classification of the host galaxy. The residual image highlights the peculiar dust lanes in the NE that are nearly perpendicular to the galactic disk \citep{Crenshaw et al.(2010)} and the contaminating effects of NLR emission in the filter bandpass close to the nucleus.

\begin{deluxetable*}{ccccccc}[ht!]
\tablecaption{GALFIT Results for Mrk~3 \label{tab:galfit}}
\tablecolumns{7}
\tablewidth{0pt}
\tablehead{
\colhead{Comp.} & \colhead{I(mag)} & \colhead{R$_e$ (pc)\tablenotemark{a}} & \colhead{n} & \colhead{b/a} & \colhead{PA(\arcdeg)}  & \colhead{frac.}
}
\startdata
1 & 13.08 & 330  &1.7 &0.85 & 23 & 0.19\\
2 & 12.93 & 2130 &1.1 &0.43 & 24 & 0.21\\
3 & 11.79 & 5260 &1.7 &0.86 & 27 & 0.60 \\
\enddata
\tablenotetext{a}{For 1\arcsec $=$ 270 pc.}
\tablecomments{Col. (1) gives the component number; Col. (2) gives the integrated I band magnitude; Col. (3) provides the effective radius; Col. (4) gives the S\'{e}rsic index; Cols. (5) and (6) give the axial ratio and position angle of the component; Col. (7) gives the fraction of the integrated flux from each component.}
\end{deluxetable*}

\subsection{Nuclear Stellar Kinematics: Gemini NIFS} \label{subsec:nuclear_stellar}

We determined the stellar kinematics of Mrk~3 within the NIFS 3\arcsec\ $\times$ 3\arcsec\ FOV using the penalized pixel-fitting (pPXF) method of \citet{Cappellari and Emsellem(2004)}. We fit the $^{12}$CO 2.29 \micron, $^{12}$CO 2.32 \micron, and $^{13}$CO 2.34 \micron\ stellar absorption features within the K-band following the procedure described in \citet{Riffel et al.(2008)}. To obtain the line-of-sight velocity distribution at each position, we used stellar templates of early-type stars \citep{Winge et al.(2009)} and included multiplicative Legendre polynomials of order 3 to fit the continuum emission.
As noted by \citet{Riffel et al.(2017)}, the CO band heads in the NIFS K-band observations of Mrk~3 are difficult to fit due to strong background continuum emission and subsequent low contrast of the absorption features, consistent with the large dilution of stellar absorption features observed in the optical \citep{Gonzalez Delgado et al.(2001)}. We were able to obtain a usable radial velocity map by smoothing it with a 5 $\times$ 5 pixel median filter.

\begin{figure}[ht!]
\plotone{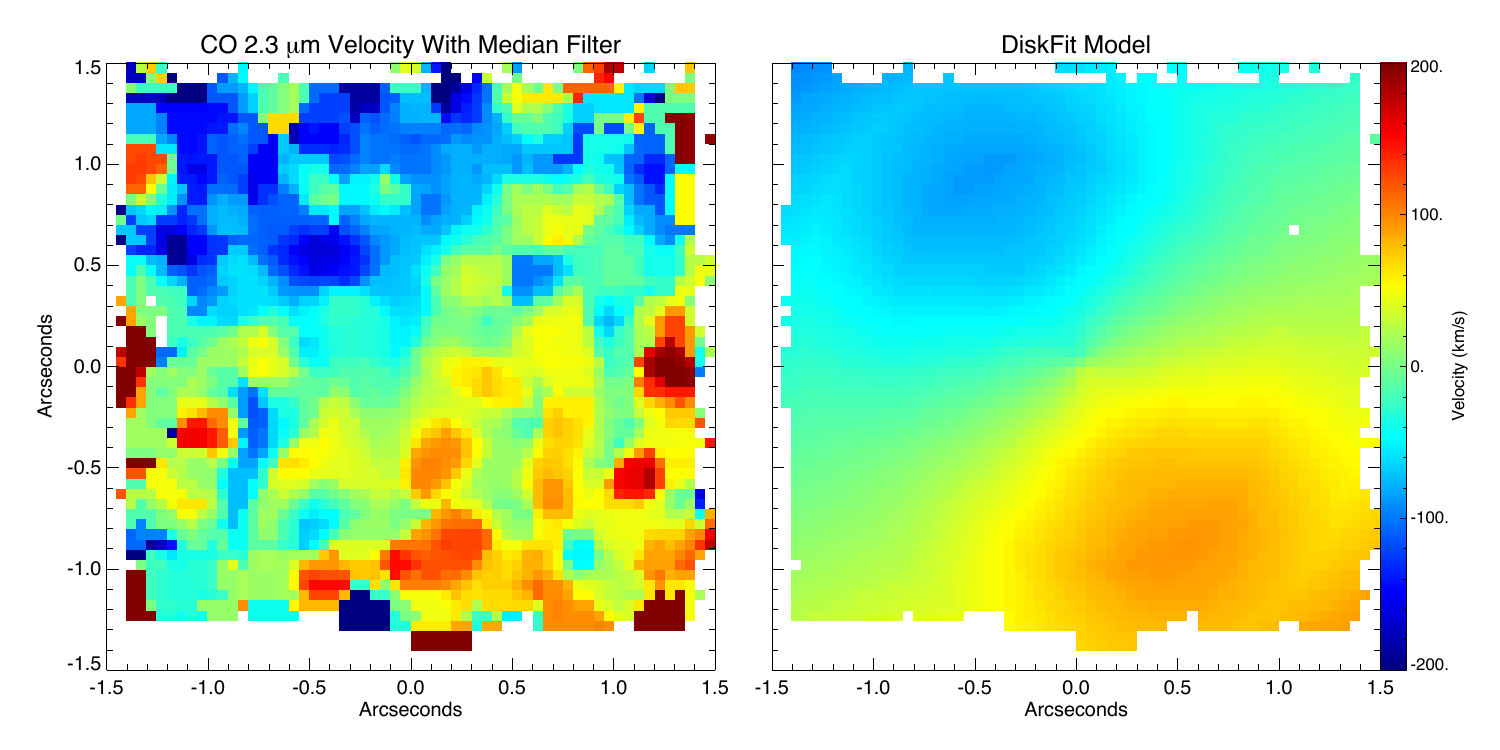}
\caption{Left: Radial velocity map from NIFS observations of the stellar CO bandheads smoothed with a 5 $\times$ 5 pixel median filter. Right: Diskfit radial velocity image from the fit to the median-filtered image with the PA and inclination fixed to 28\arcdeg\ and 33\arcdeg, respectively. The residuals span a relatively large range with an average absolute value of 40 km s$^{-1}$.
\label{fig:nifs_stellar}}
\end{figure}

Looking at the stellar radial velocity map in the left panel of Figure \ref{fig:nifs_stellar}, there is clear evidence for rotation, with blueshifts in the NE and redshifts in the SW. The observed kinematic major axis appears to be in the same direction as the large-scale photometric axis at PA $=$ 28\arcdeg. This agreement suggests that the galactic and nuclear stellar components are at the same orientation, consistent with the GALFIT results in Section \ref{subsec:photometry}.
The stellar velocity dispersion map from this analysis shows no discernible large-scale structure in the 3\arcsec\ $\times$ 3\arcsec\ FOV and an average value of $\sigma_*$ $\approx$ 230 km s$^{-1}$ $\pm$ 60 km s$^{-1}$, comparable to the value of 269 km s$^{-1}$ $\pm$ 33 km s$^{-1}$ derived from the Ca~II triplet by \citet{Nelson and Whittle(1995)}.  The large value of $\sigma_*$ is consistent with the GALFIT result that the majority of the light on large scales is dominated by a bugle or pseudobulge. Using the $M_{BH}-\sigma_*$ relation from \citet{Batiste et al.(2017)}, the SMBH has log($M_{BH}$/M$_\sun$) $=$ 8.95$\pm$0.45, consistent with the value of 8.65 from \citet{Woo and Urry(2002)} given our rather large uncertainties.

To further quantify the stellar rotation kinematics near the nucleus, we used DiskFit \citep{Spekkens and Sellwood(2007), Sellwood and Sanchez(2010), Kuzio de Naray et al.(2012)}, a publicly available code that fits non-parametric models to a given velocity field. 
The structure seen in the median-filtered radial velocity map prevented a consistent fit with all DiskFit parameters allowed to vary.
We therefore fixed the DiskFit PA to 28\arcdeg\ and inclination to 33\arcdeg, which are the average values from the I-band photometry of the host galaxy \citep{Schmitt and Kinney(2000)}, and kinematic center to that of the continuum centroid to retrieve the overall stellar velocity field. As shown in the right panel of Figure \ref{fig:nifs_stellar}, the fit results in a typical rotation pattern for the inner regions of a disk galaxy.

\subsection{Galaxy Stellar Kinematics: APO DIS} \label{subsec:galaxy_stellar}
To study the kinematics of the stellar disk on large scales, we used the stellar absorption features in the DIS far-red spectra, which include the Ca~II $\lambda\lambda\lambda$8498, 8542, 8662 triplet. As in Section \ref{subsec:nuclear_stellar}, we used pPXF on the long-slit spectra at PAs $=$ 30\arcdeg, 90\arcdeg, and 129\arcdeg. Although the Ca~II features are relatively weak in these spectra, the stellar radial velocities show a clear pattern of redshifts in the NE and E and blueshifts in the SW and W for PAs $=$ 30\arcdeg\ and 90\arcdeg, whereas they are scattered around zero km s$^{-1}$ at PA $=$ 129\arcdeg.  Furthermore, the velocity amplitudes are greatest at PA $=$ 30\arcdeg, similar to the nuclear stellar velocity field.

\begin{figure}[ht!]
\epsscale{0.8}
\plotone{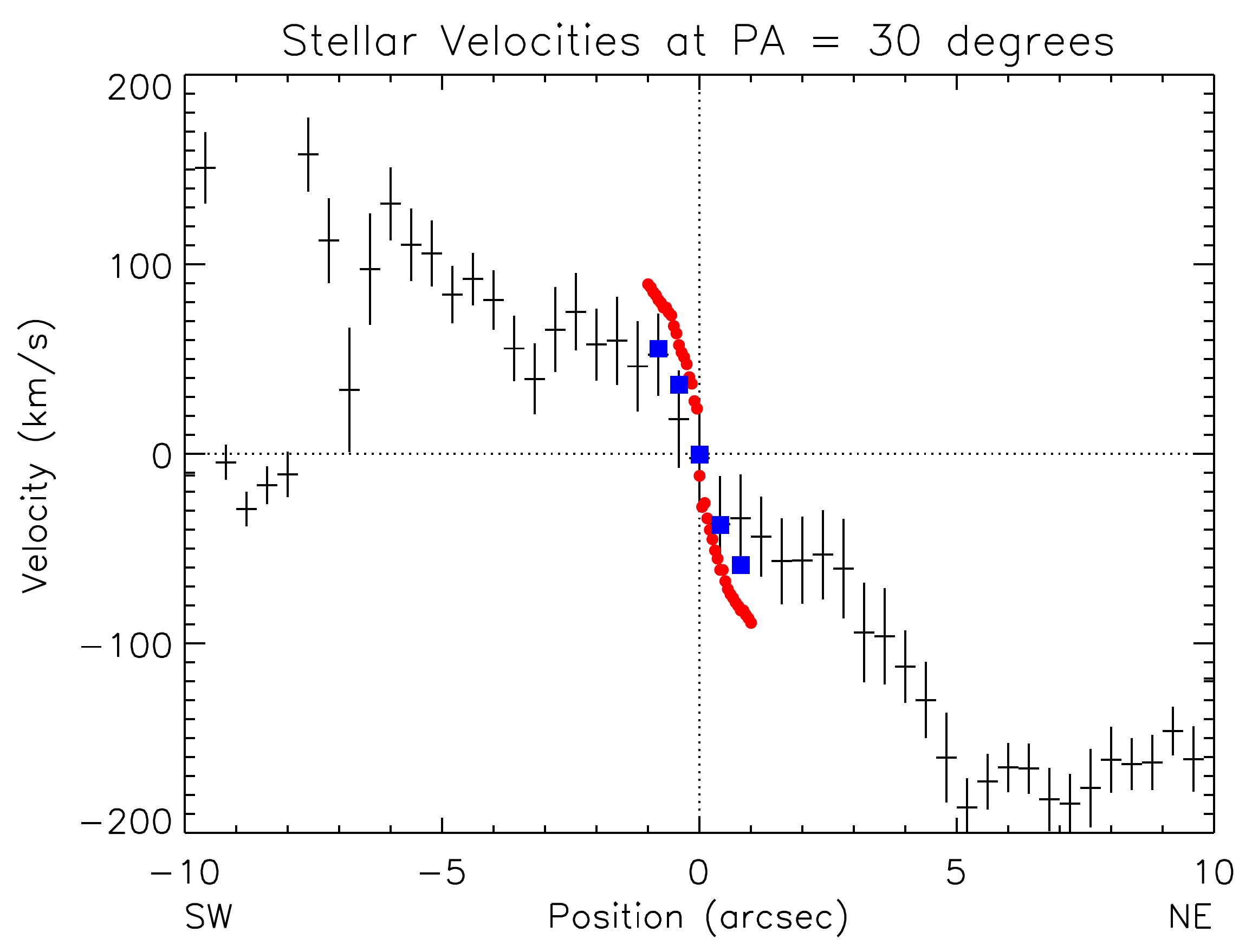}
\caption{Stellar radial velocities from the {\it APO} Ca~II triplet absorption lines at PA $=$ 30\arcdeg\ with associated uncertainties (points near zero velocity in the SW have high $\chi^2$ values and should be ignored). For comparison, NIFS velocities from a 0.\arcsec1-wide pseudo-slit at PA $=$ 30\arcdeg\ through the continuum center are plotted as red circles. NIFS velocities through a 2.\arcsec0-wide slit and sampled in 0.\arcsec4 bins along the slit to match the {\it APO} data are plotted as blue squares.
\label{fig:stellar_vels}}
\end{figure}

We show the {\it APO} DIS stellar velocity curve at PA $=$ 30\arcdeg\ in Figure \ref{fig:stellar_vels}. For comparison, we show extractions at PA $=$ 30\arcdeg\ from the NIFS velocity field in Figure \ref{fig:nifs_stellar} through 0.\arcsec1- (red circles) and 2.\arcsec0-wide (blue squares) pseudo-slits with the latter sampled in 0.\arcsec4 bins along the slit to match the {\it APO} data. The resampling reduces the NIFS velocity amplitude significantly and provides a reasonably good match between {\it APO} DIS optical and {\it Gemini} IR velocities.
The slight difference in amplitude may be due to additional instrumental effects or to a kinematic difference in the stellar populations sampled by the Ca II triplet and CO features \citep{Riffel et al.(2015a)}.
Nevertheless, the close match in both PA and velocity indicates a smooth transition between the nuclear and stellar kinematics, with no evidence for a different stellar component on small scales. We conclude that the nuclear and galactic stellar velocity fields in Mrk~3 have rotation kinematic major axes that are aligned with the photometric axis and show no evidence for kinematic disturbances.

\subsection{NLR Gas Kinematics: {\it HST} STIS} \label{subsec:nlr_gas_stis}

Spatially-resolved spectra of the NLRs in AGN often show multiple kinematic components at each location as evidenced by distinct separations, bumps, or inflections in the emission-line profiles \citep{Das et al.(2005), Das et al.(2006), Fischer et al.(2013)}, indicating further subdivision of the emission-line knots seen in NLR images. To separate these components in the emission lines from all of our spectroscopic observations, we fit multiple Gaussian profiles to each line of interest at each spatial location to obtain the radial velocity centroid ($v_r$), full width at half maxiumum (FWHM), and integrated flux of each component. We employed an automated Bayesian fitting routine that starts with a continuum followed by zero, one, two, and three successive Gaussians to determine the number of significant Gaussian components and their parameters, as described in detail by \citet{Fischer et al.(2017), Fischer et al.(2018)}, who provide several examples of the profile fits. We note that automated fitting routines are essential for objective determinations of the number of kinematic components and their parameters in large datasets (e.g., up to 3600 spectra for one NIFS observation). We also note that that this procedure effectively separates the outflow and rotation components as shown by \citet{Fischer et al.(2017)} and later in this paper, but it does not account for intrinsically non-Gaussian and/or asymmetric profiles; thus weak or tertiary components should be interpreted with some caution.

We chose to sort the components by integrated flux, except where otherwise noted, to gauge their relative contributions to the total flux. This is useful, for example, to isolate the brightest components that contribute most to the mass outflow. The emission-line knots are spatially resolved with many pixels or spaxels clustered together, so their properties can be identified regardless of the sorting method. Line flux ratios may differ between different kinematic components, so we do not use the automated fitting and sorting routines to match kinematic components when fitting different emission lines independently, but rather rely on $v_r$ and FWHM plots of the components in each line.

We required an emission-line component to have a signal to noise ratio (SNR) $>$ 3 in its integrated flux for a positive detection. To obtain $v_r$ values in the rest frame of the galaxy, we subtracted the systemic velocity. We determined the intrinsic FWHM of each component by subtracting the FWHM of the line spread function for each grating (i.e., the velocity resolution in km s$^{-1}$) from the observed FWHM in quadrature. Uncertainties in the kinematic measurements of the emission lines come primarily from three sources as described by \citet{Das et al.(2005)}: 1) deviation of the emission-line components from pure Gaussian functions, 2) displacment of the emission-line knots from the center of the slit or spaxel in the dispersion direction, and 3) photon noise. We estimated the uncertainties in the radial velocities and FWHM by adding the values from above in quadrature for each spectrograph.

\begin{figure}[ht!]
\plottwo{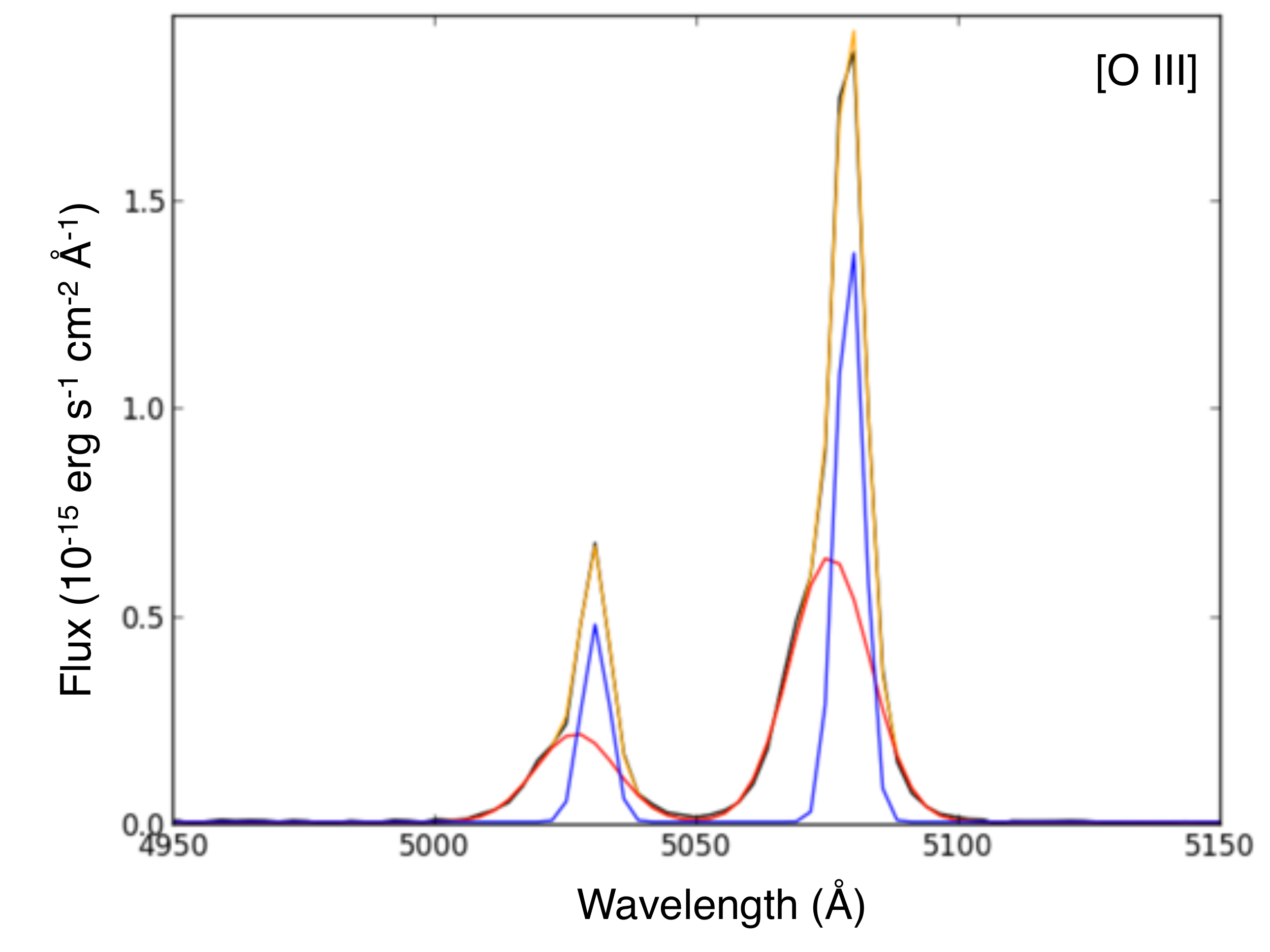}{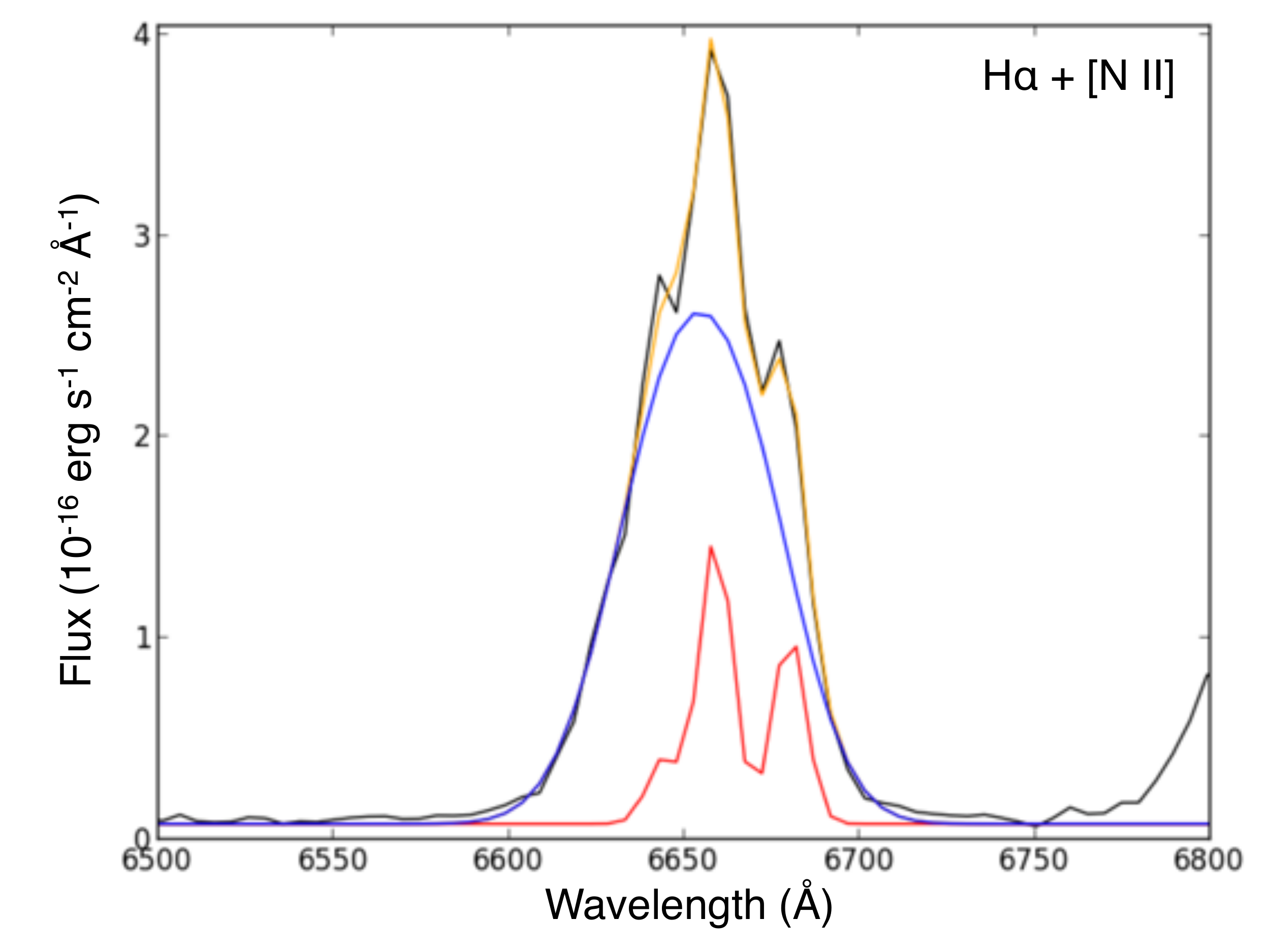}
\caption{Independent spectral fits to the [O~III] $\lambda\lambda$4959, 5007 (left) and H$\alpha$ plus [N~II] $\lambda\lambda$6548, 6583 lines (right) from the STIS spectrum of the nucleus of Mrk~3 (central 0.\arcsec05 $\times$ 0.\arcsec1 bin), showing the two kinematic components in blue and red. The observed continuum-subtracted spectra are in black and the sums of the two components are shown in orange.  The components are sorted by integrated flux (blue is higher), so the blue-colored component in [O~III] is associated in velocity and width with the red-colored component in H$\alpha$ $+$ [N~II] and vice-versa.
\label{fig:stis_fits}}
\end{figure}

To assist in the interpretation of the NLR and ENLR kinematics from {\it Gemini} NIFS and {\it APO} STIS, we revisited the kinematics of the NLR in Mrk~3 observed by {\it HST} STIS \citep{Ruiz et al.(2001)}. We refit the strong emission lines in the STIS G430L and G750L long-slit spectra in 0.\arcsec05 bins along the linear portion of the NLR at PA $=$ 71\arcdeg. We fit the strong [O~III] $\lambda\lambda$4959, 5007 lines simultaneously with essentially no constraints on the parameters except to fix the velocity centroid and width of the former to that of the latter and its flux to 1/3 of the latter according to their radiative transition probabilities \citet{Osterbrock and Ferland(2006)}). We also constrained the FWHM of all components to be $\geq$ the spectral resolution and $\leq$ 2000 km s$^{-1}$ as appropriate for a Seyfert 2 galaxy. We independently fit the H$\alpha$ and [N~II] $\lambda\lambda$6548, 6584 complex with the same constraints as a check on the [O~III] kinematics, as discussed below. In order to obtain emission-line ratios for each kinematic component, we also used the [O~III] fits as templates for all of the other lines (including H$\alpha$ and [N~II]) in 0.\arcsec05 bins along the slit by fixing the velocity centroids and widths to the [O~III] values and allowing the fluxes to vary (keeping the doublet constraints) to provide the best matches to the observed lines and blends. In Figure \ref{fig:stis_fits}, we show independent two-component fits to the [O~III] doublet and the H$\alpha$ plus [N~II] complex at the nucleus, demonstrating the  goodness of the fits despite the relatively low spectral resolution of the STIS spectra.

\begin{figure}[ht!]
\plottwo{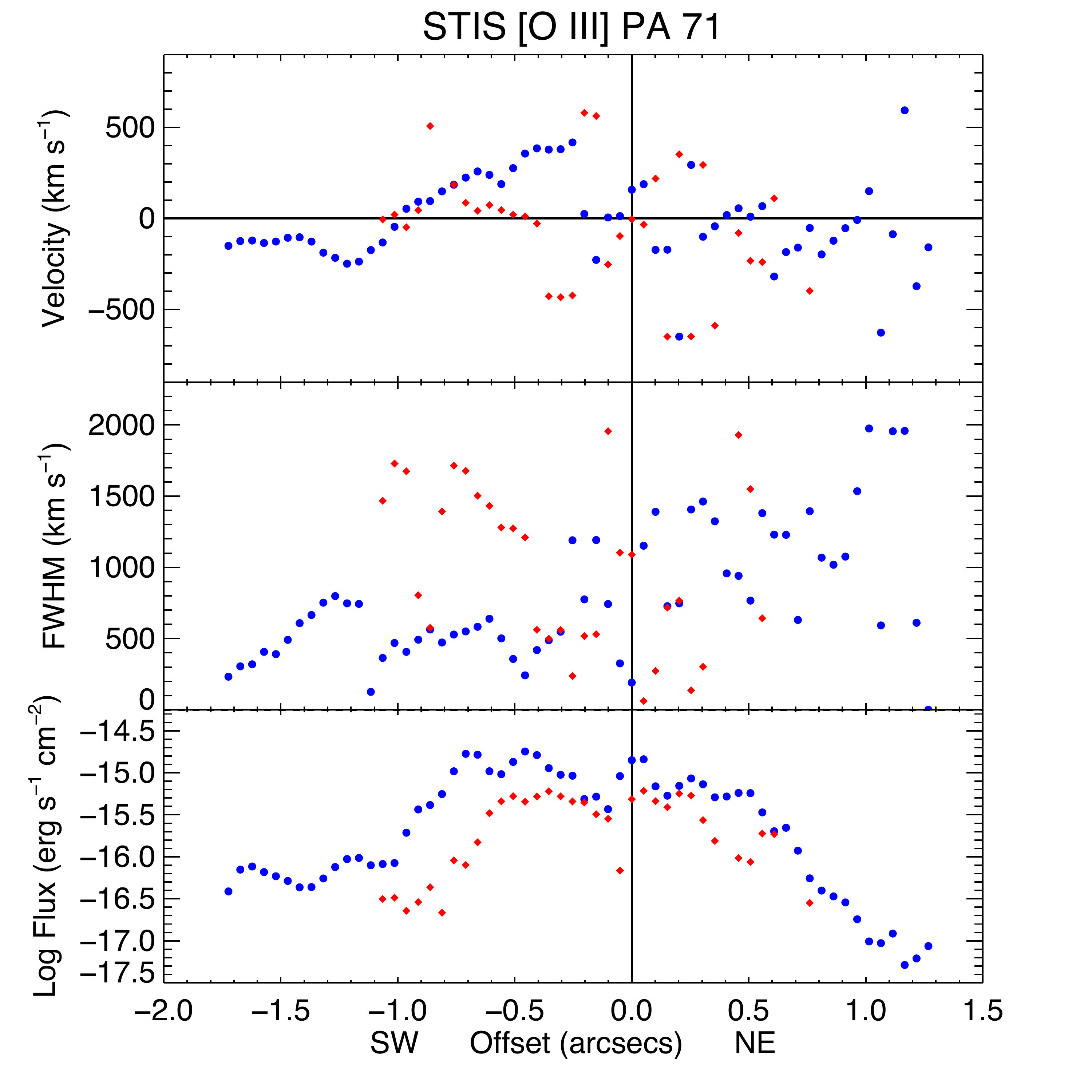}{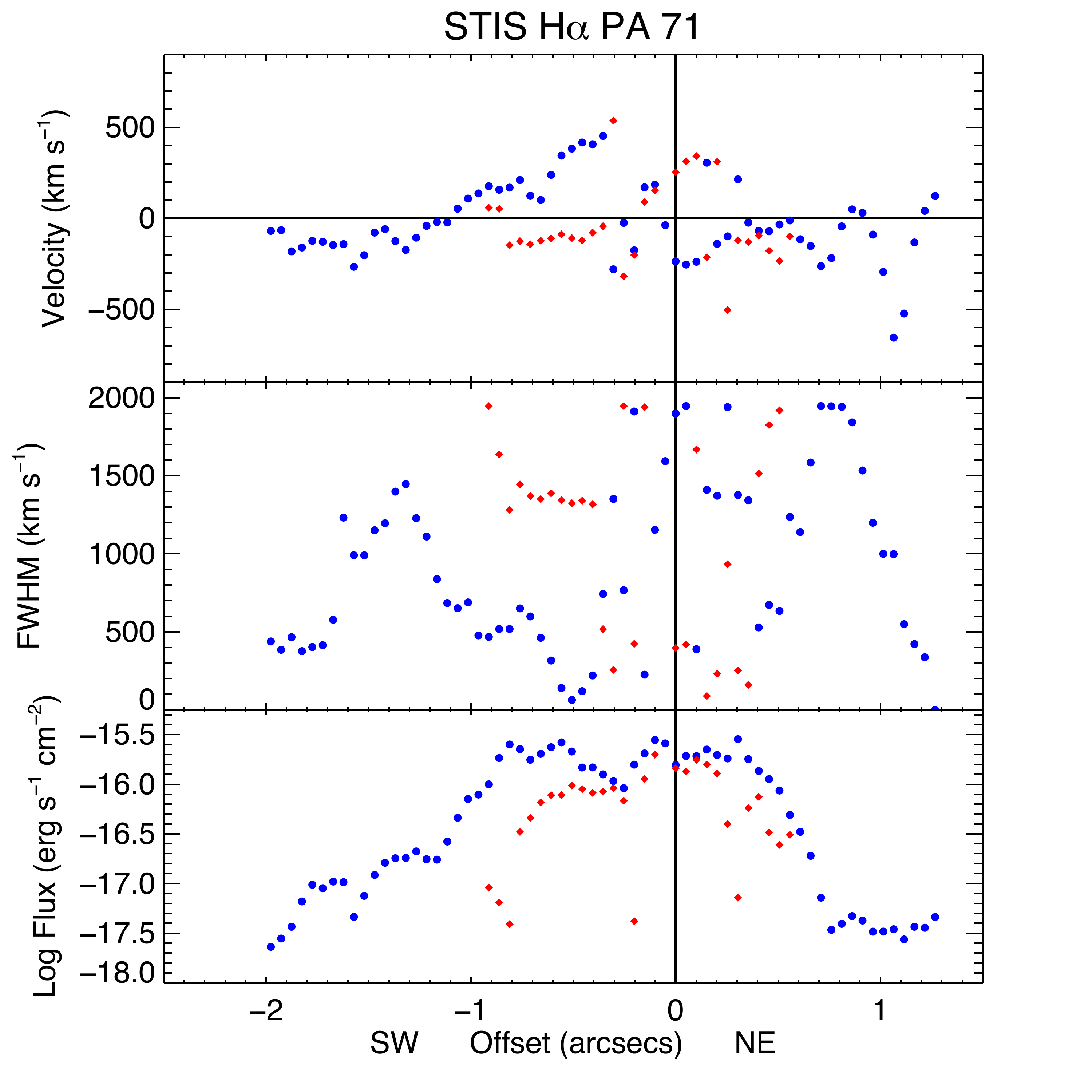}
\caption{Radial velocity centroid, FWHM (corrected for the line-spread function), and integrated flux for each emission-line component from our kinematic fits to the [O III] $\lambda$5007 (left) and H$\alpha$ (right) emission lines along the {\it HST} STIS slit at PA = 71\arcdeg. Data points are marked as blue and red circles corresponding to higher and lower integrated line fluxes at each location, respectively. The vertical solid line at zero position gives the location of the continuum centroid. The maximum uncertainties in radial velocity and FWHM are $\pm$60 km s$^{-1}$.
\label{fig:stis_kinematics}}
\end{figure}

In Figure \ref{fig:stis_kinematics}, we show the radial velocity, FWHM, and integrated flux for each [O~III] $\lambda$5007 and H$\alpha$ component as a function of position along the slit, where a position of zero corresponds to the galaxy continuum peak in the STIS slit. In general, the [O~III] and H$\alpha$ lines are each fit well by two Gaussian components within $\pm$1\arcsec\ of the continuum peak and one Gaussian component at greater distances. The [O~III] and H$\alpha$ kinematic components in Figure \ref{fig:stis_kinematics} show similar trends (the colors don't always match due to flux sorting and different line ratios in the components), giving us confidence that our automated fitting routine works well. The [O~III] radial velocities are also similar to those in \citet{Ruiz et al.(2001)} who used a more interactive fitting routine on the same dataset.

Looking at the overall velocity pattern for both flux components in the STIS data, the radial velocities peak at about $+$600 km s$^{-1}$ in the SW and $-$600 km s$^{-1}$ in the NE, both at about 0.\arcsec2 (54 pc projected distance) from the center. This symmetry is maintained with a peak radial velocity of about $-$400 km s$^{-1}$ in the SW and $+$300 km s$^{-1}$ in the NE at 0.\arcsec2 $-$ 0.\arcsec25 from the center. The velocities decrease outward to reach zero km s$^{-1}$ at about 1.\arcsec2 from the center. For the single component fits between 1.\arcsec2 and 1.\arcsec5, there is a systematic departure from zero km s$^{-1}$ in the SW and mostly scattered points in the NE that are due to the low SNR in this region. The FWHMs in Figure \ref{fig:stis_kinematics} range from 200 km s$^{-1}$ to our set upper limit of 2000 km s$^{-1}$. Although a few points reach the FWHM limit of 2000 km s$^{-1}$ and should therefore be treated with some caution, they appear to be continuous in flux and $v_r$ with adjacent points and removing the upper limit did not significantly improve the fits.

Although there are significant intrinsic variations in $v_r$ attributable to resolved emission-line knots with their own peculiar velocities \citep{Das et al.(2005), Das et al.(2006)}, \citet{Ruiz et al.(2001)} were able to fit the overall pattern above with a biconical outflow model with the bicone axis inclined towards the Earth by 5\arcdeg\ in the NE and minimum and maximum opening angles of 15\arcdeg\ and 25\arcdeg. \citet{Crenshaw et al.(2010)} argued that the maximum opening angle should be increased to 51\arcdeg\ to encompass the emission-line gas in both the ENLR, but this does not otherwise change the NLR kinematic model of \citet{Ruiz et al.(2001)}.

The radial velocities and FWHMs in Figure \ref{fig:stis_kinematics} peak at much larger values than expected for gas dominated by gravitational motions at these locations \citep{Ruiz et al.(2001), Crenshaw et al.(2010)}. Furthermore, both redshifts and blueshifts are seen in the NLR on each side of the AGN, with amplitudes much larger than those of the rotation component seen in the ENLR (see Section \ref{subsec:enlr_gas_apo}). In fact, we find no clear continuation of the rotation component into the inner regions. 
We confirm that outflow dominates along the linear portion of the backwards ``S''-shaped NLR in Mrk~3 to at least 1.\arcsec2\ (320 pc) from the SMBH.

\bigskip\bigskip

\subsection{NLR Gas Kinematics: {\it Gemini} NIFS} \label{subsec:nlr_gas_nifs}

For the NIFS emission lines, we used the Bayesian fitting routine described in Section \ref{subsec:nlr_gas_stis} to fit multiple kinematic components to each line of interest in each 0.\arcsec05 $\times$ 0.\arcsec05 spaxel. We began by fitting the brightest emission line in the NIFS data, [S~III] 0.9533 \micron\ in the Z band. The best fits required up to three kinematic components near the nucleus and one or two components closer to the edges of the FOV. We also fit the strongest warm molecular line, H$_2$ 2.122 \micron\ in the K band, which required one or two components. Analysis of additional lines in the NIFS data will be discussed in a future paper. 

\begin{figure}[ht!]
%\vspace{-8pt}
\centering
\subfigure{
\includegraphics[scale=0.42]{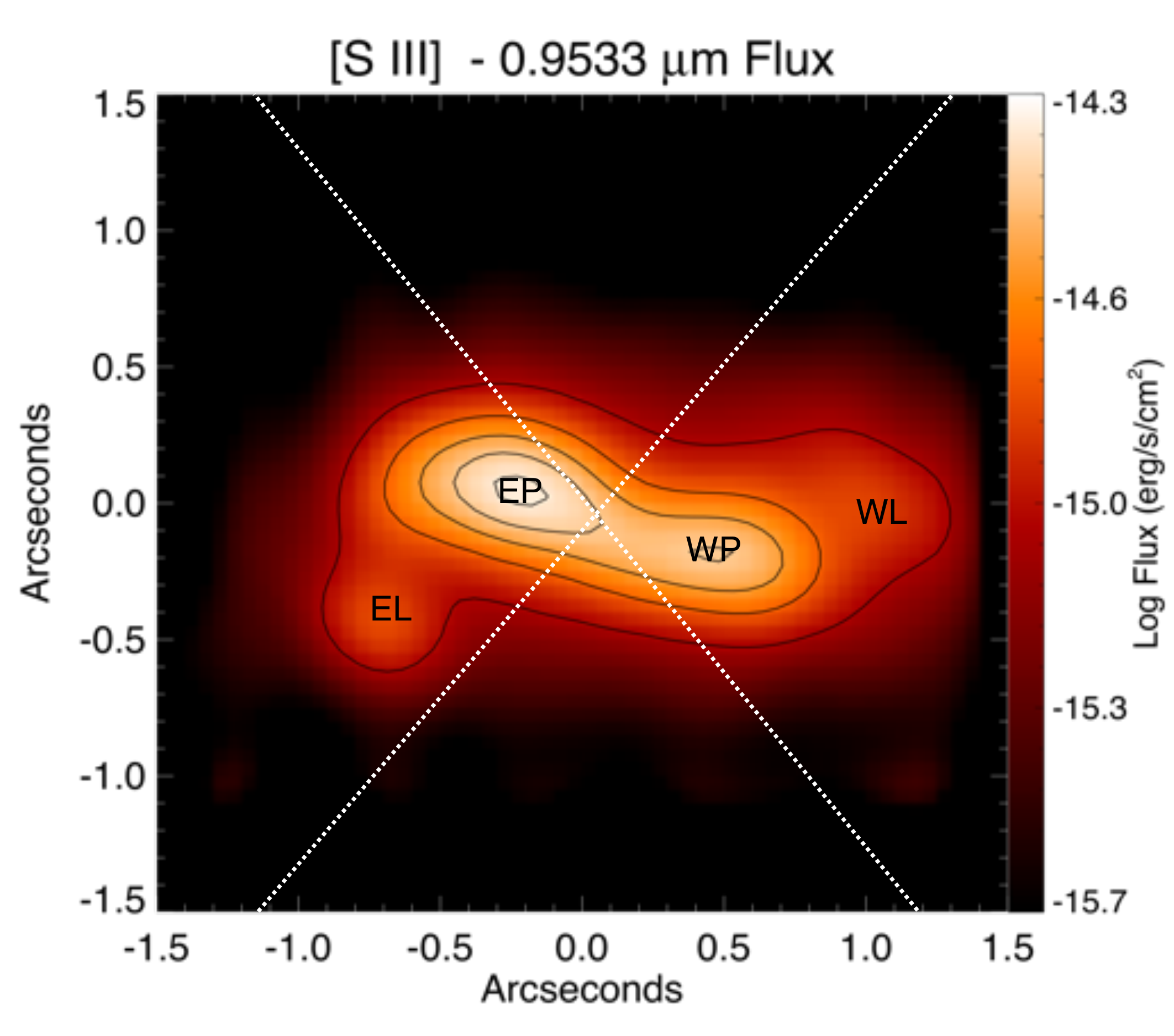}}%\hspace{5ex}
\subfigure{
\includegraphics[scale=0.42]{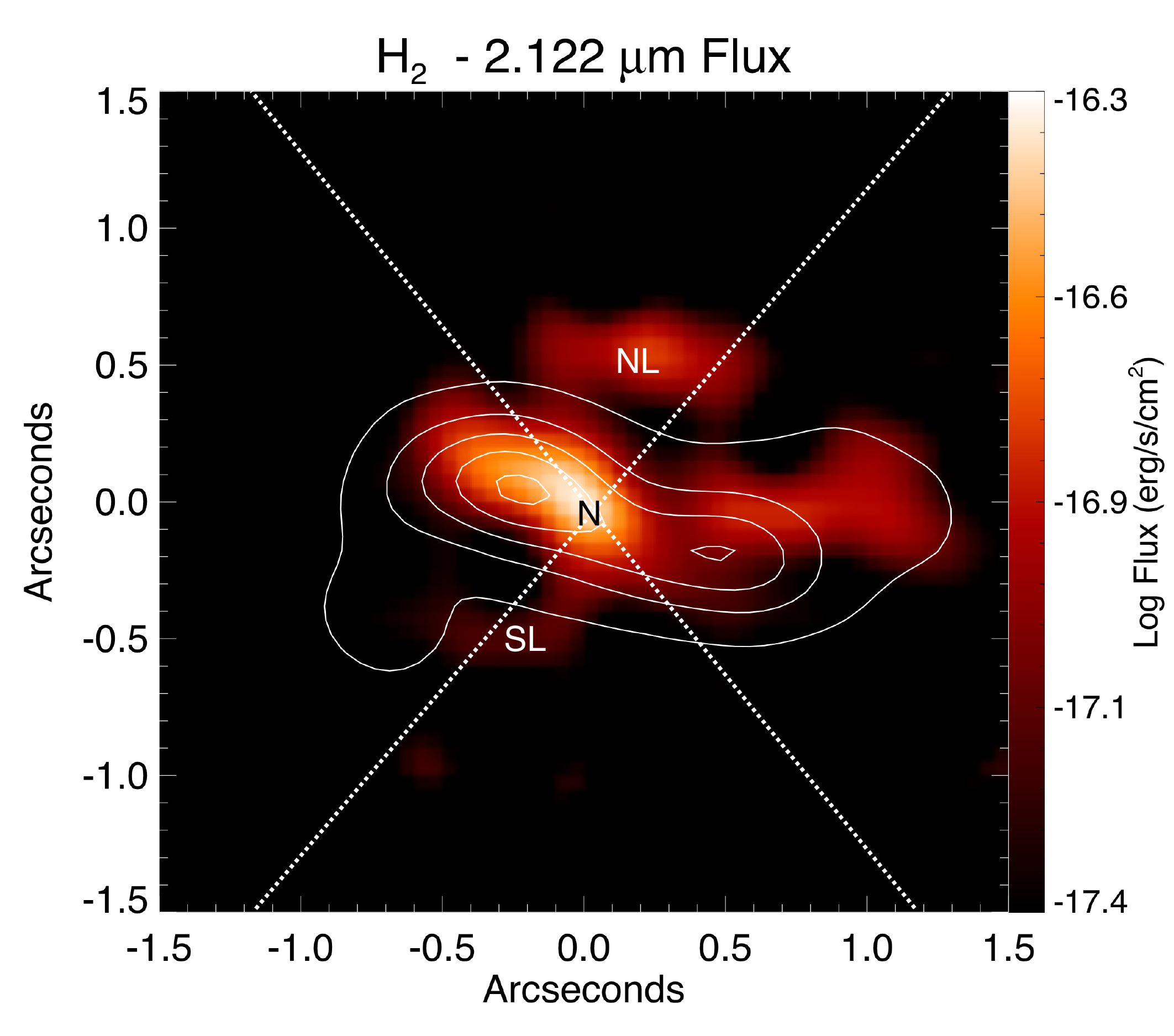}}
\caption{Continuum-subtracted flux maps of the [S~III] and H$_2$ emission lines from NIFS observations of the NLR in Mrk~3. North is up and east is to the left. Contour levels for the [S~III] emission are given in both images beginning at the 3$\sigma$ background level and increasing in steps of 3$\sigma$. Outer edges of the bicone defined by \citet{Crenshaw et al.(2010)} with axis at PA $=$ 89\arcdeg\ are given as dashed white lines. Four areas of interest for discussion are labeled on the [S~III] image: the eastern peak (EP) and western peak (WP) along the linear portion of the NLR, and the eastern lobe (EL) and western lobe (WL) that define the ends of the backwards ``S''- shaped NLR. Areas of interest in the H$_2$ image include the nucleus (N), northern lane (NL), and southern land (SL).
\label{fig:nifs_fluxes}}
\end{figure}

Before we examine the kinematic fits, we take a look at the morphology of the emission-line regions. In Figure \ref{fig:nifs_fluxes}, we show images of the [S~III] and H$_2$ emission lines extracted from the NIFS data cubes over their full wavelength extents, with contours of the [S~III] emission superimposed on both. We also show the outer edges of the AGN ionization bicone with an axis at PA $=$ 89\arcdeg, defined by \citet{Crenshaw et al.(2010)} to encompass the [O~III] emission seen in {\it HST} images of the NLR in Mrk~3 \citep{Capetti et al.(1995), Capetti et al.(1996), Capetti et al.(1999), Kukula et al.(1999), Schmitt and Kinney(1996), Schmitt et al.(2003)}.

As shown in Figure \ref{fig:nifs_fluxes}, the [S~III] emission shows the same backwards ``S'' shape as that of [O~III], consisting of a nearly linear E-W portion centered on the central continuum source and extending over $\sim$1.\arcsec5, a distinct eastern lobe (EL), and a more diffuse western lobe (WL), albeit with lower angular resolution than {\it HST} due to the lower Strehl ratio of the {\it Gemini} observations. The brightest knot of [S~III] emission (EP) is offset from the continuum peak by 0.\arcsec25 to the NE and the second brightest (WP) is 0.\arcsec4\ to the west.

\begin{figure}[ht!]
%\vspace{-8pt}
\centering
\subfigure{
\includegraphics[scale=0.53]{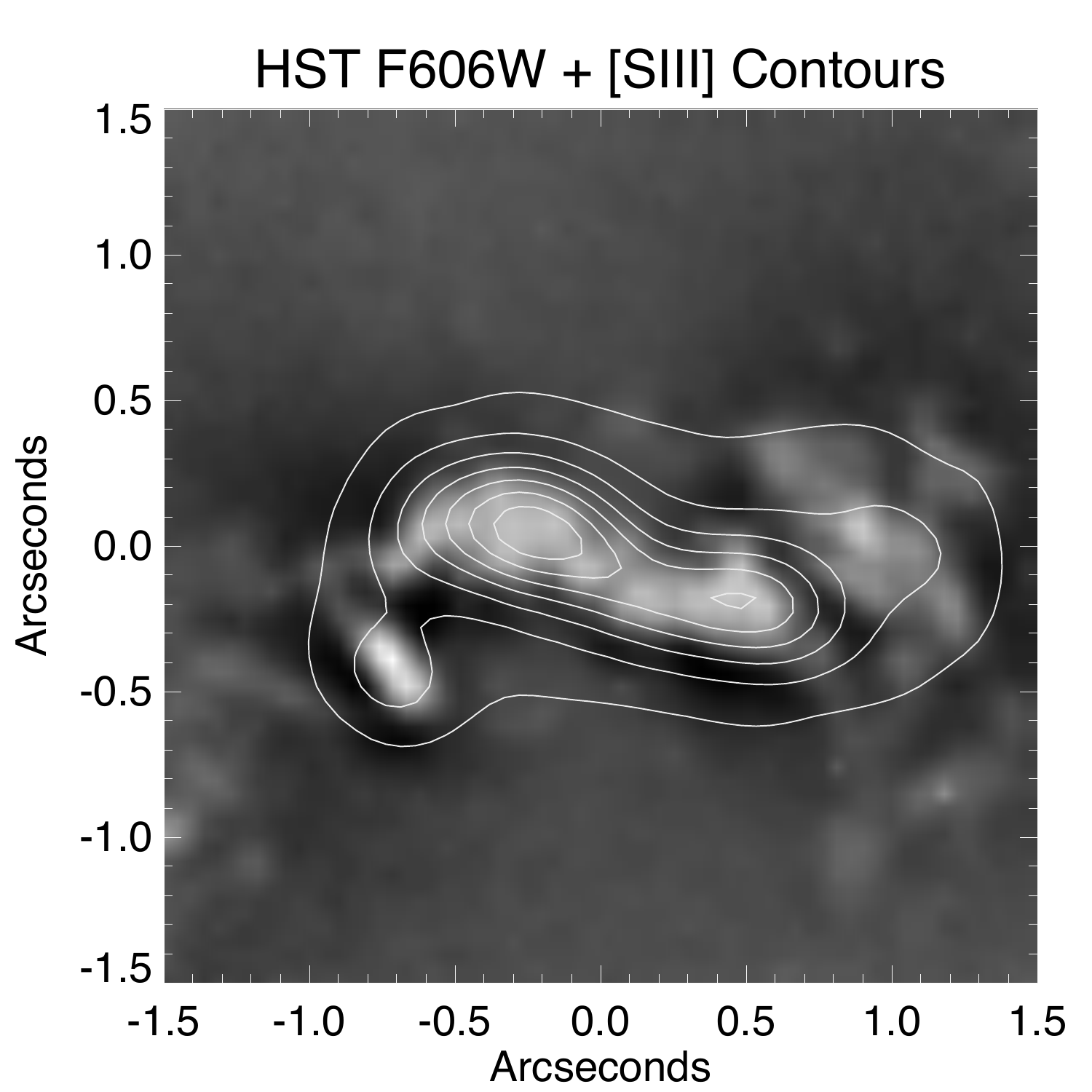}}\hspace{5ex}
\subfigure{
\includegraphics[scale=0.53]{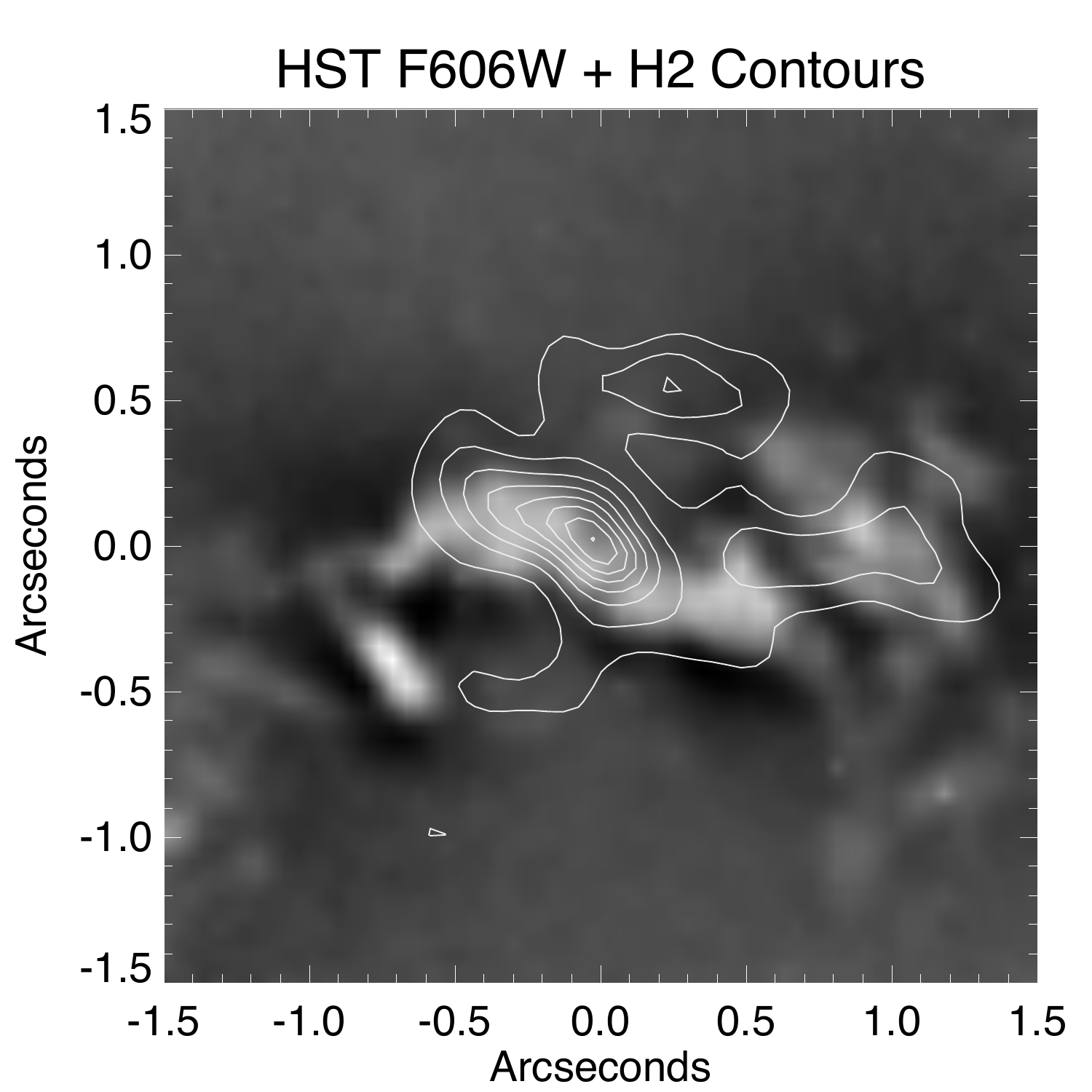}}
\caption{{\it HST} F606W structure map with [S~III] and H$_2$ contours superimposed.
\label{fig:sm_nifs}}
\end{figure}

The H$_2$ flux map shows a significantly different morphology from that of [S~III]. Within the projected bicone, the strong H$_2$ emission matches the overall extent of the linear portion and WL of [S~III] emission, but there is little if any H$_2$ emission in the EL.  Even within the overlapping regions, the small-scale structure differs and the strong lanes of emission are not exactly coincident in the two lines. In contrast to the [S~III] peak, the brightest peak of H$_2$ emission is nearly coincident with the optical continuum peak (or ``N'' for nucleus), and the gap between the [S III] linear feature and western lobe (seen more clearly in Figure \ref{fig:sm_nifs}) contains significant H$_2$ emission. There is also H$_2$ emission outside of the ionization bicone, including strong emission to the NW of center (northern lane, NL) as well as weaker emission to the SE (southern lane, SL), with both appearing to connect to the nucleus. This morphology is similar to that of Mrk~573 \citep{Fischer et al.(2017)} in that there are arcs or lanes of H$_2$ emission in addition to those seen in [S~III] that lie just outside of the nominal bicone. These can be explained by X-ray excitation of H$_2$ from radiation that is filtered by gas closer to the central AGN, as suggested by \citet{Collins et al.(2009)} to explain the strong [O~II] emission from low-ionization gas outside of the nominal bicone in Mrk~3. As with Mrk~573, the H$_2$ arcs appear to curve into the nucleus, possibly indicating the current fueling flow to the AGN.

Figure \ref{fig:sm_nifs} shows the {\it HST} F606W image with the [S~III] and H$_2$ contours superimposed. The left image demonstrates the close correspondence between the [S~III] and optical emission-line structure. Although there is not a one-to-one match between the dust lanes and H$_2$ emission in the right image, as we found in Mrk~573 \citep{Fischer et al.(2017)}, the H$_2$ emission crosses over the gap between the [S~III] linear structure and western lobe (likely a dust lane), and the NL feature NW of center may be a continuation of a dust lane to its west.

\begin{figure}[ht!]
%\vspace{-8pt}
\centering
\subfigure{
\includegraphics[scale=0.33]{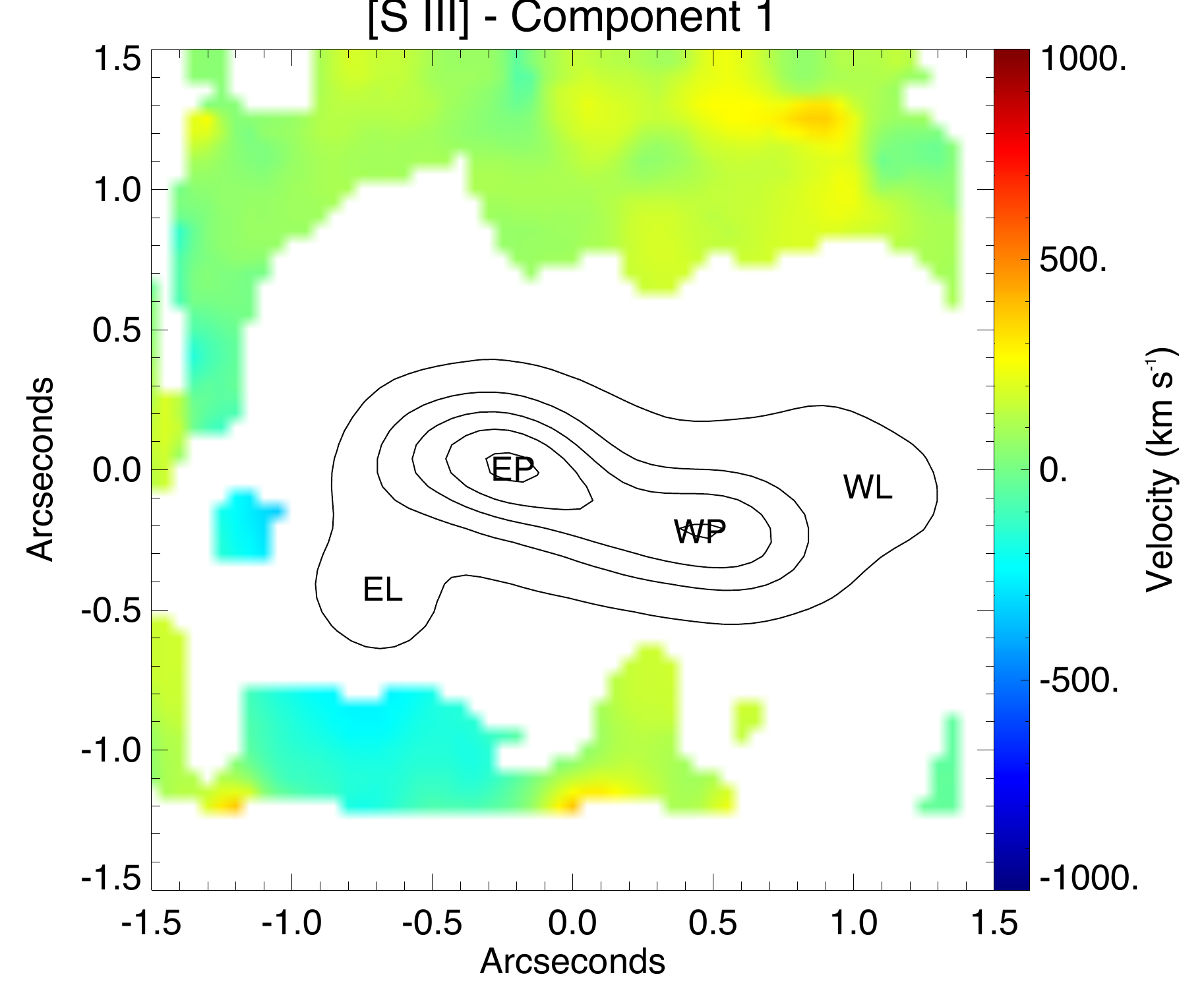}}%\hspace{5ex}
\subfigure{
\includegraphics[scale=0.33]{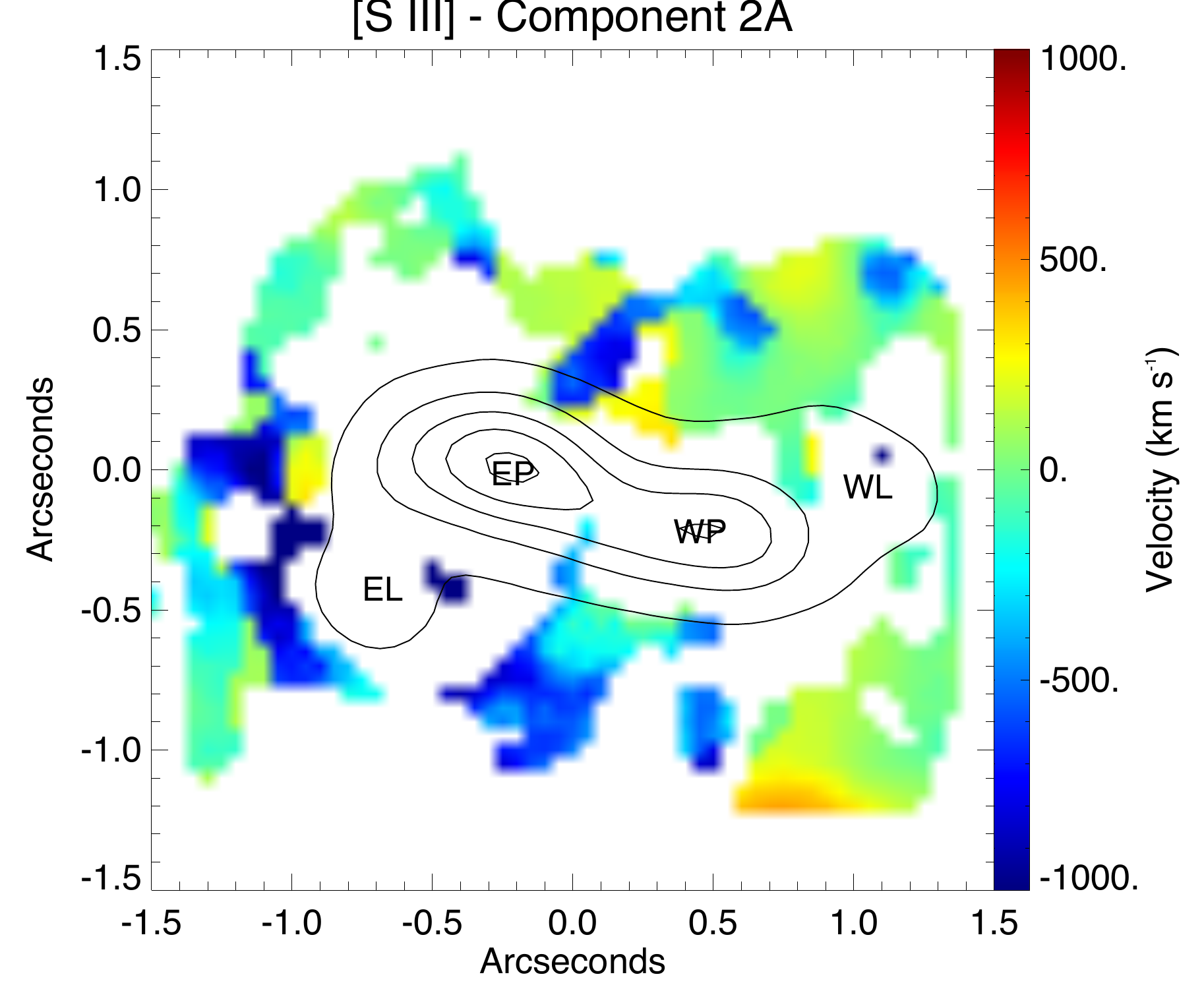}}%\hspace{5ex}
\subfigure{
\includegraphics[scale=0.33]{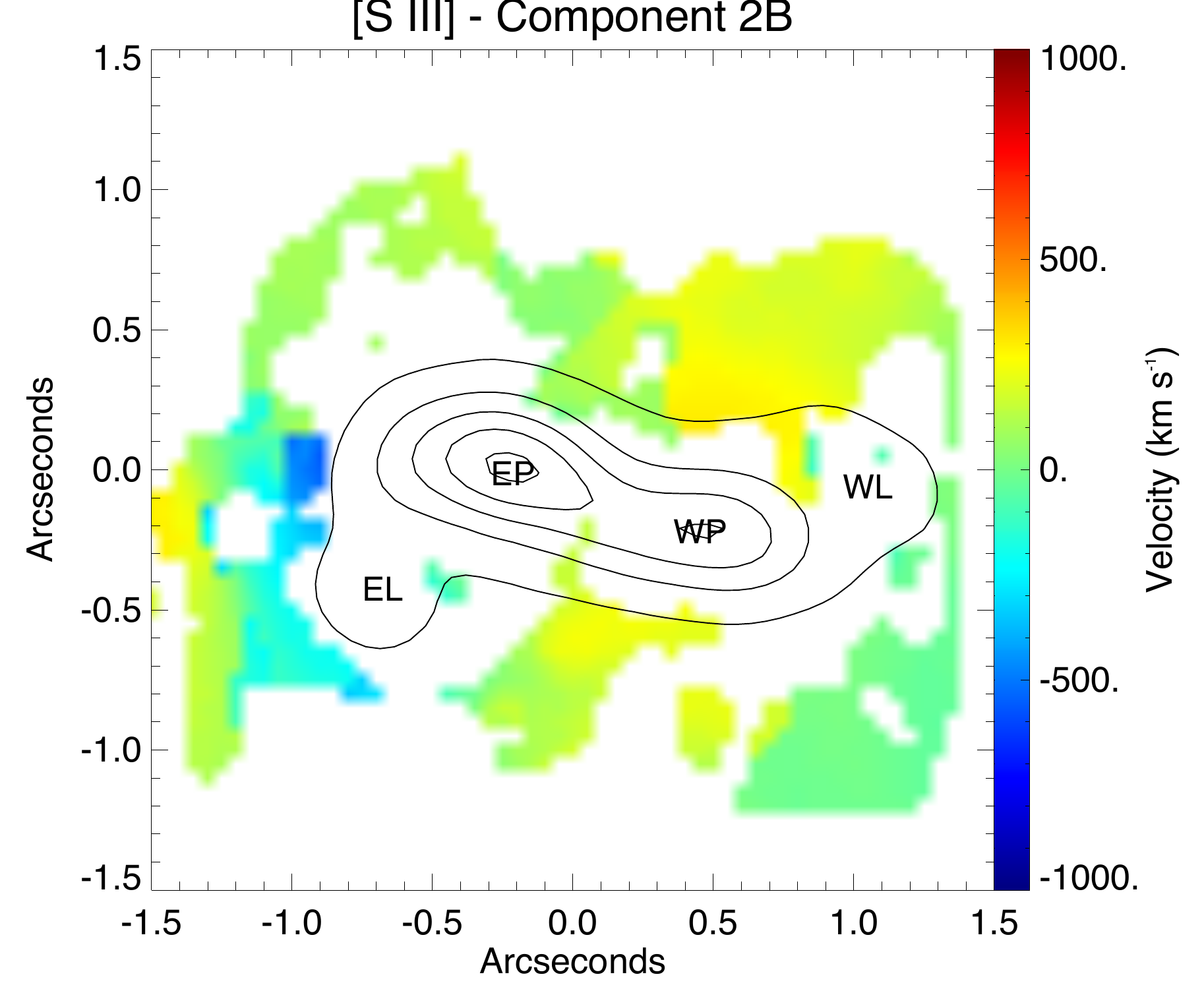}}
\subfigure{
\includegraphics[scale=0.33]{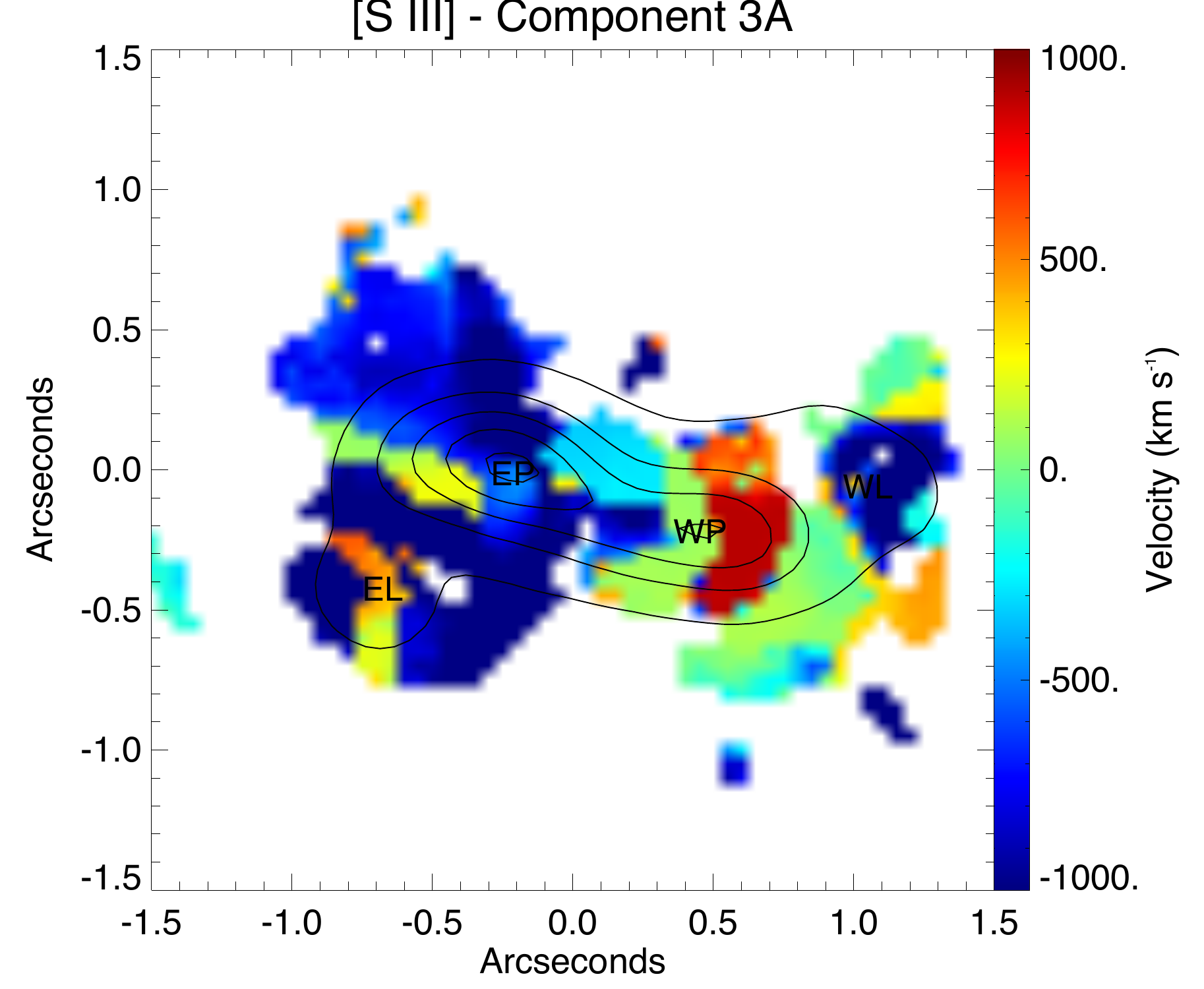}}%\hspace{5ex}
\subfigure{
\includegraphics[scale=0.33]{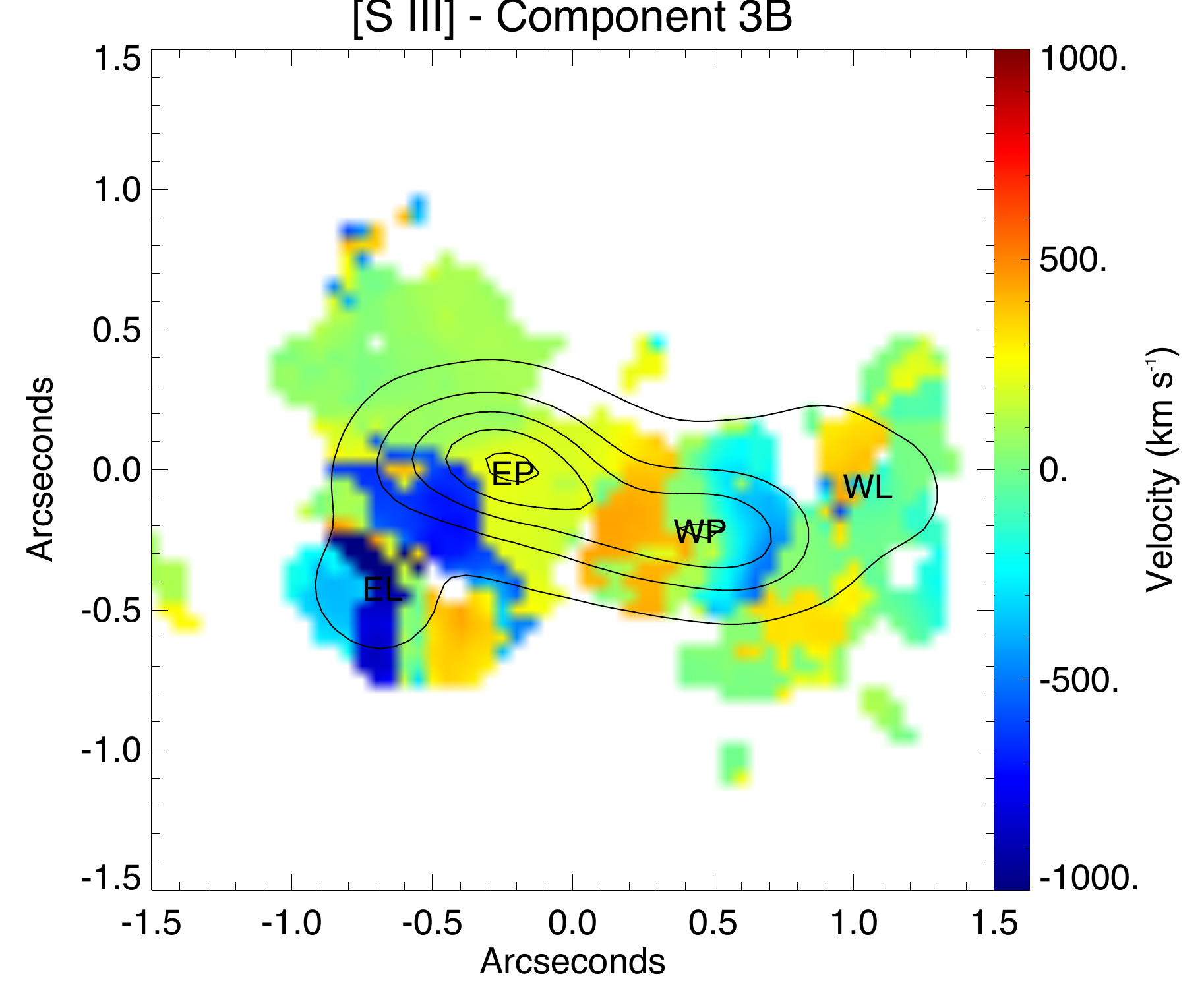}}%\hspace{5ex}
\subfigure{
\includegraphics[scale=0.33]{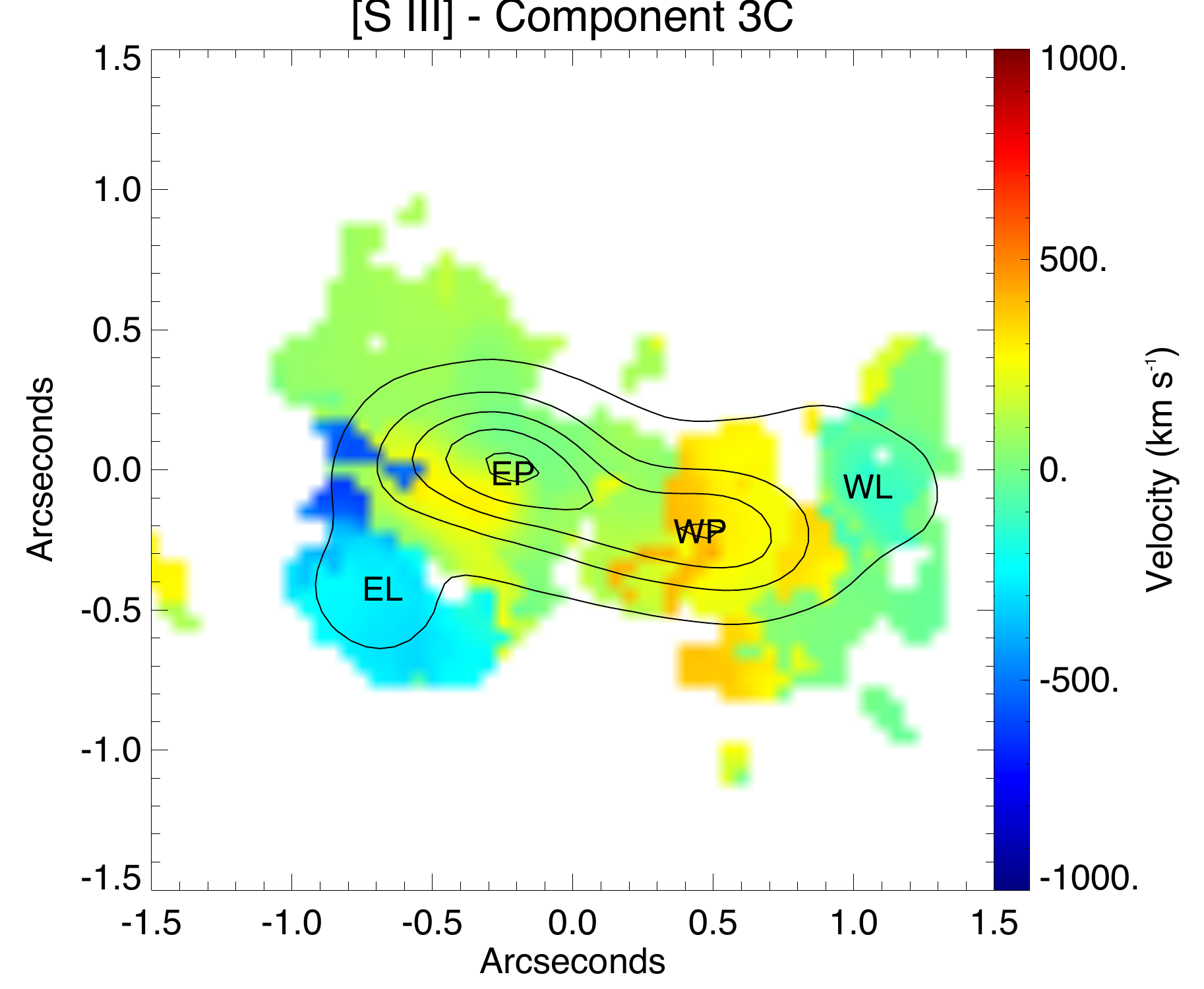}}
\subfigure{
\includegraphics[scale=0.33]{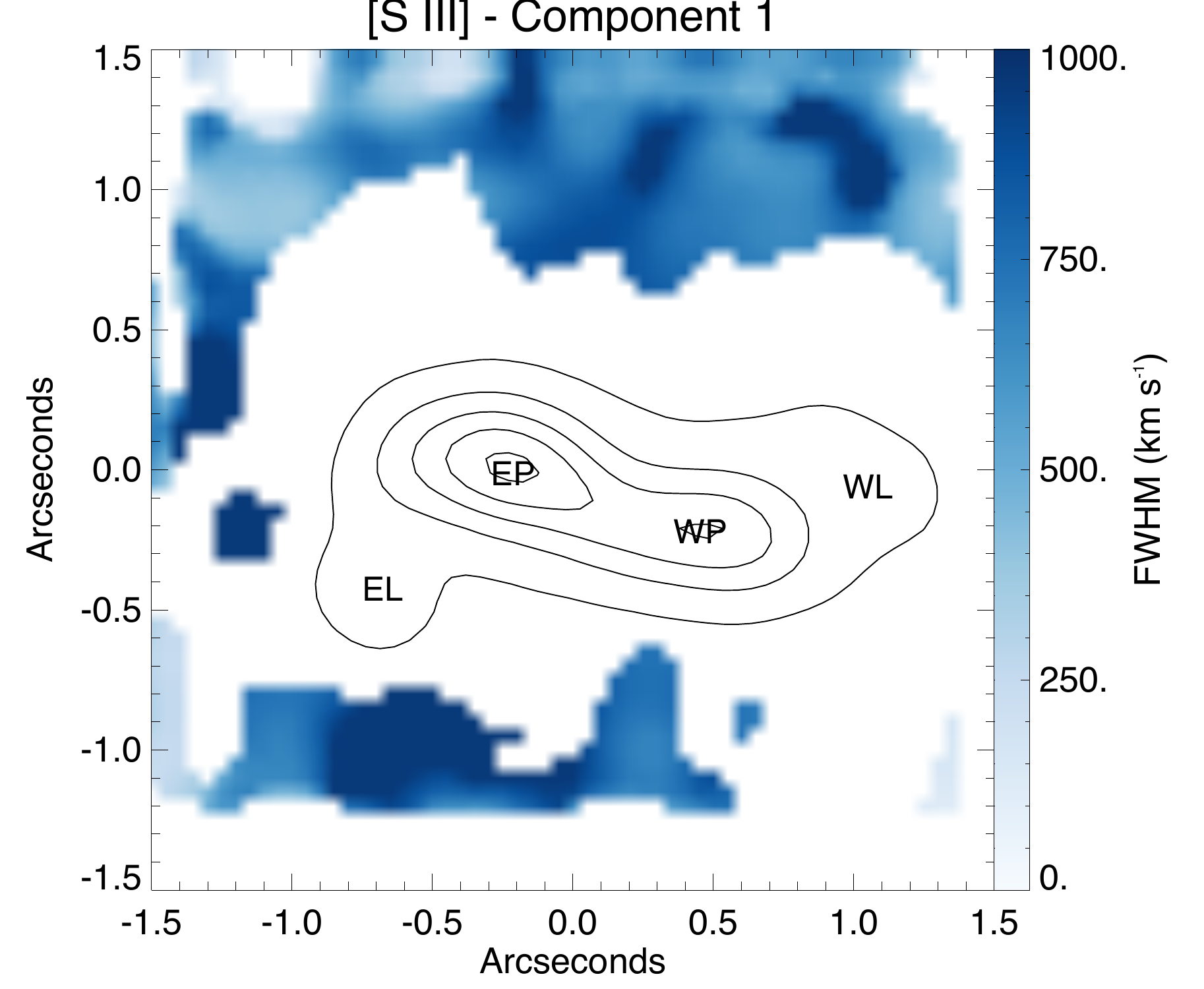}}%\hspace{5ex}
\subfigure{
\includegraphics[scale=0.33]{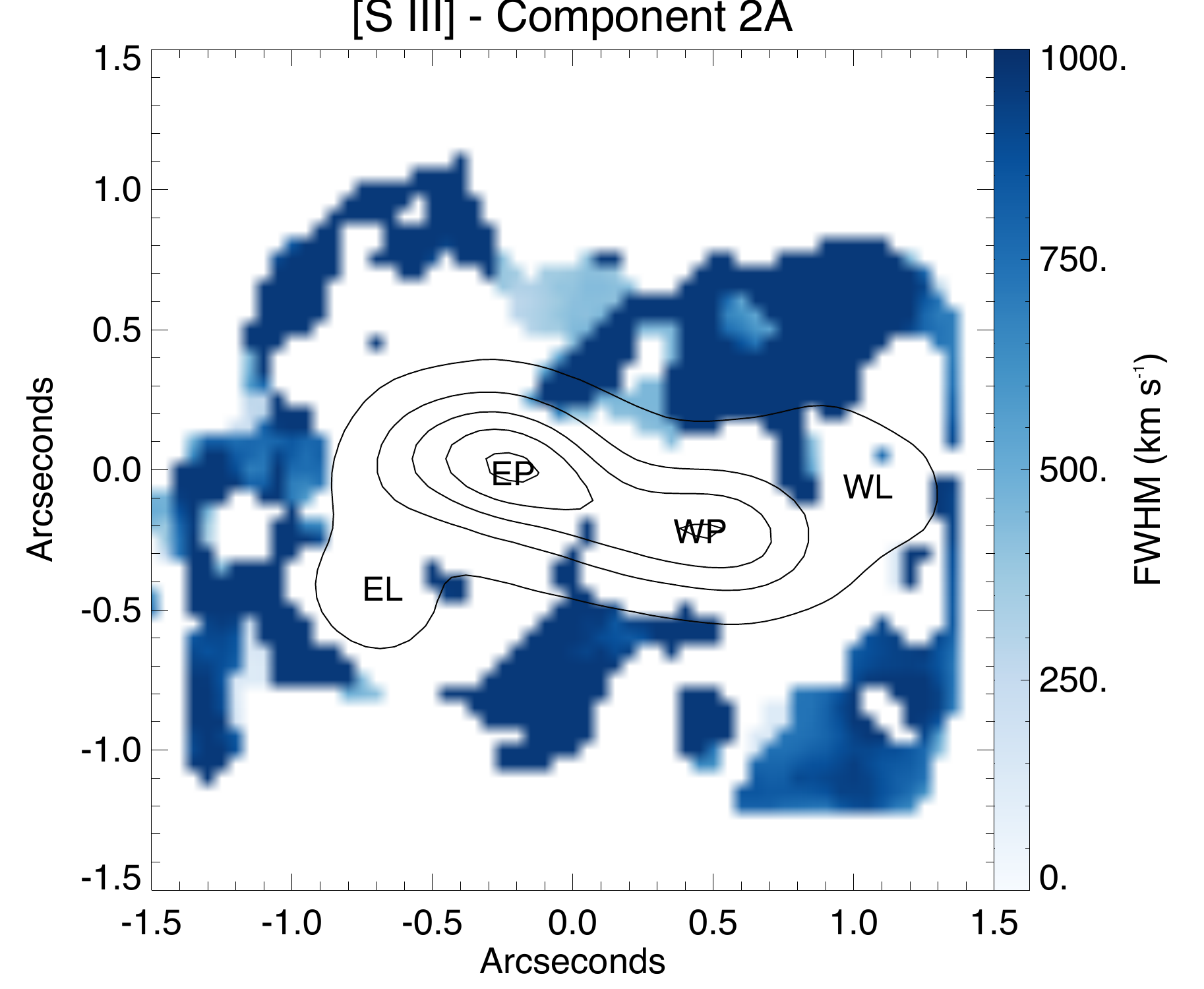}}%\hspace{5ex}
\subfigure{
\includegraphics[scale=0.33]{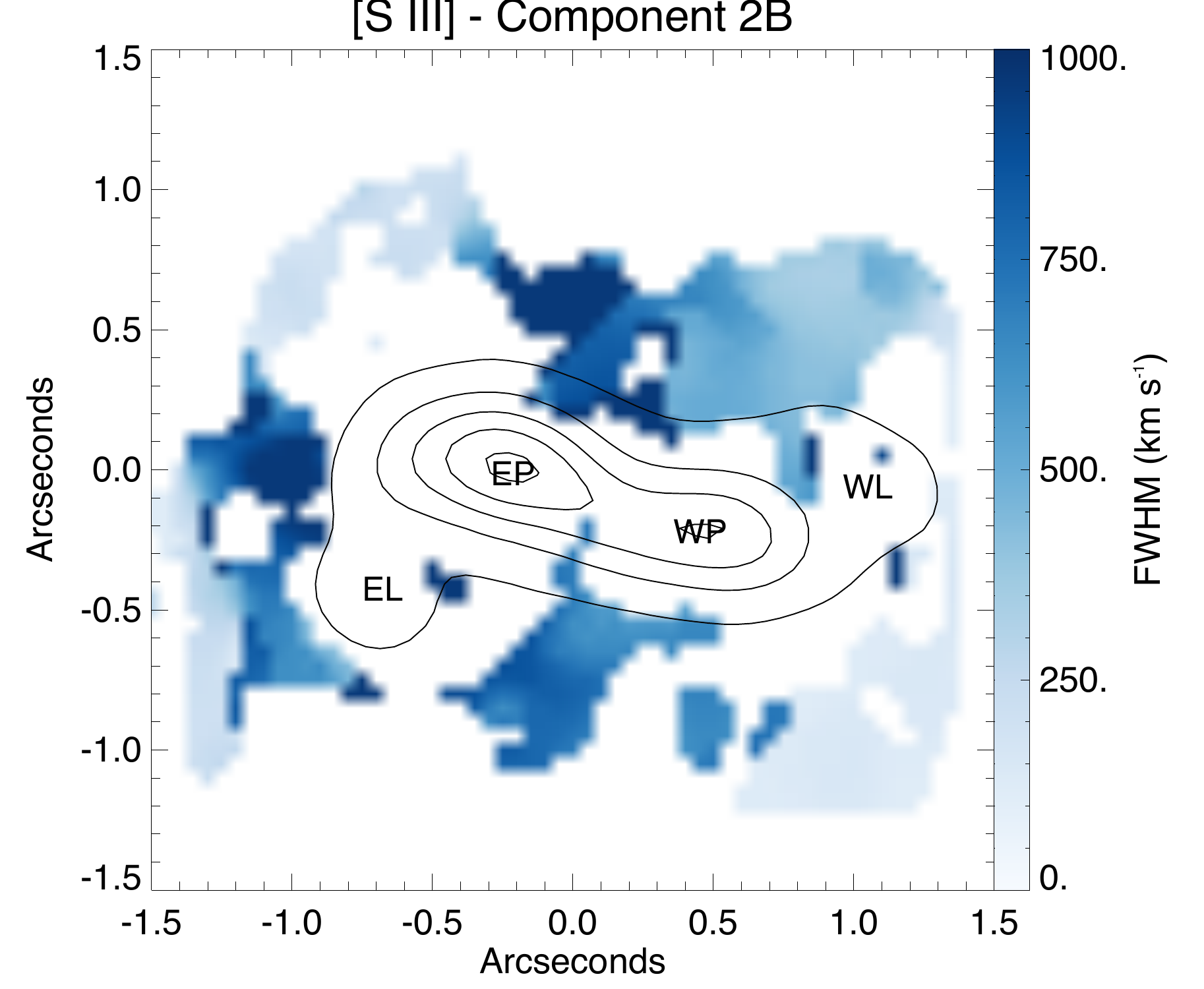}}
\subfigure{
\includegraphics[scale=0.33]{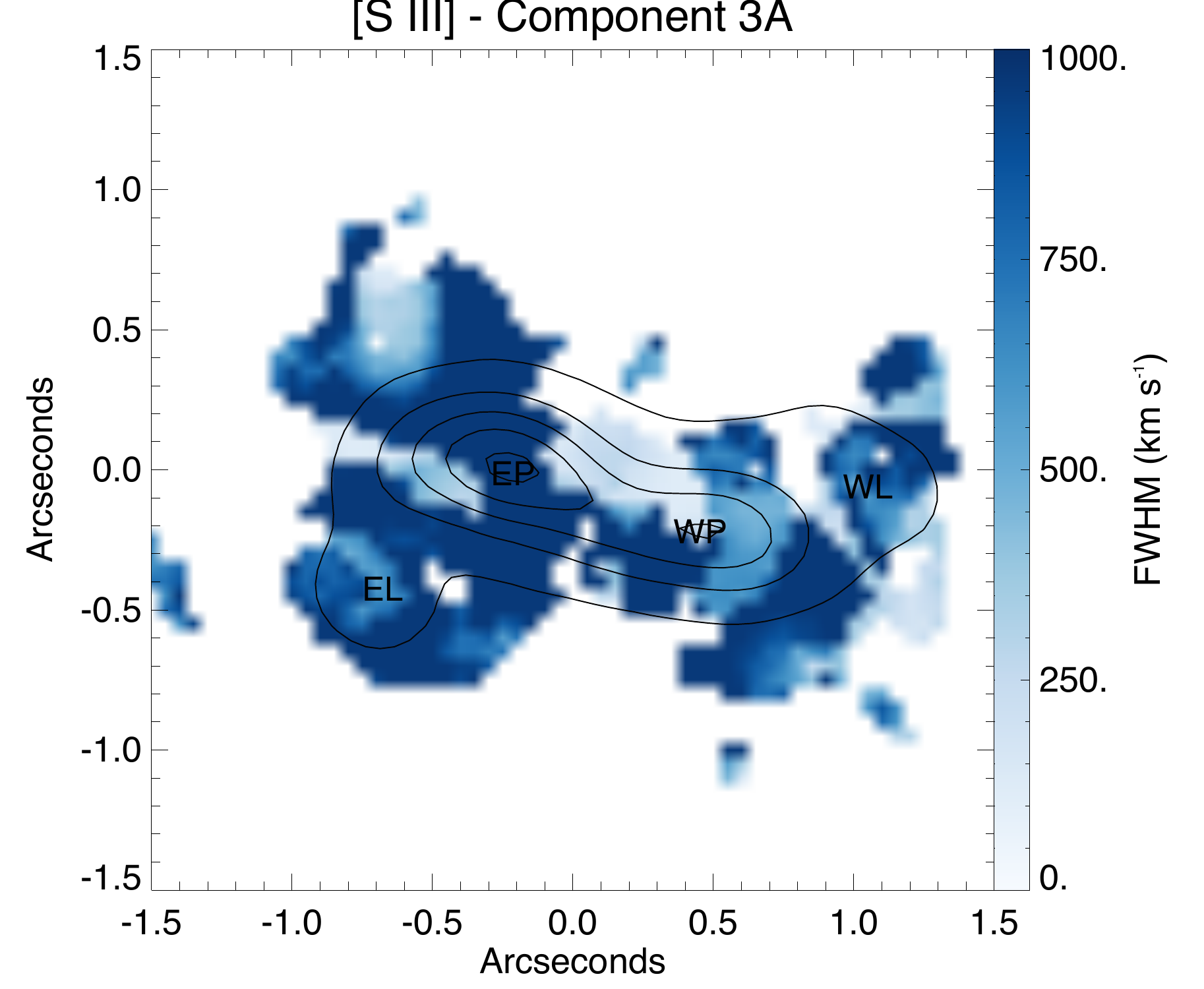}}%\hspace{5ex}
\subfigure{
\includegraphics[scale=0.33]{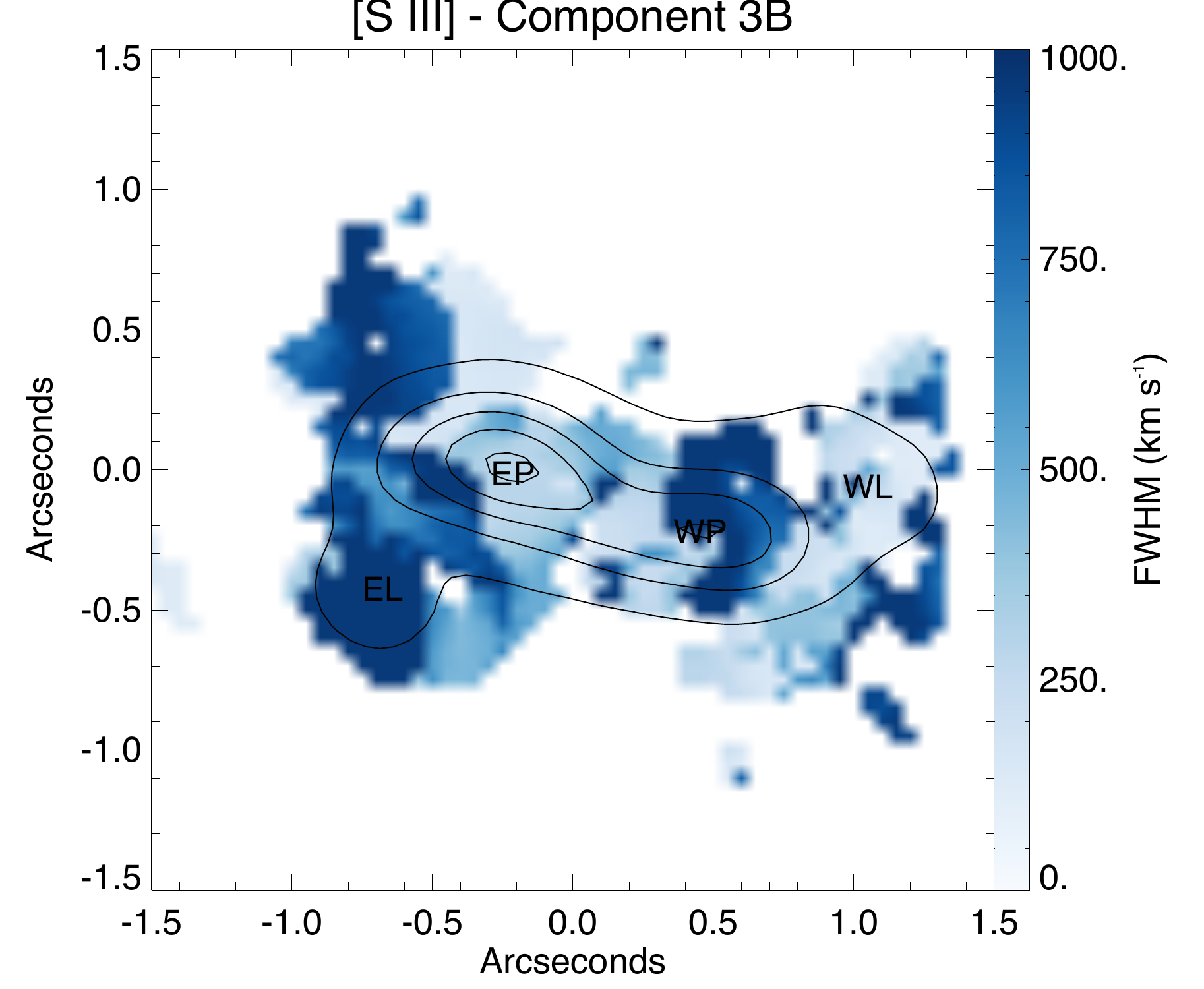}}%\hspace{5ex}
\subfigure{
\includegraphics[scale=0.33]{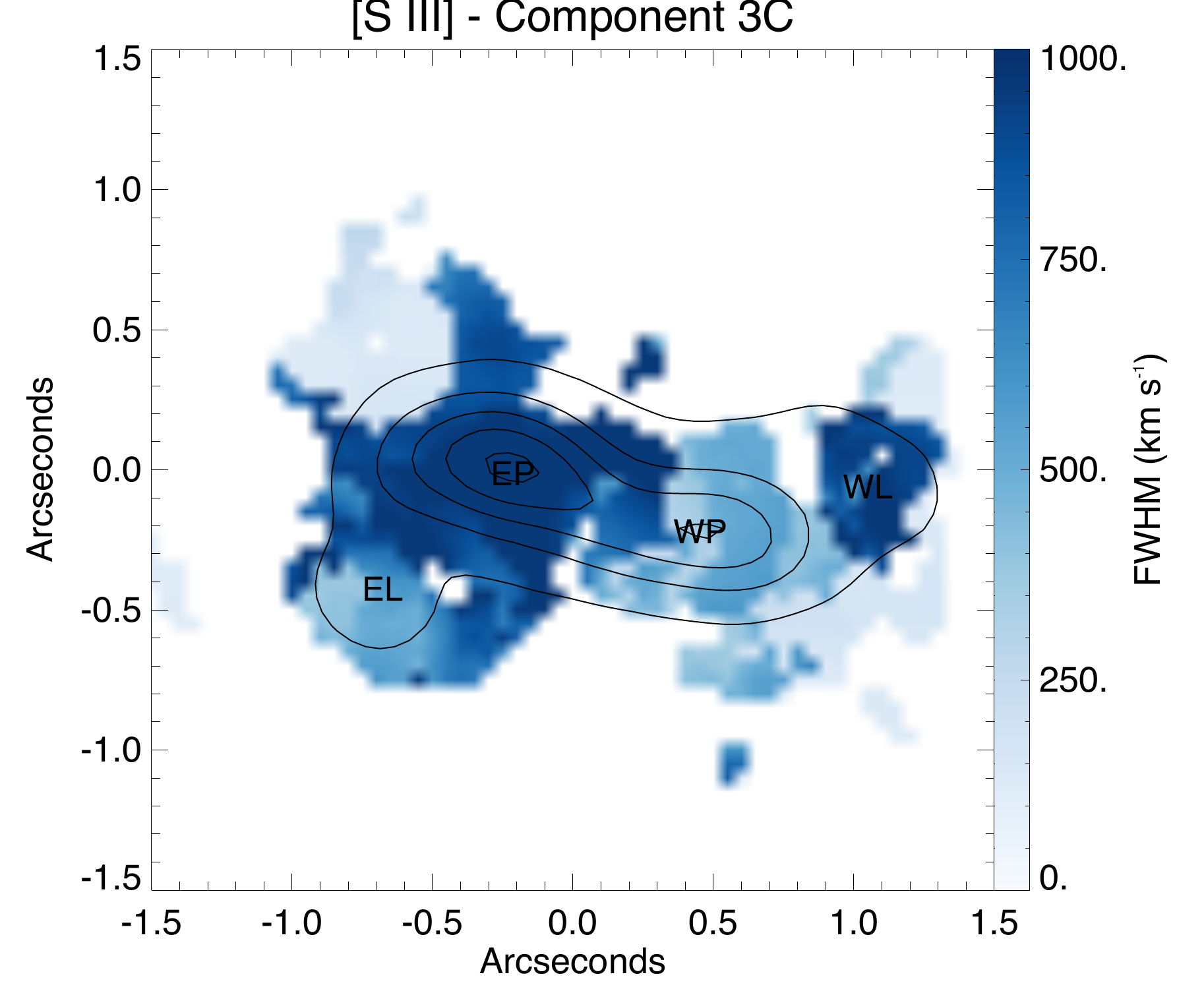}}
\caption{Kinematic maps from multi-component Gaussian fits to the [S~III] emission line, separated by number of components (1 -- 3) and relative peak flux decreasing from A to C: top -- radial velocity, bottom -- FWHM. [S~III] flux contours are superimposed. North is up and east is to the left. The maximum uncertainties in radial velocity and FWHM are $\pm$50 km s$^{-1}$.
\label{fig:nifs_siii_kinematics}}
\end{figure}

In Figure \ref{fig:nifs_siii_kinematics}, we show the radial velocities from the multi-component fits. First, we consider the brightest [S~III] regions, which all required 3 kinematic (Gaussian) components (Components 3A, 3B, and 3C) for a good fit. The EP along the linear portion of the NLR shows mostly blue-shifted gas with radial velocities around $-$600 km s$^{-1}$ and a weaker component of redshifted gas at about $+$300 km s$^{-1}$, in agreement with the STIS observations of this region. The extended region just to the north of the EP shows even higher blueshifts (up to $-$1500 km s$^{-1}$) as well as emission within $\pm$100 km s$^{-1}$  of systemic (zero) velocity. The WP in the linear portion around 0.\arcsec4 from the central continuum source shows the strongest redshifts in the NIFS observations, between $+$400 and $+$800 km s$^{-1}$), as well as lower velocity blueshifted components, also in reasonable agreement with the STIS observations.

The EL and WL at the ends of the backwards ``S'' have not been previously isolated spectroscopically. The EL shows mostly blueshifted emission with radial velocities in the range $-$300 to $-$1500 km s$^{-1}$ plus small regions of redshifted emission at $+$300 to $+$500 km s$^{-1}$. The WL shows distinct regions of blueshifted, systemic, and redshifted gas spanning $-$1200 to $+$400 km s$^{-1}$. The more extended regions of [S~III] emission with one or two components (1, 2A, 2B) follow the same general trend of mixed positive and negative radial velocities on either side of the nucleus, with blueshifted emission dominating in the east and redshifted emission more prominent in the west. 

As shown in in Figure \ref{fig:nifs_siii_kinematics}, the FWHMs of the emission-line components extends over a large range after correction for the LSF, from close to zero to 1000 km s$^{-1}$, and tend to decrease somewhat at the extremes of the NIFS FOV. Regions with high radial velocities ($| v_r | \geq$ 500 km s$^{-1}$) , particularly the blueshifted ones in the east, tend to have high FWHM ($\geq$ 500 km s$^{-1}$), although there are a few regions with low $| v_r |$ and high FWHM. We detect regions of low FWHM and low $| v_r |$ to the NE and SW of the nucleus, which we will focus on later in this section. 

\begin{figure}[ht!]
%\vspace{-8pt}
\centering
\subfigure{
\includegraphics[scale=0.33]{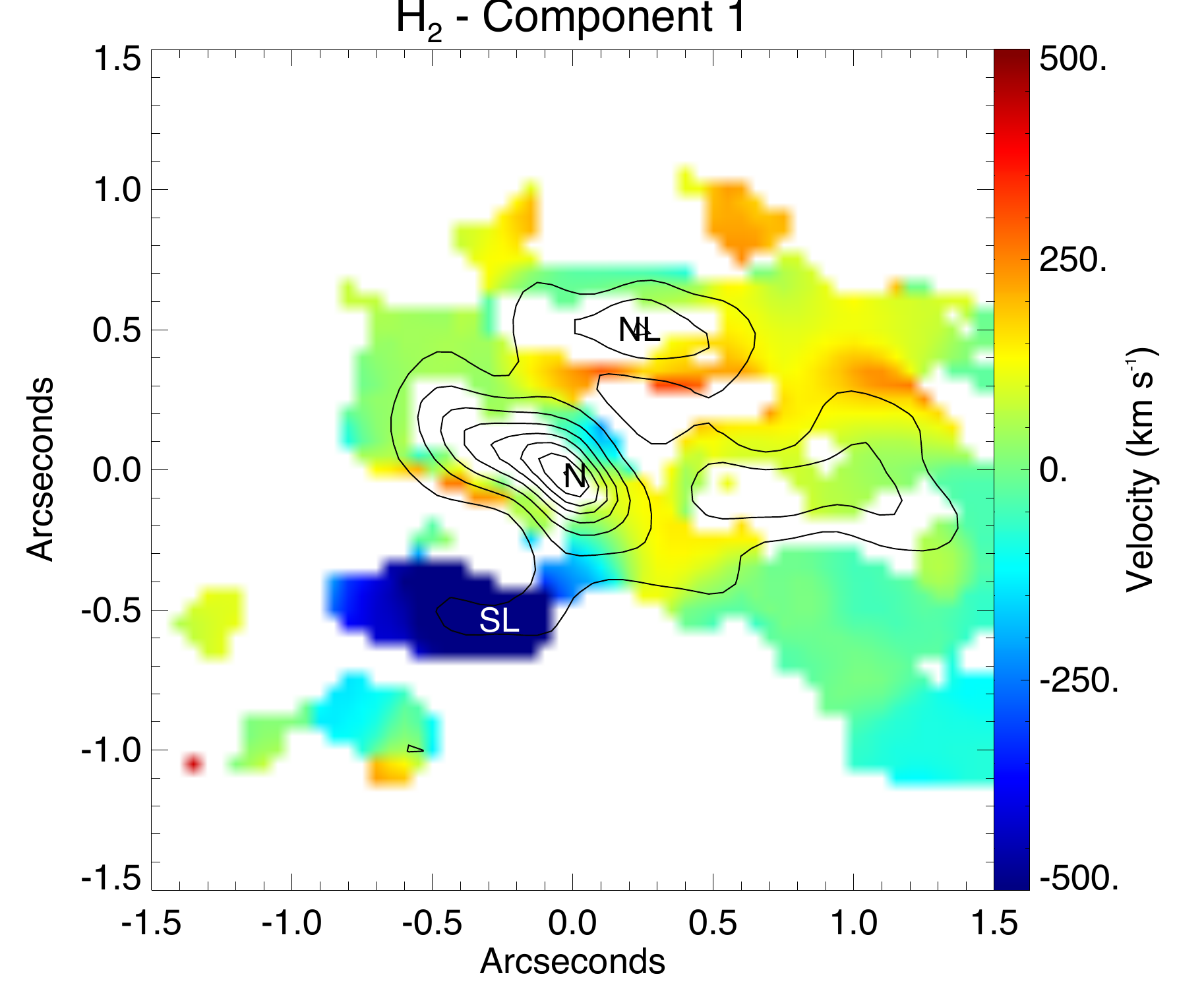}}%\hspace{5ex}
\subfigure{
\includegraphics[scale=0.33]{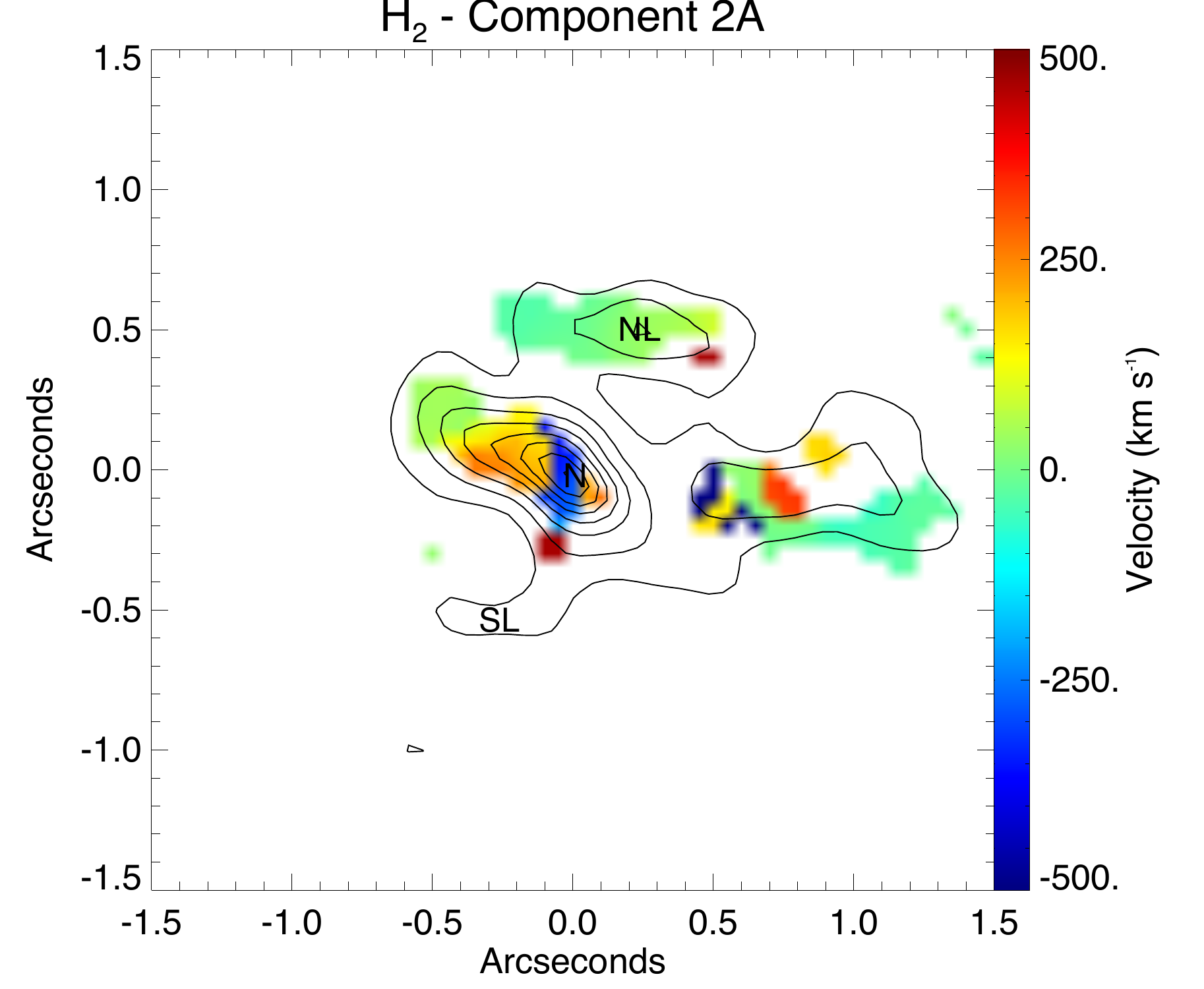}}%\hspace{5ex}
\subfigure{
\includegraphics[scale=0.33]{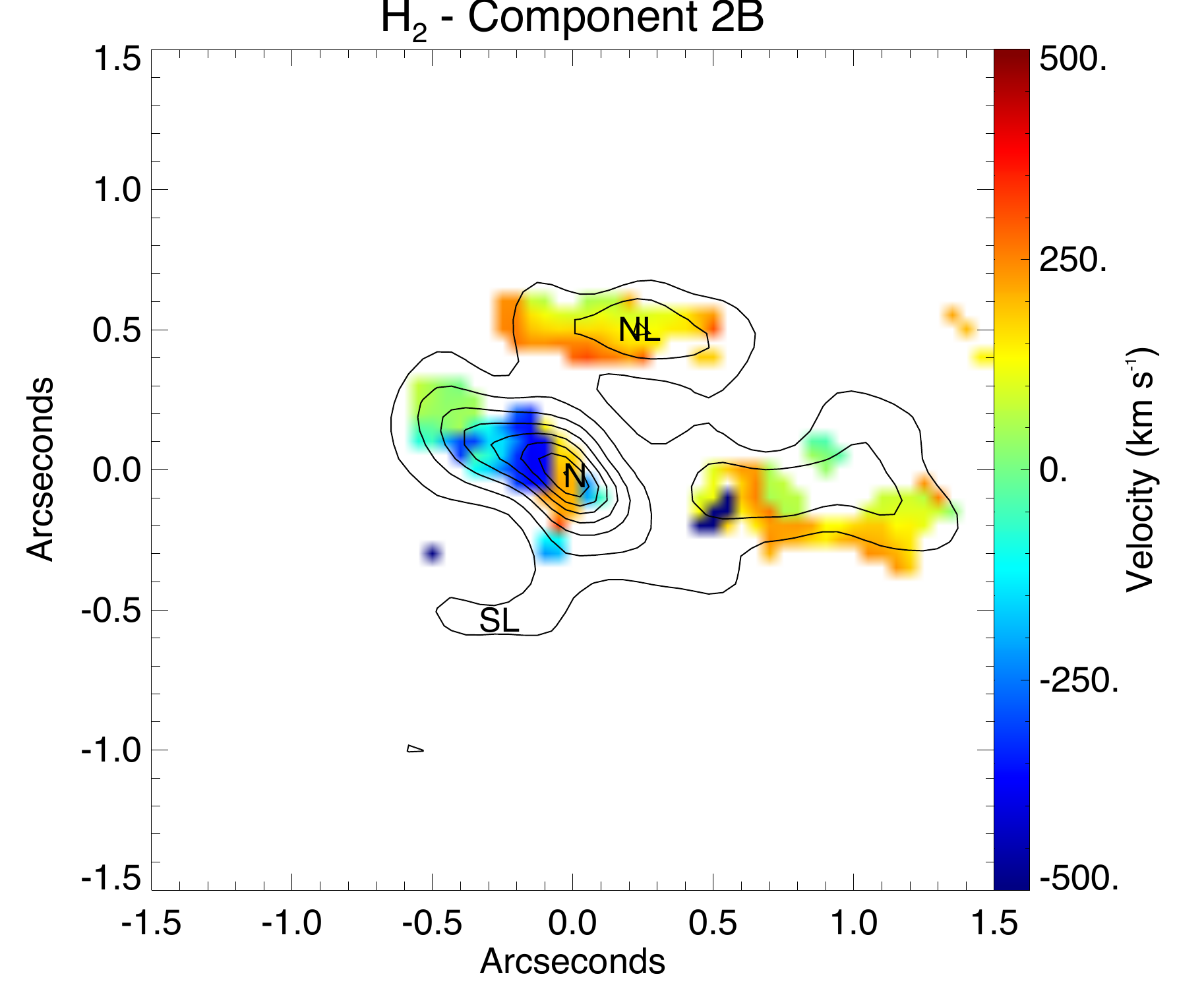}}
\subfigure{
\includegraphics[scale=0.33]{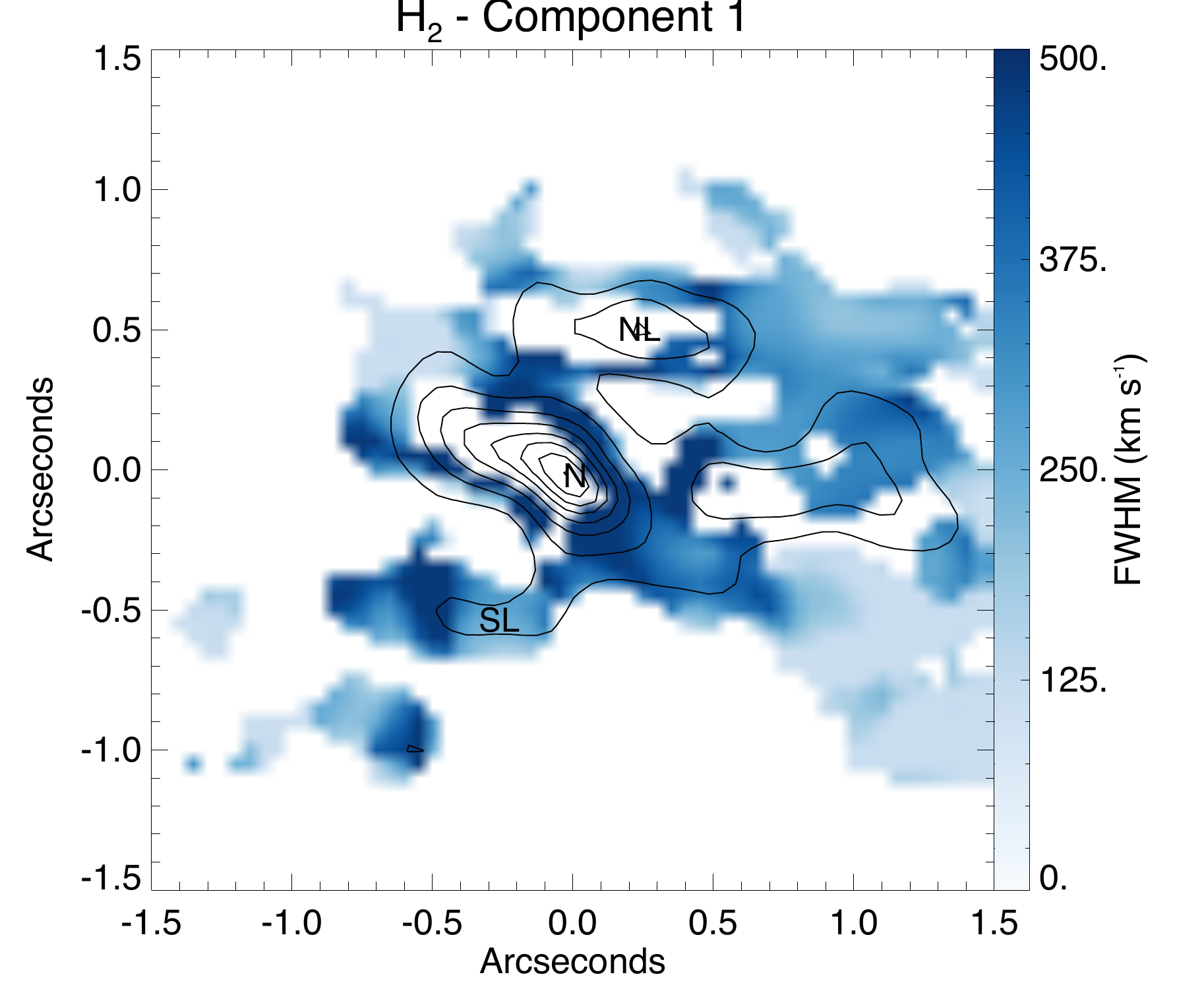}}%\hspace{5ex}
\subfigure{
\includegraphics[scale=0.33]{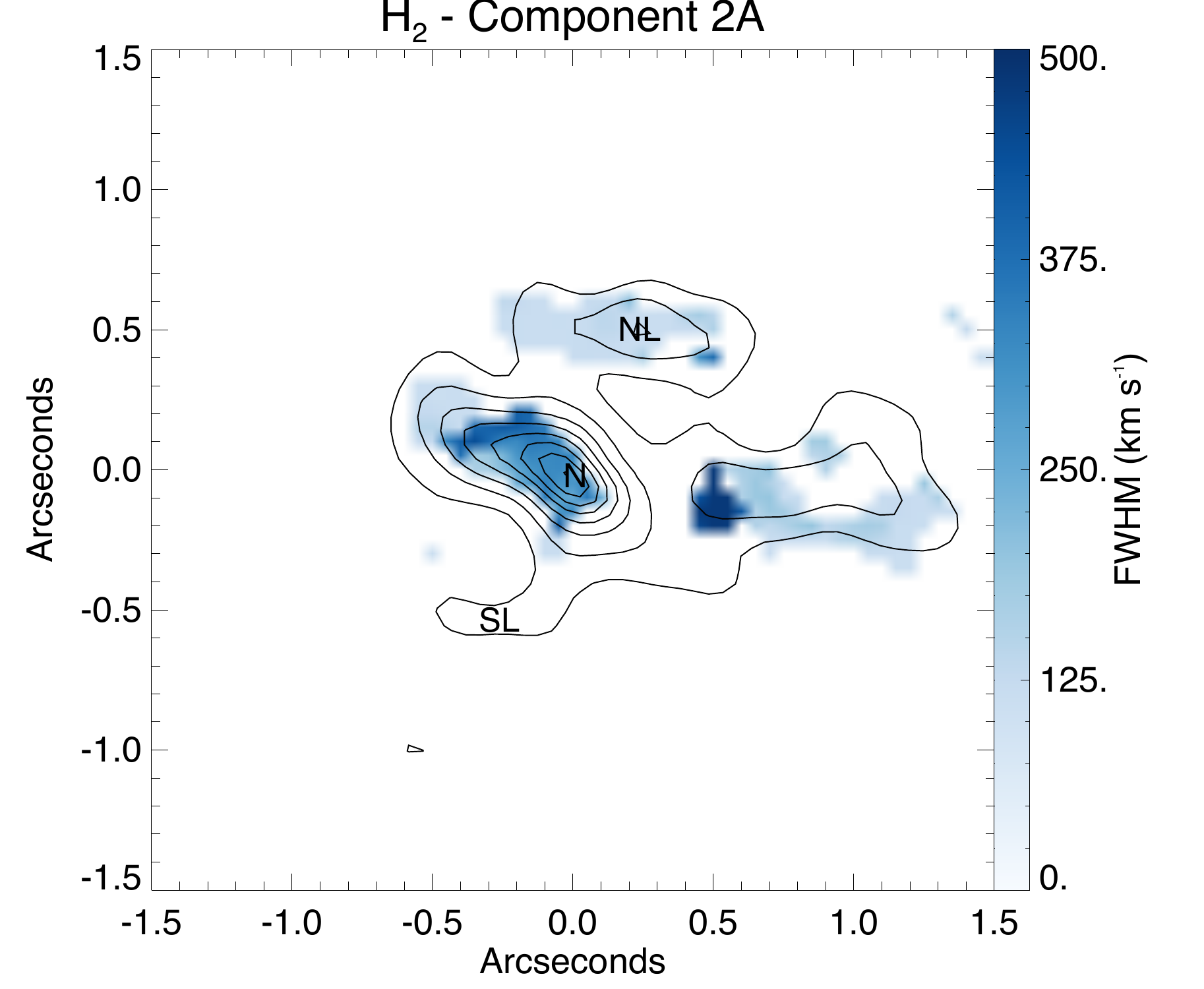}}%\hspace{5ex}
\subfigure{
\includegraphics[scale=0.33]{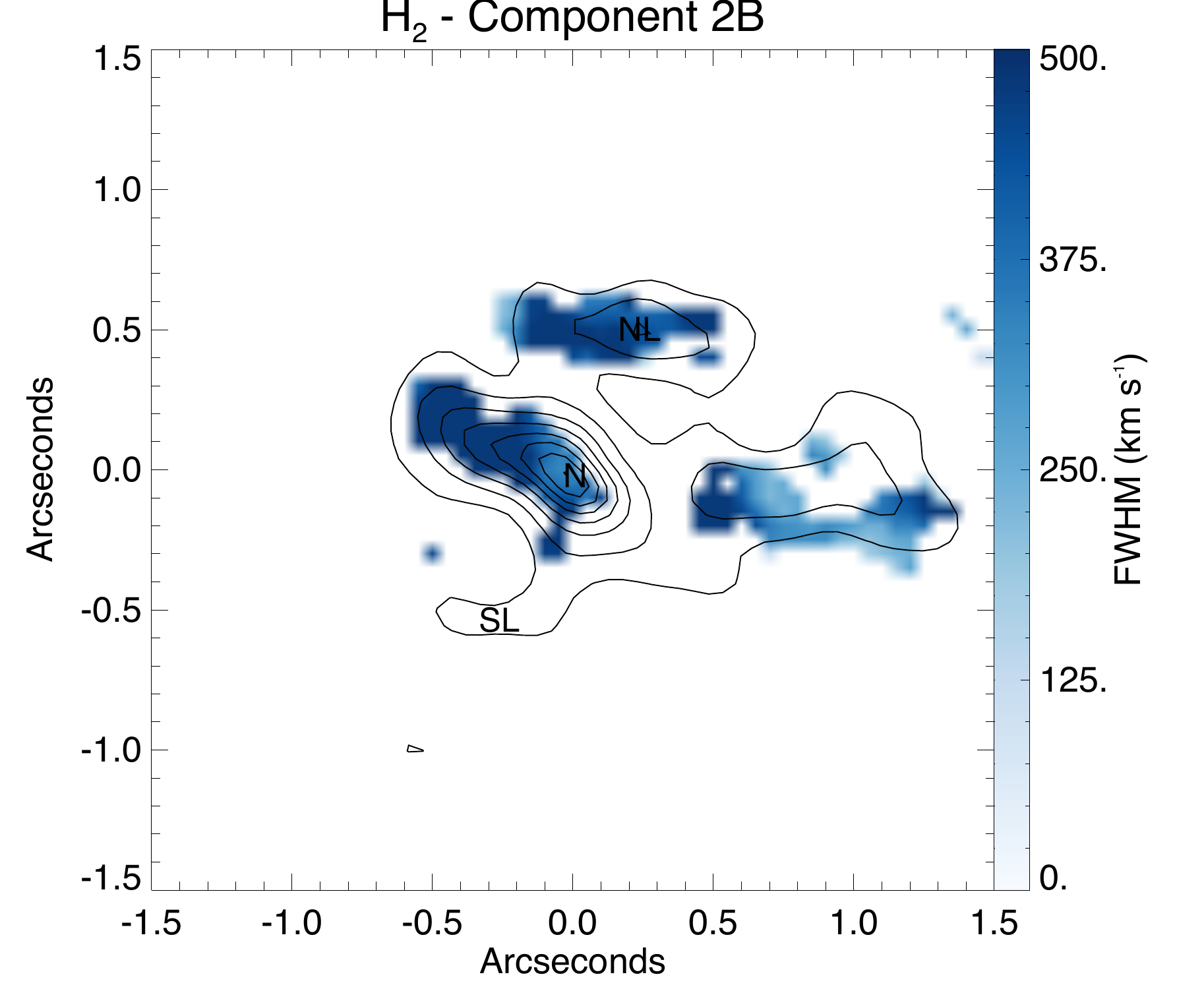}}
\caption{Kinematic maps from multi-component Gaussian fits to the H$_2$ emission line, separated by number of components (1 --2) and relative peak flux decreasing from A to B: top -- radial velocity, bottom -- FWHM. H$_2$ flux contours are superimposed. North is up and east is to the left.
\label{fig:nifs_h2_kinematics}}
\end{figure}

In Figure \ref{fig:nifs_h2_kinematics}, we show the single- and double-component fits to the kinematics of the H$_2$ 2.122 \micron\ emission line. Overall, the magnitudes of the radial velocities are lower than those for [S~III]. The south lane (SL) shows the highest values, with single components of blueshifted emission around $-$600 km s$^{-1}$. This region is just outside of the nominal bicone, coincident with [S~III] emission at $-$1500 km s$^{-1}$, and adjacent to the [S~III] EL, which is mostly blueshifted as well, but does not have detected H$_2$ emission. The north lane (NL) consists of two kinematic components: one is near zero (systemic) and one is redshifted up to $+$350 km s$^{-1}$. It is also just outside of the nominal bicone, coincident with redshifted emission with velocities up to $+$800 km s$^{-1}$), and adjacent to strong systemic and redshifted emission in the [S~III] WL. Similar to the [S~III] kinematics, the nucleus (N) shows strong blueshifted and redshifted components of H$_2$ emission, albeit at smaller amplitudes of $-$300 to $+$300 km s$^{-1}$.

The FWHMs of the H$_2$ components ranges from near zero to $\sim$500 km s$^{-1}$, which is a broad range but with a lower maximum than the [S~III] FWHMs, reminiscent of the observed decrease in FWHM of the narrow-emission lines as a function of ionization potential in Mrk~3 and a number of other AGN \citep{De Robertis and Osterbrock(1986)}. As with [S~III], high $| v_r |$ regions all show high FWHM but some low $| v_r |$ regions show high FWHM as well. The high $| v_r |$ and FWHM values of the warm H$_2$ compared to the stellar values shown in Figure \ref{fig:nifs_stellar} and the coincidence with [S~III] emission with similar blueshifts or redshifts, with the latter often at somewhat higher radial velocities, indicate that most of the H$_2$ emission within $\pm$1.\arcsec5 of the nucleus in Mrk~3 is outflowing. However, we also detect regions of low FWHM and low $| v_r |$ in the single H$_2$ component map to the NE and SW of the nucleus, similar to those seen in [S~III].

\begin{figure}[ht!]
%\vspace{-8pt}
\centering
\subfigure{
\includegraphics[scale=0.45]{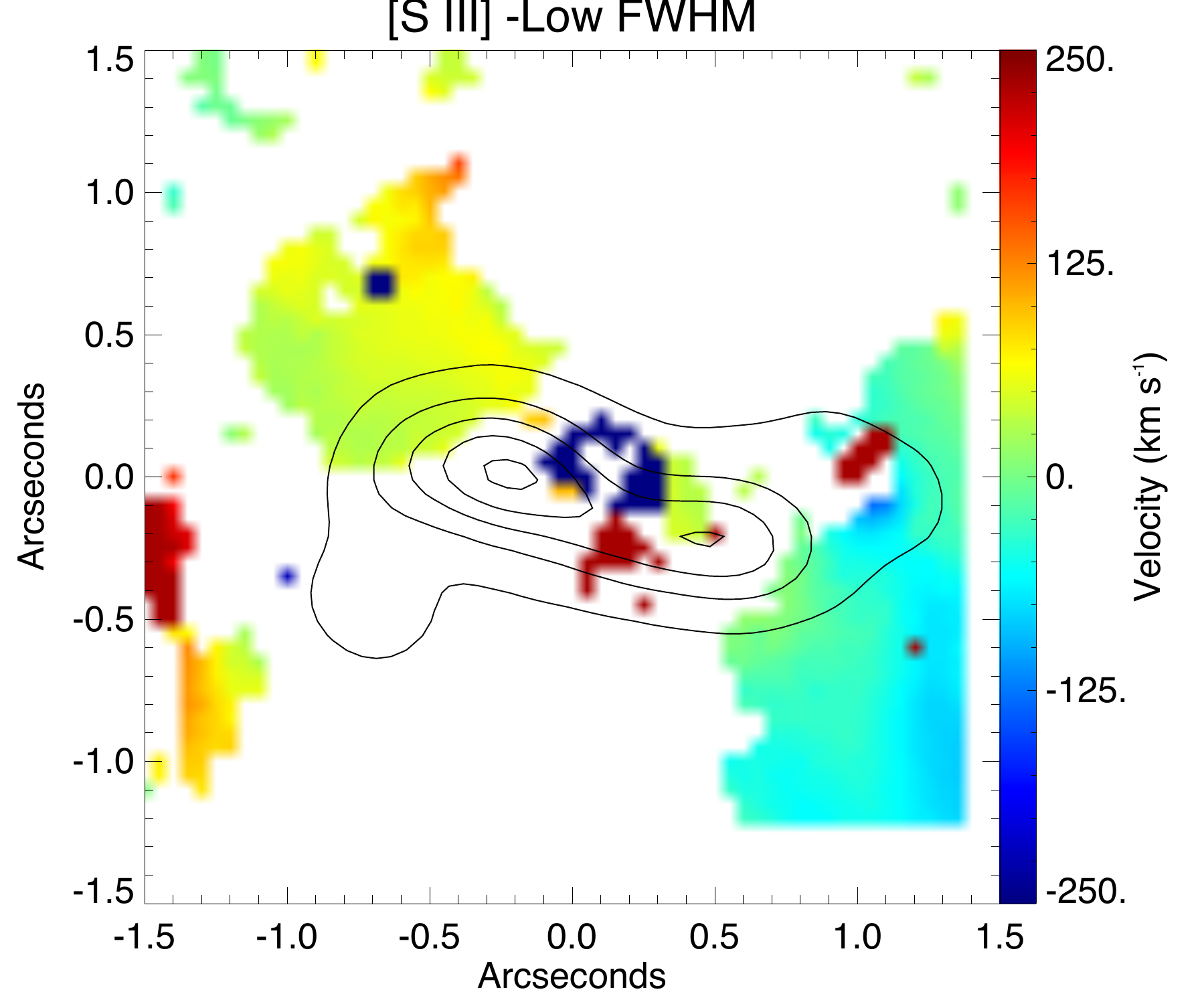}}%\hspace{5ex}
\subfigure{
\includegraphics[scale=0.45]{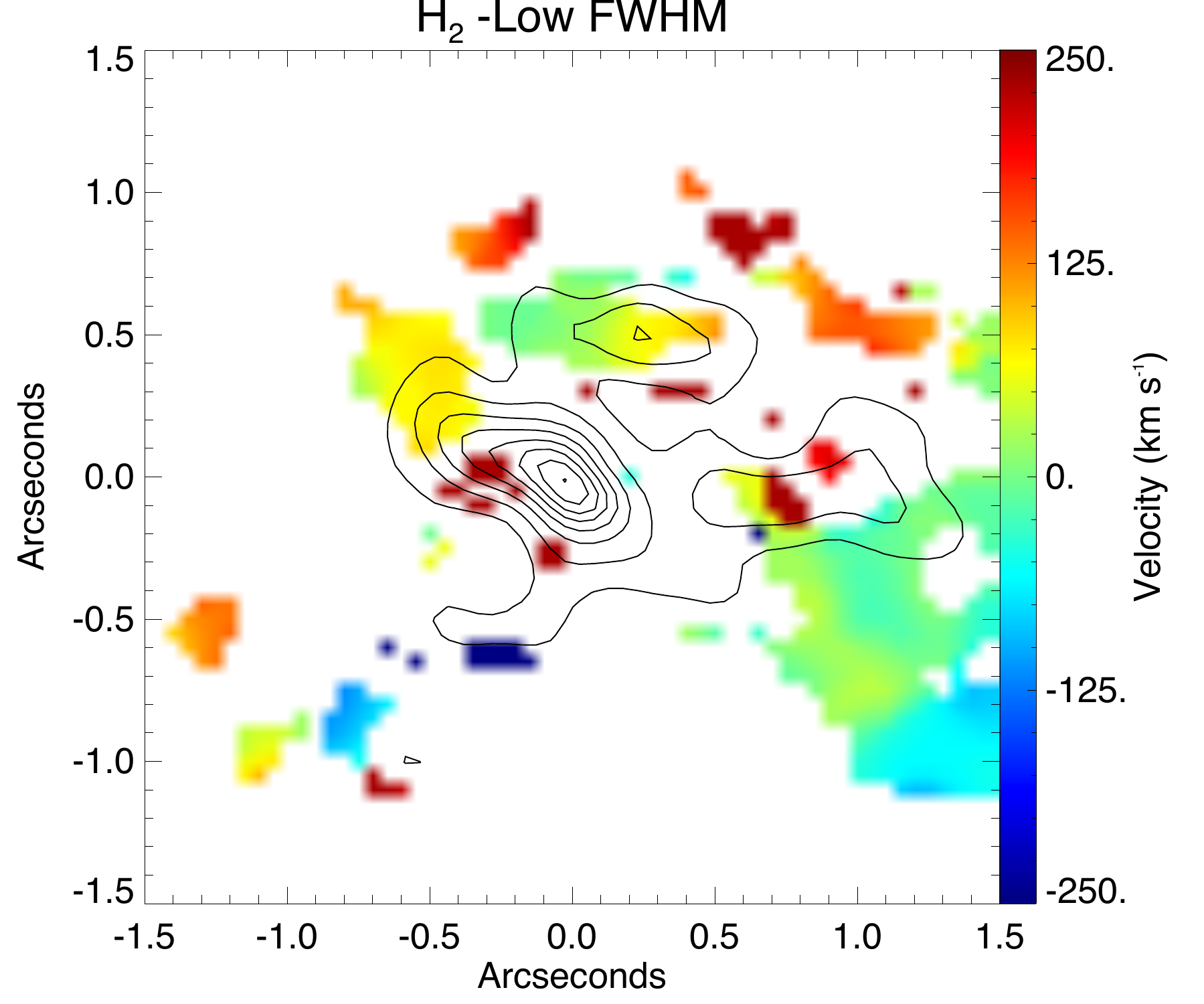}}
\subfigure{
\includegraphics[scale=0.45]{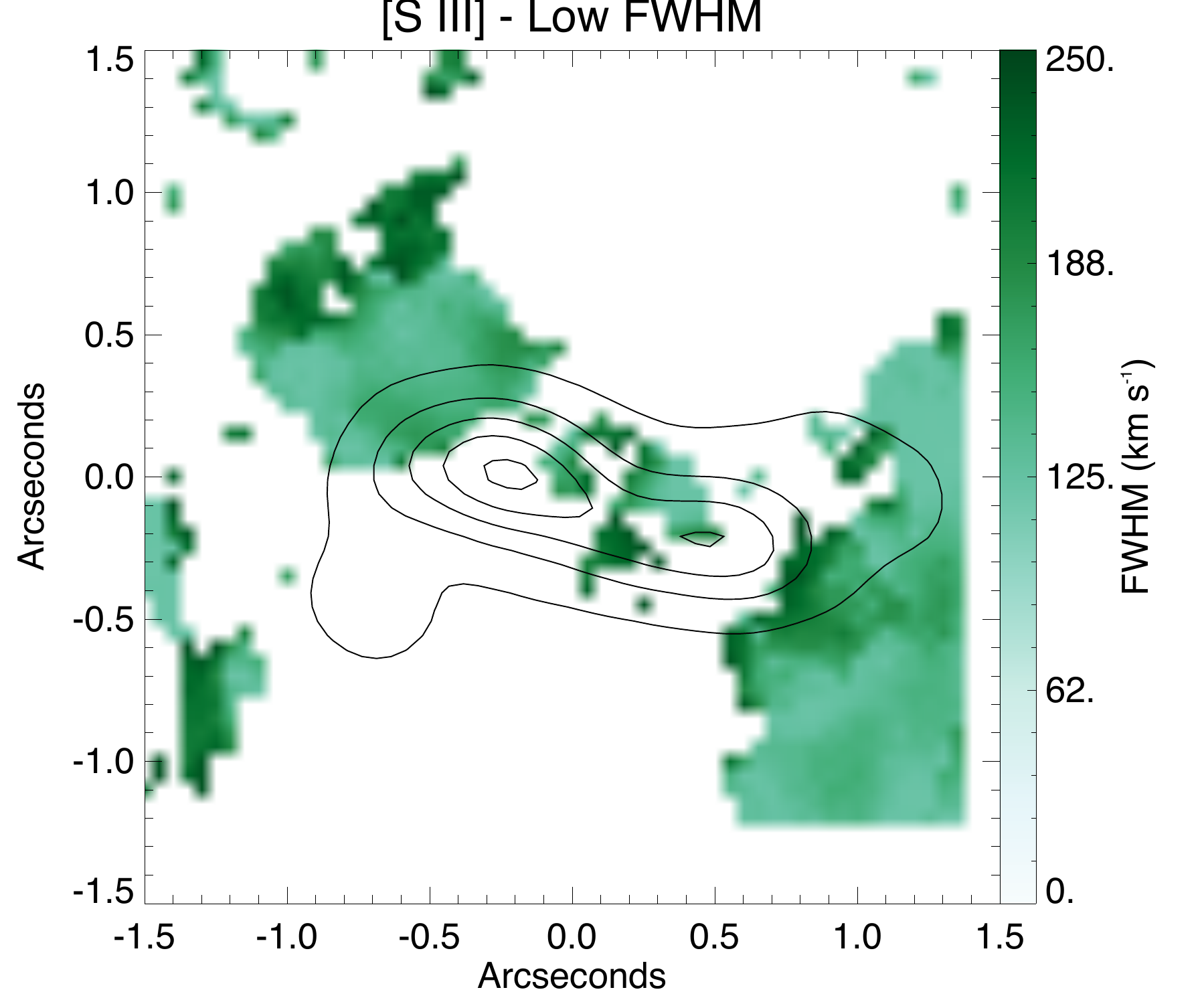}}%\hspace{5ex}
\subfigure{
\includegraphics[scale=0.45]{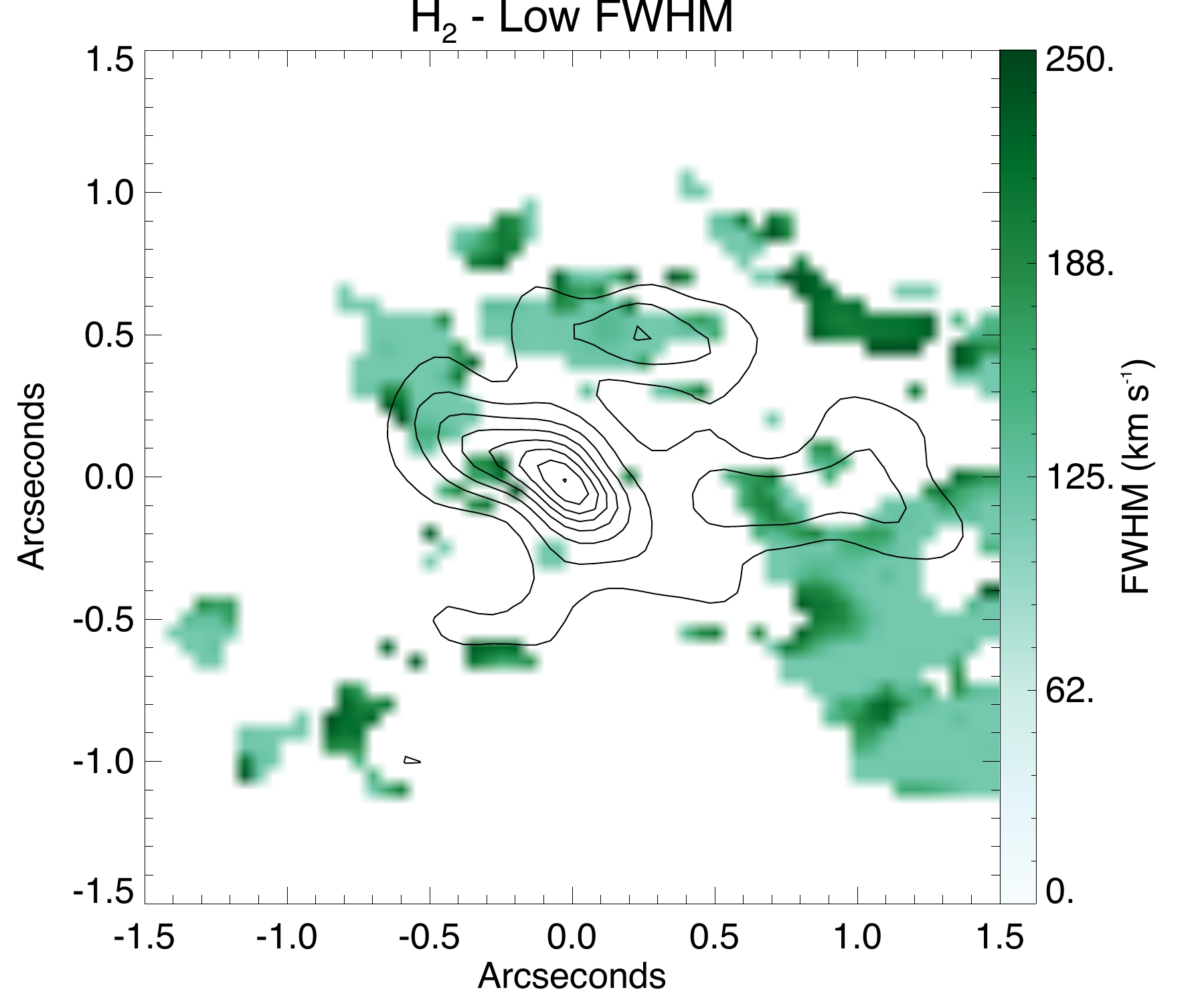}}
\caption{NIFS radial velocity and FWHM maps of the low width (FWHM $\leq$ 250 km s$^{-1}$) components of [S~III] (left) and H$_2$ emission (right). Both v$_r$ maps shows redshifted emission to the NE (upper left) and blueshifted emission to the SW (lower right), with maximum amplitudes of $\pm$70 km s$^{-1}$, indicating a rotating ionized and warm molecular gas disk within $\sim$1.\arcsec5 ($\sim$400 pc) of the nucleus.
\label{fig:low_fwhm}}
\end{figure}

We isolated the regions of low $| v_r |$ and low FWHM  in both [S~III] and H$_2$ by combining the lowest width component in each spaxel with the single component fits and retaining only those components with FWHM $\leq$ 250 km s$^{-1}$. We show the resulting FWHM and radial velocity maps in Figure \ref{fig:low_fwhm}. The majority of the low FWHM emission has relatively low radial velocity ($<$70 km s$^{-1}$) and is oriented in the NE-SW direction (although there are small pockets of high-velocity emission likely associated with outflow).
A reasonable interpretation of the low FWHM, low $v_r$ component is that it shows rotation with radial velocity amplitudes of $\pm$70 km s$^{-1}$ in both [S~III] and H$_2$ at distances up to $\sim$1.\arcsec5 ($\sim$400 pc) from the SMBH, similar to those of the stellar velocity field shown in Figure \ref{fig:stellar_vels}.
It appears that this structure is a rotating ionized and warm molecular gas disk at a PA $\approx$ 45\arcdeg, which is relatively close to the value of PA $\approx$ 30\arcdeg\ for the NIFS stellar velocity field. The nuclear gas disk is counter-rotating with respect to the stellar disk shown in Figure \ref{fig:nifs_stellar}, which seems unusual, but counter-rotating gas disks have been detected in the nuclear regions of other AGN \citep{Hicks et al.(2013), Davies et al.(2014), Raimundo et al.(2017)}.

Overall, the kinematics of the [S~III] emission from NIFS are consistent with those from STIS long-slit observations of [O~III] in the regions of overlap. The NIFS observations provide kinematics of emission-line knots outside of the STIS slit, including those in the eastern and western lobes, and show that the radial velocities extend to significantly higher blueshifted velocities, up to $-$1500 km s$^{-1}$. The prevalence of blueshifted emission, particularly in the east, suggests a significant amount of extinction of the redshifted emission, presumably by dust in the galactic disk, assuming a symmetric distribution of material in the NLR (see Section \ref{subsec:feedback}). This finding is consistent with the reddening measured along the STIS slit by \citet{Collins et al.(2005)}, which increases substantially along the slit from W to E.

Although the overall morphology of the bright H$_2$ emission from warm molecular regions is different than that of [S~III], their kinematics are similar in overlap or adjacent regions , with the H$_2$ and [S~III] gas moving in the same directions. For the gas at high FWHM, the H$_2$ $| v_r |$'s are somewhat lower than those of [S~III], but much higher than the stellar values. The H$_2$ north lane (NL) and south lane are outside of the nominal bicone, but share similar directions of motion with nearby [S~III] knots of emission. Thus, most of the H$_2$ gas within $\pm$1.\arcsec5 of the nucleus is outflowing. However, we have identified a low radial velocity, low FWHM component that is likely a nuclear ionized and molecular disk that is counter-rotating with respect to the stellar disk. We return to these issues in Section \ref{sec:discussion}.

\bigskip

\subsection{ENLR Gas Kinematics: {\it APO} DIS} \label{subsec:enlr_gas_apo}

To determine the kinematics of the emission-line gas observed by {\it APO} DIS on large scales, we used the same Bayesian fitting routine and the same constraints for the optical emission lines described in Section \ref{subsec:nlr_gas_stis}. Once again, independent fits to the [O~III] and H$\alpha$ emission lines yielded very similar kinematics, and we therefore used the [O~III] kinematic components as templates for fitting the other lines (including H$\alpha$ and [N~II]) in 0.\arcsec4 steps along the 2\arcsec-wide slits. We note that the {\it APO} PSF encompasses at least several bins according to the seeing measurements in Table \ref{tab:hst} and thus the kinematic measurements are not completely independent of their neighbors. 

\begin{figure}[!ht]
%\vspace{-8pt}
\centering
\subfigure{
\includegraphics[scale=0.2]{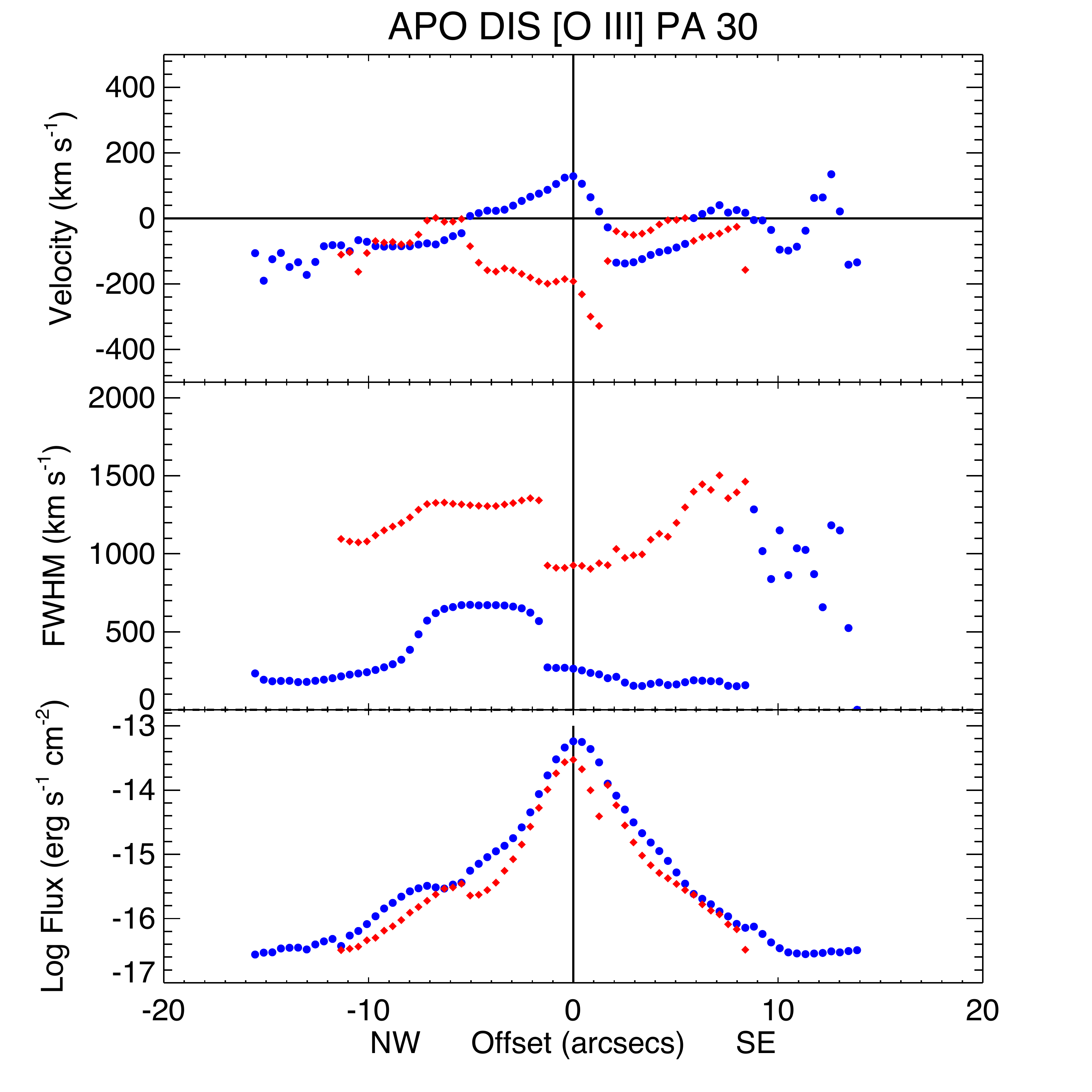}}\hspace{-3ex}
\subfigure{
\includegraphics[scale=0.2]{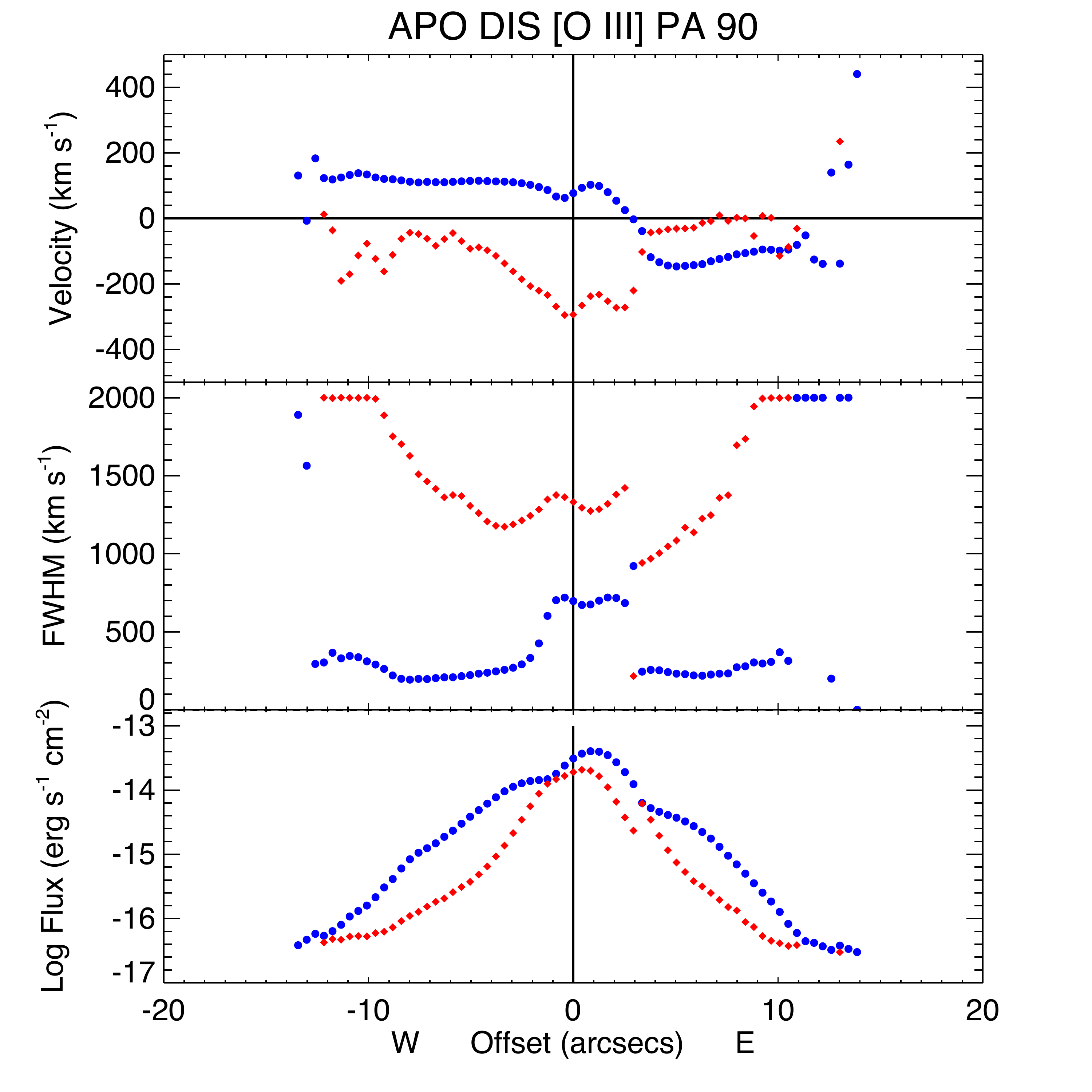}}\hspace{-3ex}
\subfigure{
\includegraphics[scale=0.2]{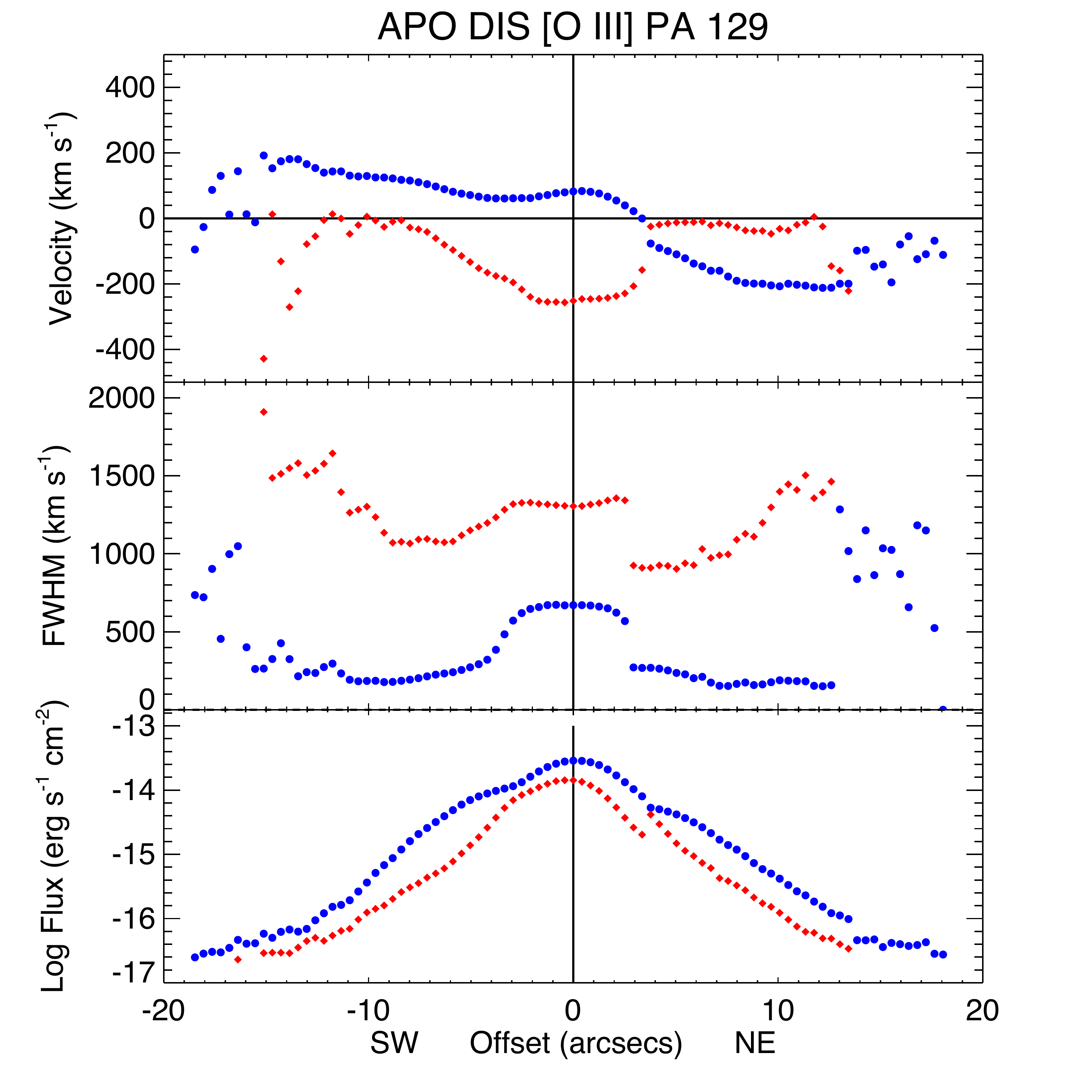}}
\caption{Radial velocity centroid, FWHM (corrected for line-spread function), and integrated flux from our kinematic fits to the [O~III] $\lambda$5007 emission lines along the {\it APO} DIS slit positions at PA = 30\arcdeg, 90\arcdeg, and 129\arcdeg\ from left to right. Data points are marked as blue and red circles corresponding to higher and lower integrated line fluxes at each location, respectively. The vertical solid line at zero position gives the location of the continuum centroid. The maximum uncertainties in radial velocity and FWHM are $\pm$40 km s$^{-1}$ within $\pm$10\arcsec\ and $\pm$100 km s$^{-1}$ at greater distances. Values associated with the FWHM upper limit of 2000 km s$^{-1}$ should be treated with some caution.
\label{fig:apo_kinematics}}
\end{figure}

In Figure \ref{fig:apo_kinematics}, we show the radial velocity, FWHM, and integrated flux for each [O~III] $\lambda$5007 component as a function of position along the slit, where a position of zero corresponds to the galaxy continuum peak. In general, the [O~III] line was well fit by two Gaussian components, except at large radii where one Gaussian was sufficient. We are able to trace [O~III] emission out to a projected distance of $\sim$20\arcsec ($\sim$5.4 kpc) from the AGN, which is about half of the visible semi-major axis of the galaxy.

Comparing the [O~III] kinematics at different {\it APO} slit positions, the higher flux component shows radial velocities that reflect a characteristic disk rotation pattern, except for the inner $\pm$4\arcsec\ (1.1 kpc) where the outflow components seen in the STIS spectra appear to contribute strongly. The low FWHMs between 4\arcsec\ and 12\arcsec\ are also consistent with the bright component arising primarily from disk rotation.  The radial velocity and FWHM amplitudes inside 4\arcsec\ are significant, but lower than those in the STIS data, presumably due to averaging of a number of velocities over a much larger area. The brighter component shows a bump in its flux profile within $\pm$4\arcsec, further emphasizing the transition from outflow to rotation. However, we note that this boundary may have been stretched due to scattered-light contributions from the very bright outflowing clouds in the NLR within 1.\arcsec5 of the SMBH (see Figure \ref{fig:mrk3_slits}), and therefore only claim that outflow dominates to at least $\sim$320 pc from the central SMBH.

The kinematics of the fainter {\it APO} component in Figure \ref{fig:apo_kinematics} show a different pattern. Within $\pm$4\arcsec, this component includes a strong, and likely dominant, contribution from the blueshifted outflow components seen in the STIS data. Beyond $\sim$4\arcsec, the radial velocities are low but the FWHMs are high. This component is very similar to that identifed as``disturbed gas'' in an {\it HST} STIS study of Type 2 quasars by \citet{Fischer et al.(2017)}, which does not show a strong component of outflow but is nevertheless disturbed by the AGN in some fashion to produce the large FWHMs.

The kinematics beyond $\sim$12\arcsec\ (3.2 kpc) show considerable scatter due to lower SNRs and their origin is therefore more difficult to determine.  At PA $=$ 129\arcdeg\ the radial velocities appear to extend the rotation curve, albeit with higher FWHMs. Otherwise, the kinematics of the primarily single component beyond $\sim$12\arcsec\ resemble those of the disturbed component closer to the SMBH, possibly mixed with pure rotation.

For the rotation component, the slit at PA $=$ 129\arcdeg\ has the highest amplitude radial velocities and the greatest extent of [O~III] emission. The slit at PA $=$ 90\arcdeg\ has lower amplitudes radial velocities and the same sense of rotation, with redshifts in the west and blueshifts in the east. At PA $=$ 30\arcdeg, the rotational component is close to zero km s$^{-1}$ and the [O~III] flux beyond 4\arcsec\ drops much more sharply because this position is outside of the nominal bicone \citep{Crenshaw et al.(2010)}. Thus, this rotational component is consistent with the presence of an inclined disk of gas and dust postulated by \citet{Crenshaw et al.(2010)} at PA$=$ 129\arcdeg, which is offset in PA from the stellar rotation component by $\sim$100\arcdeg.

\begin{figure}
%\vspace{-8pt}
\centering
\subfigure{
\includegraphics[scale=0.38]{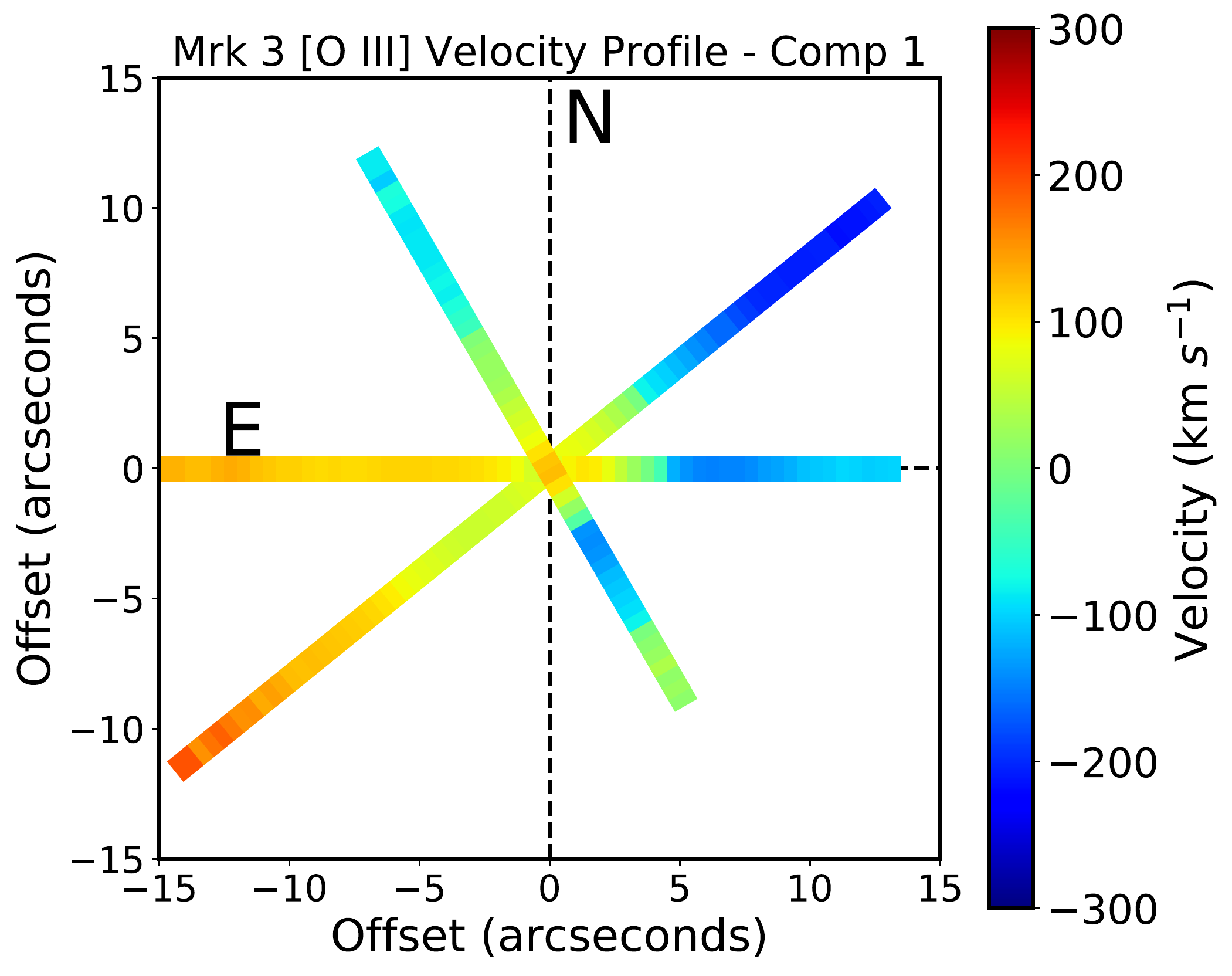}}\hspace{5ex}
\subfigure{
\includegraphics[scale=0.38]{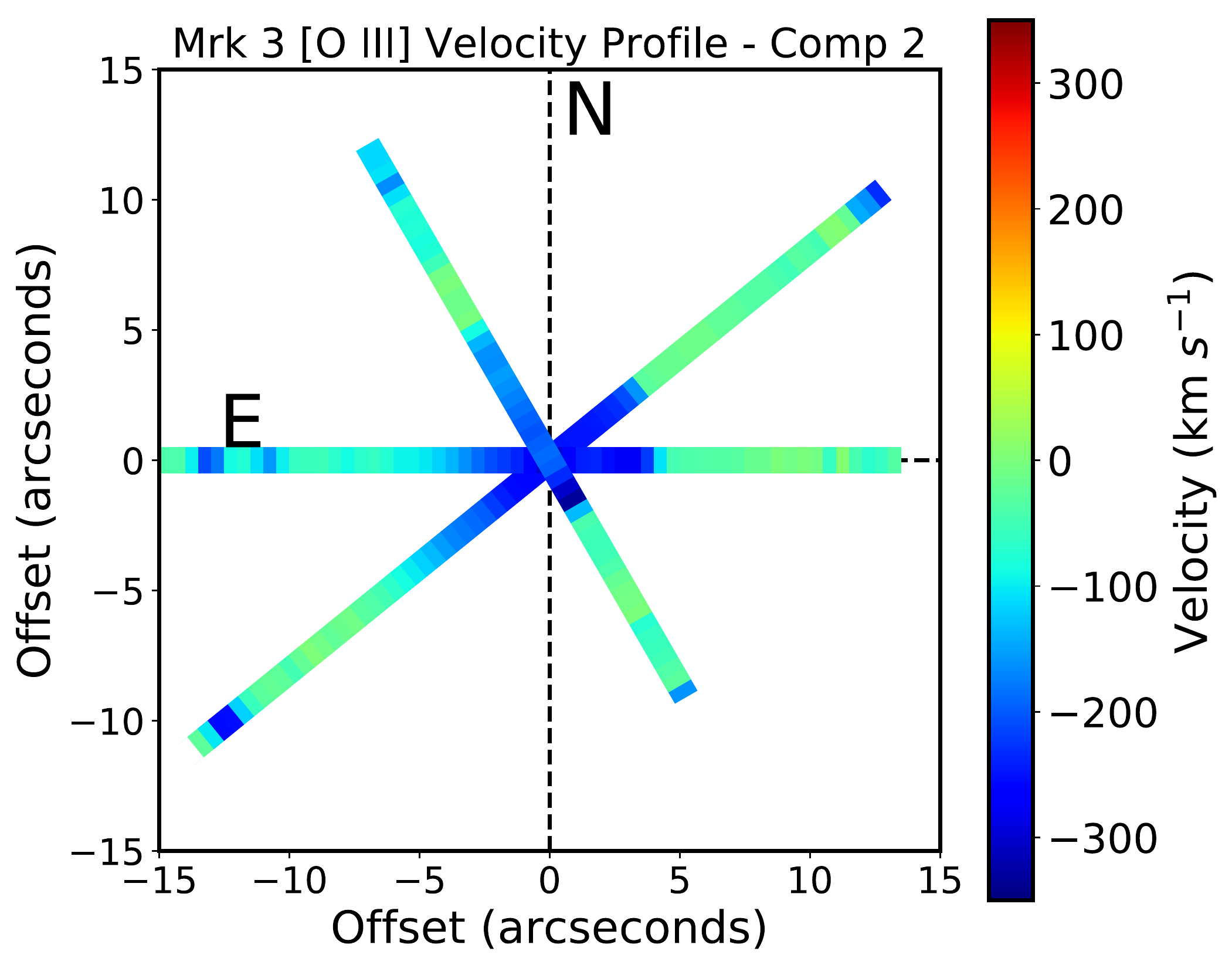}}\hspace{5ex}
\subfigure{
\includegraphics[scale=0.38]{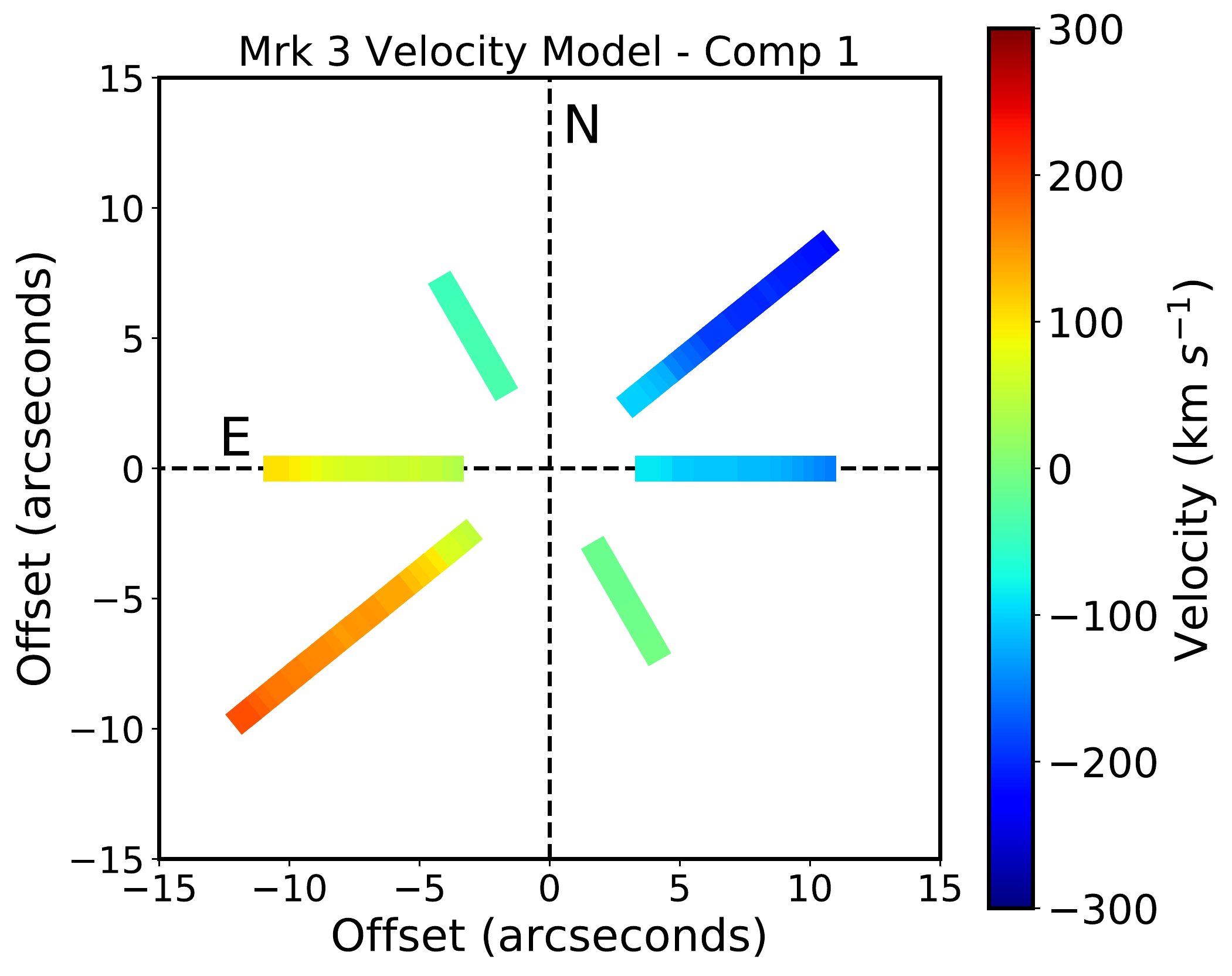}}\hspace{5ex}
\subfigure{
\includegraphics[scale=0.38]{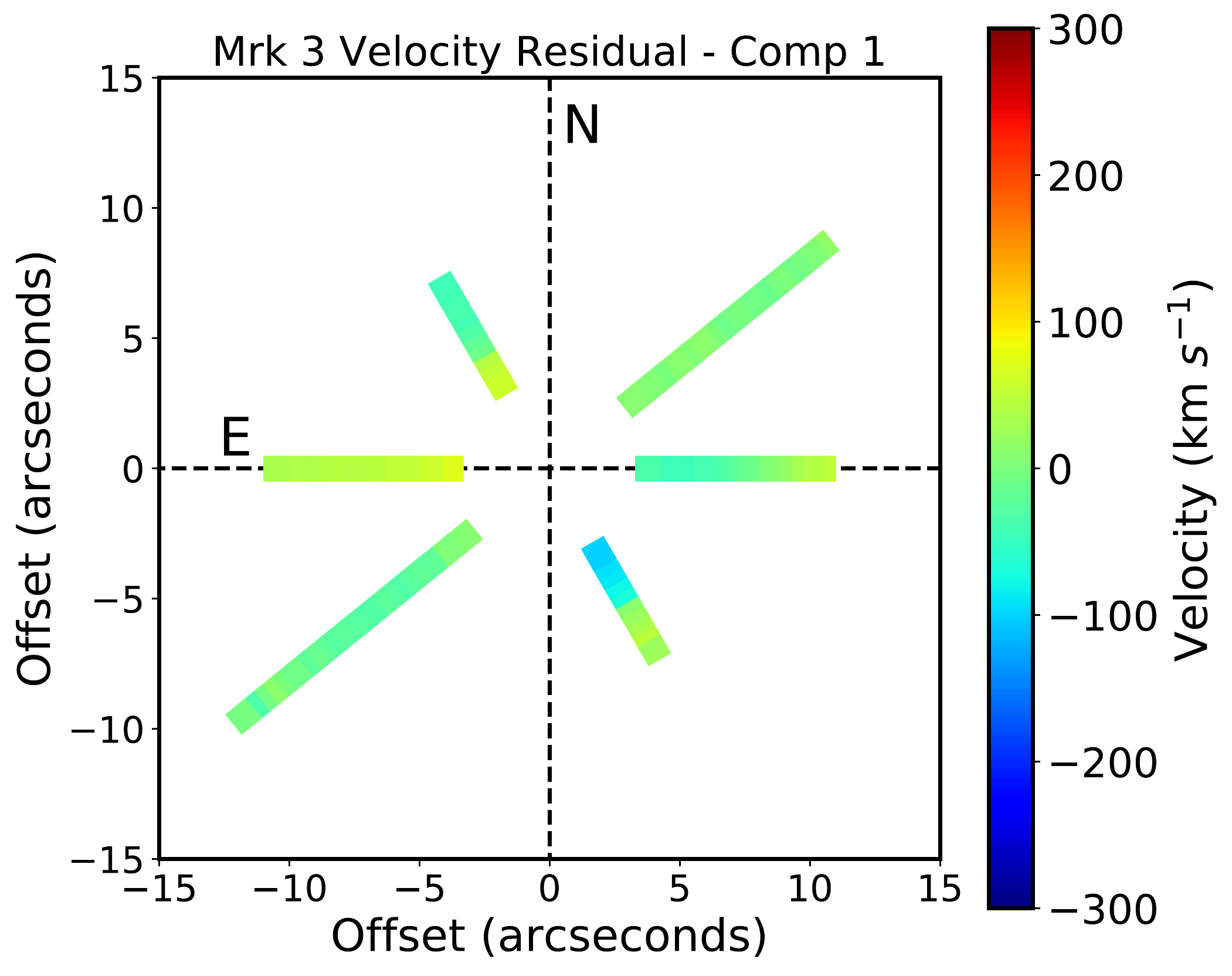}}
\caption{Radial velocity maps for the three {\it APO} DIS slit positions at PA $=$ 30\arcdeg, 90\arcdeg, and 129\arcdeg\ for the higher (upper left) and lower (upper right) flux kinematic components of [O~III]. Velocities shown in Figure \ref{fig:apo_kinematics} within 5\arcsec\ of the ends of the slits have been removed. The DiskFit model (lower left) and residuals (lower right) are shown for the higher flux (rotation) component.
\label{fig:apo_spider}}
\end{figure}

Once again, we used DiskFit to model the observed velocities. With only three PA's, we were not able to obtain a good fit with all parameters allowed to vary, so we fixed the PA to 129\arcdeg, inclination to 64\arcdeg, and center to the continuum centroid using the imaging values from \citet{Crenshaw et al.(2010)}. 
In Figure \ref{fig:apo_spider}, we show color-coded maps of the observed radial velocities for the two flux components in all three slit positions, truncating the noisy points in Figure \ref{fig:apo_kinematics} at the ends of the slits in the last $\sim$5\arcsec.
For the DiskFit model of rotation in the high-flux component, we also removed the points within 4\arcsec\ of the nucleus to avoid the outflows.
We show the DiskFit model velocities and residuals for the high-flux (rotation) component, which demonstrate a good fit for the fixed photometric constraints.
Thus, the observed radial velocities are consistent with the proposed gas/dust disk from the photometry and allows us to determine the true velocity field of the disk.
At $\sim$10\arcsec\ from the nucleus, the radial velocities peak at $\pm$180 km s$^{-1}$  which correspond to a true maximum rotational velocity of 200 km s$^{-1}$.
This value is close to the peak velocity of the stellar kinematic field shown in Figure \ref{fig:stellar_vels} as expected for a similar gravitational potential despite the $\sim$100\arcdeg\ offset in PA.

\subsection{NLR and ENLR Gas Ionization: {\it APO} DIS} \label{subsec:nlr_enlr_ionization}

\begin{figure}
%\vspace{-8pt}
\centering
\subfigure{
\includegraphics[scale=0.4]{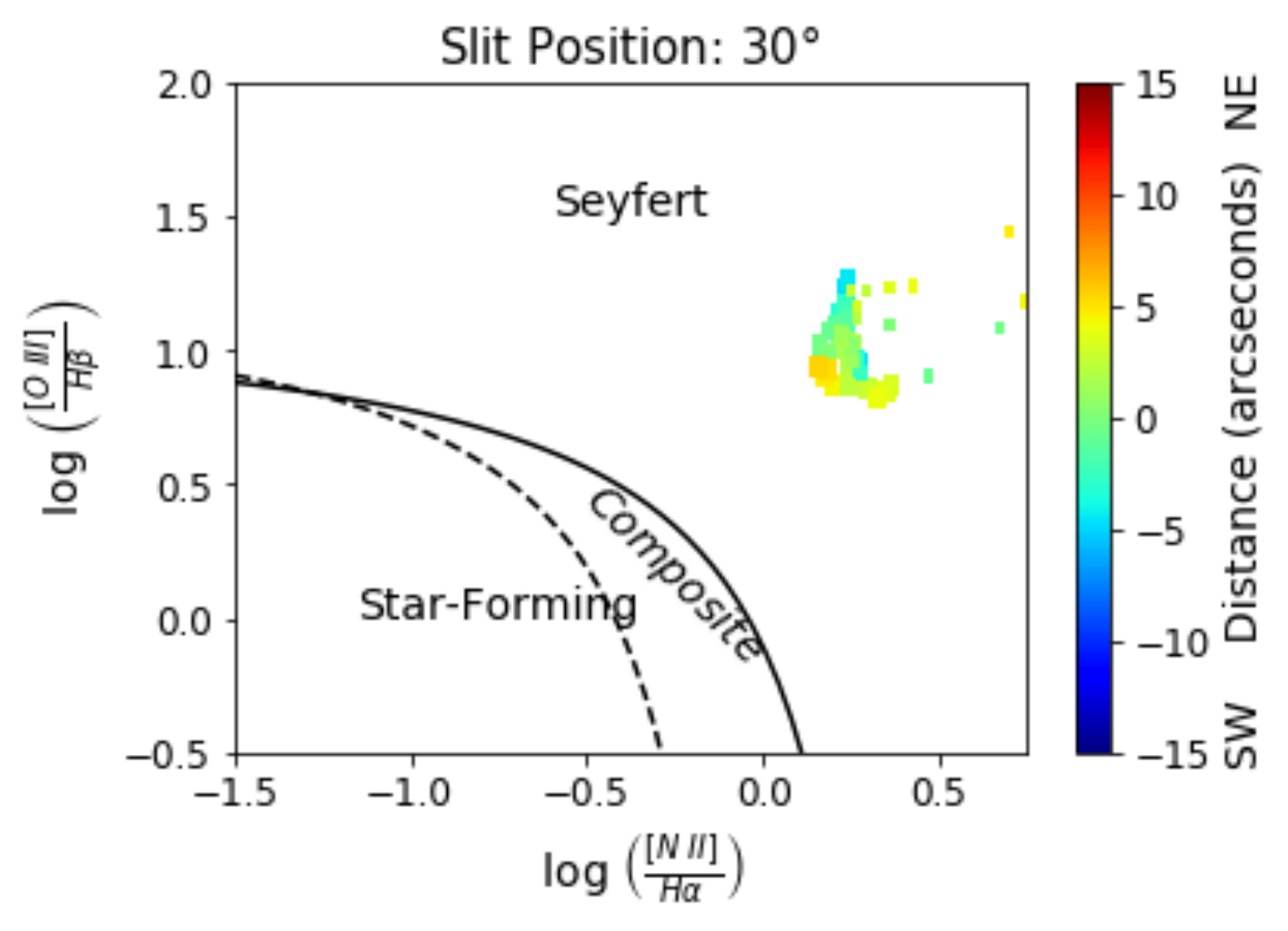}}%\hspace{-5ex}
\subfigure{
\includegraphics[scale=0.4]{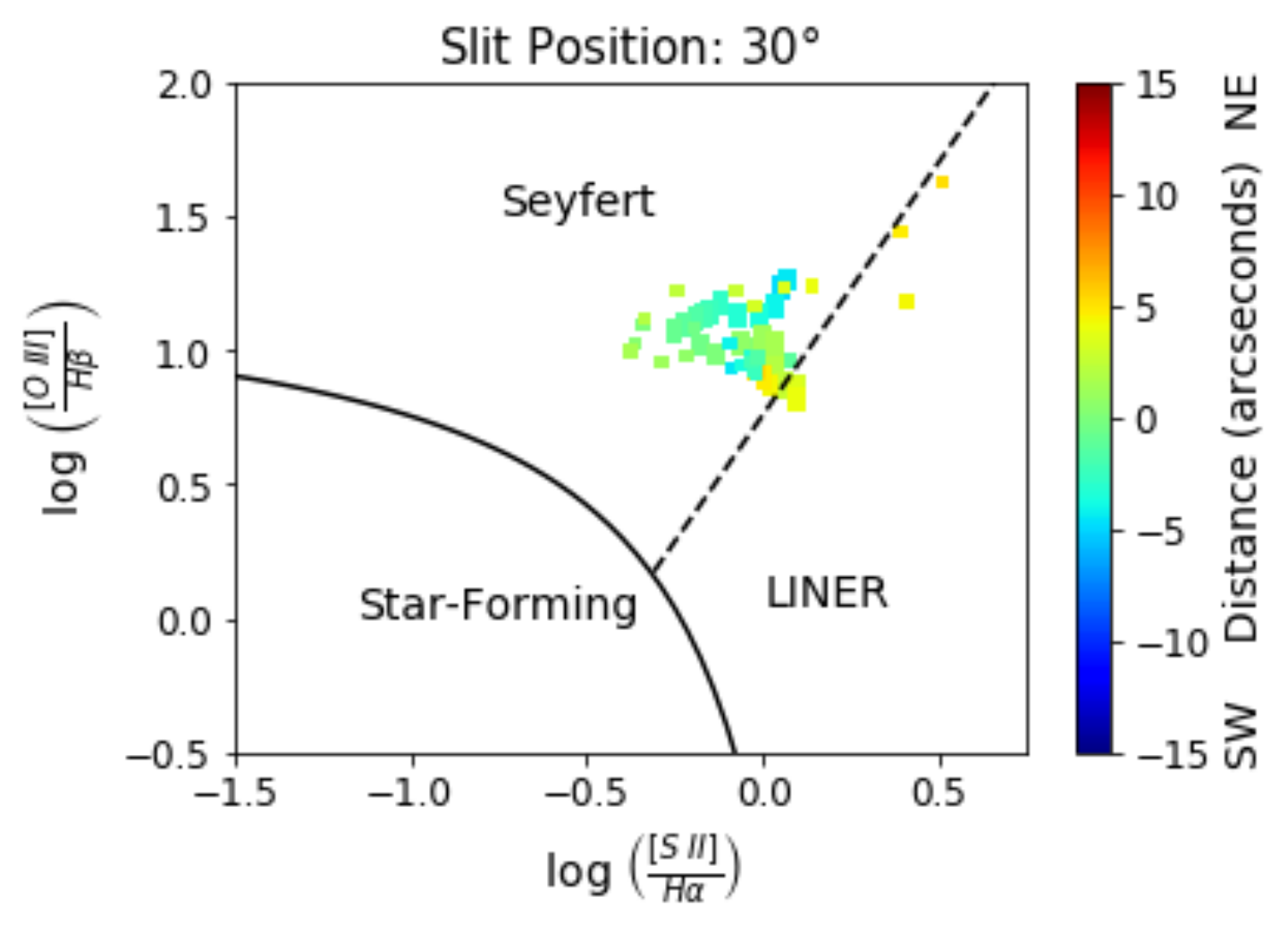}}%\hspace{-5ex}
\subfigure{
\includegraphics[scale=0.4]{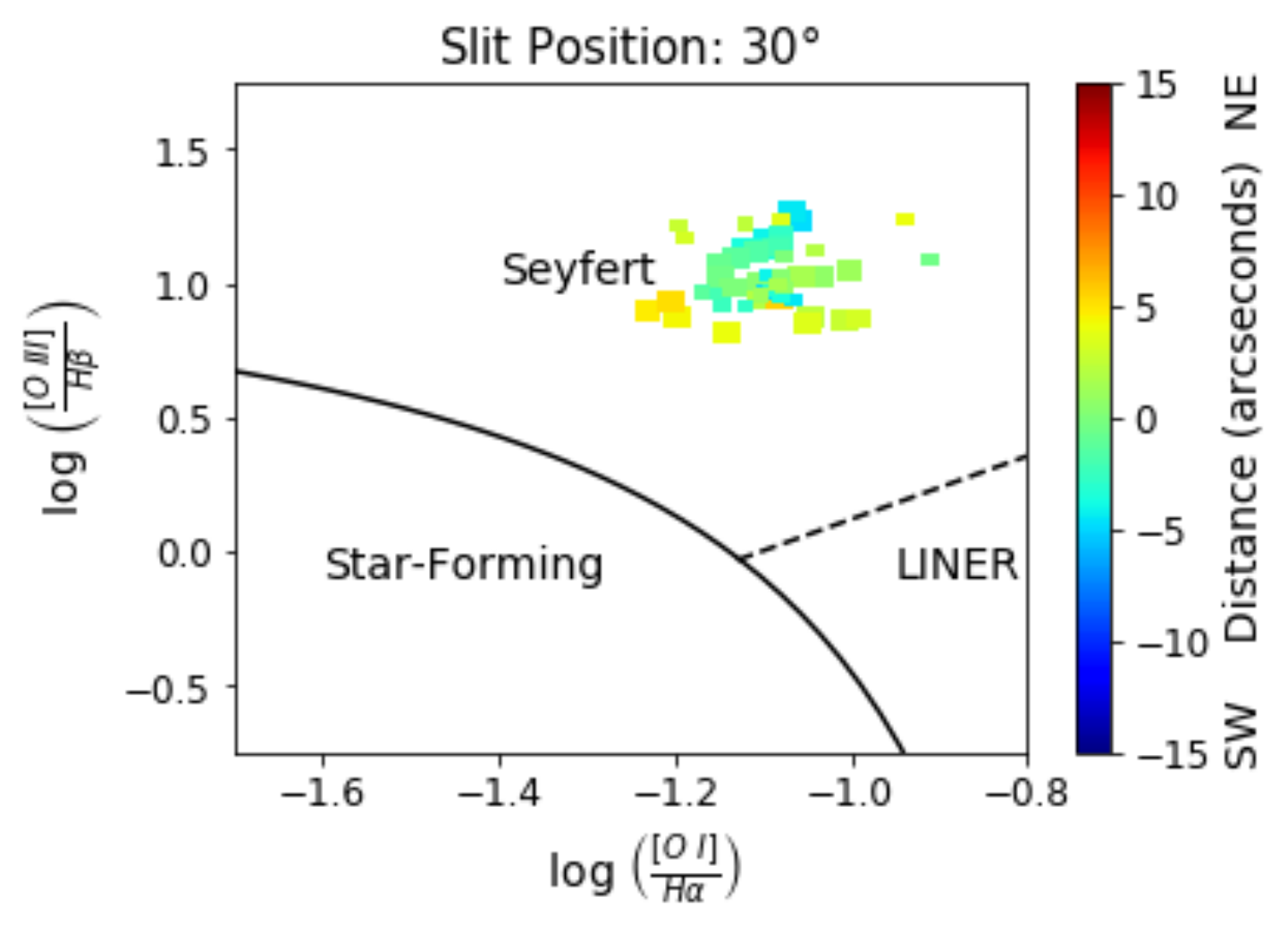}}

\subfigure{
\includegraphics[scale=0.4]{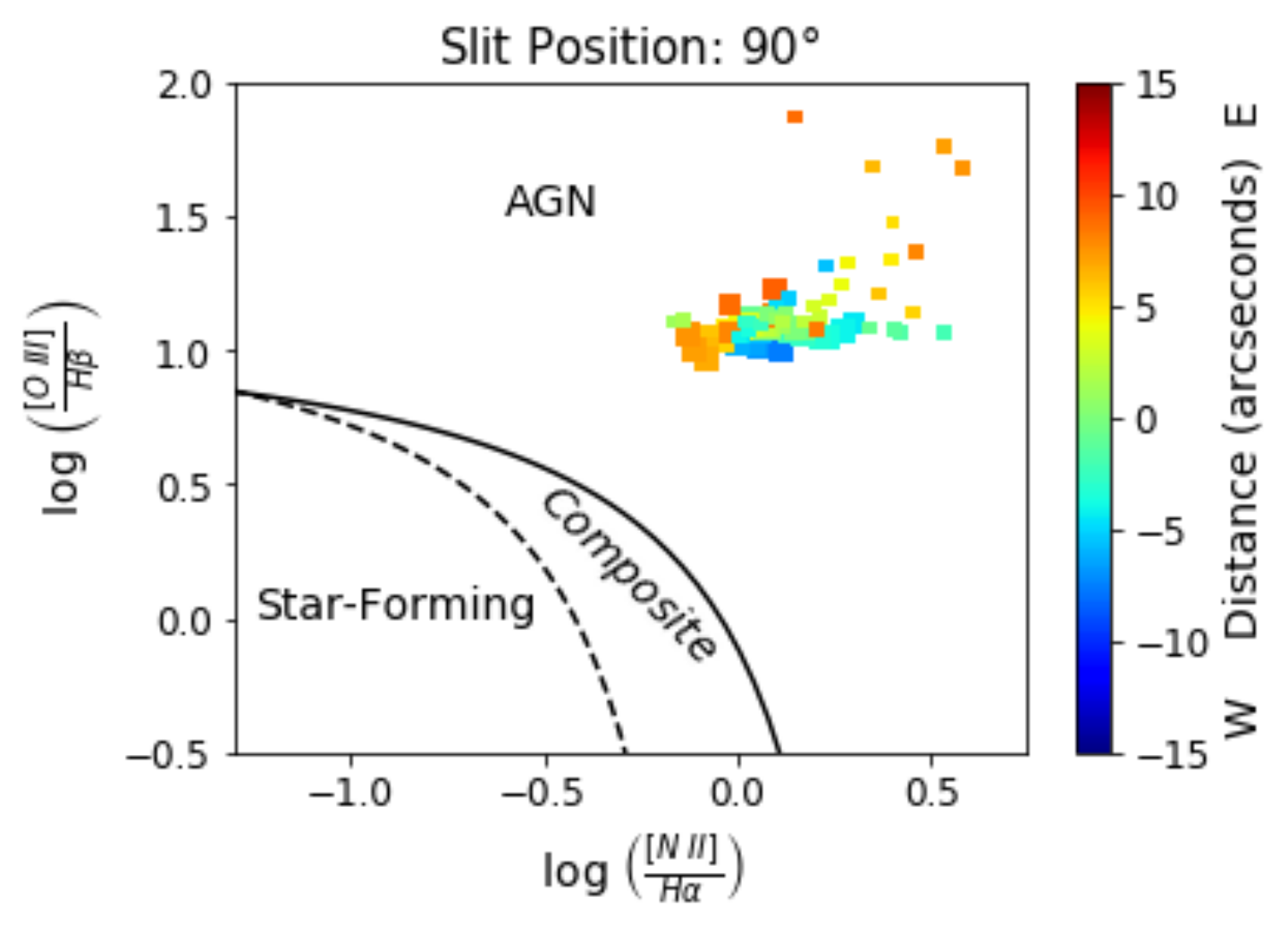}}%\hspace{-5ex}
\subfigure{
\includegraphics[scale=0.4]{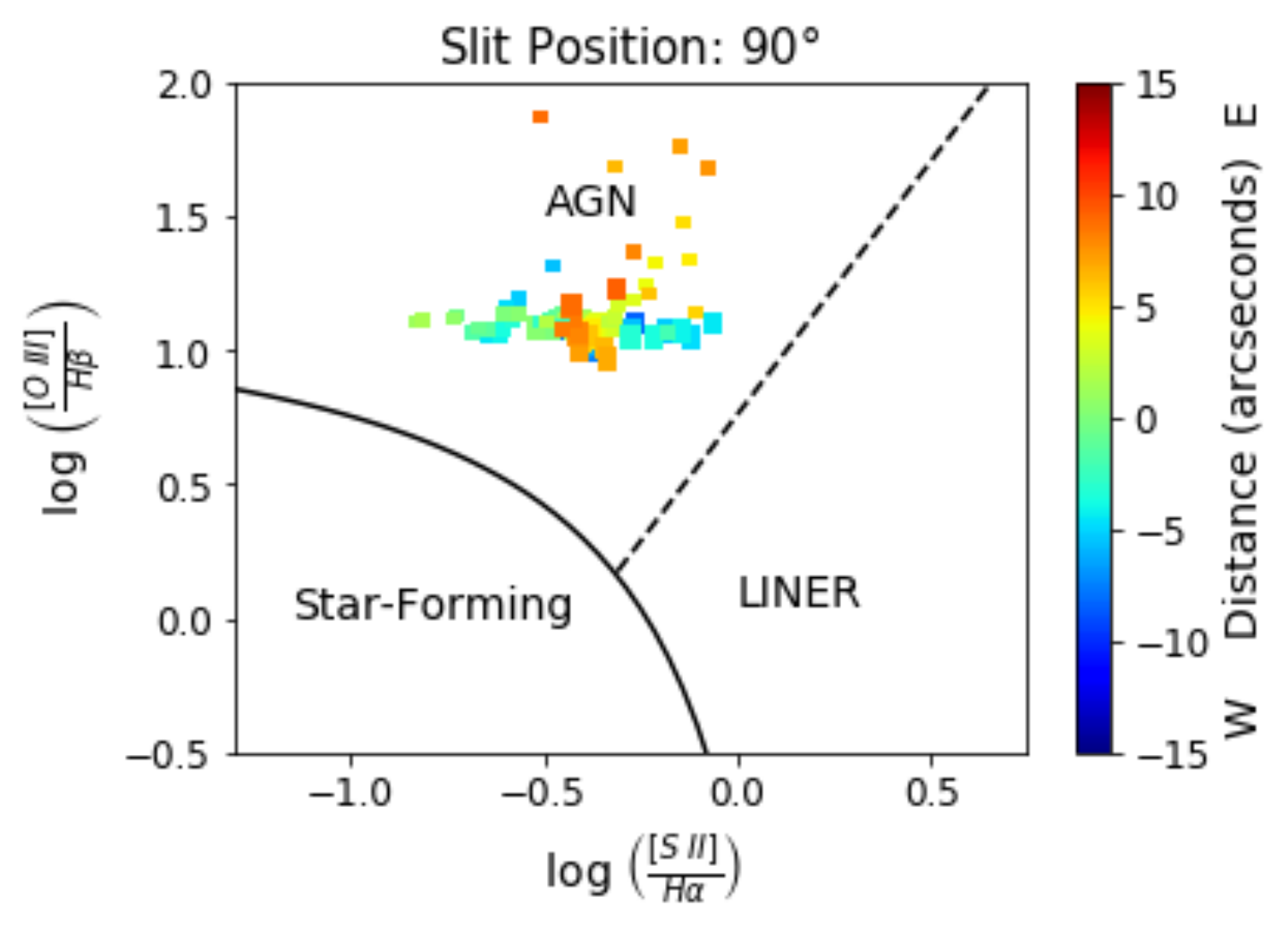}}%\hspace{-5ex}
\subfigure{
\includegraphics[scale=0.4]{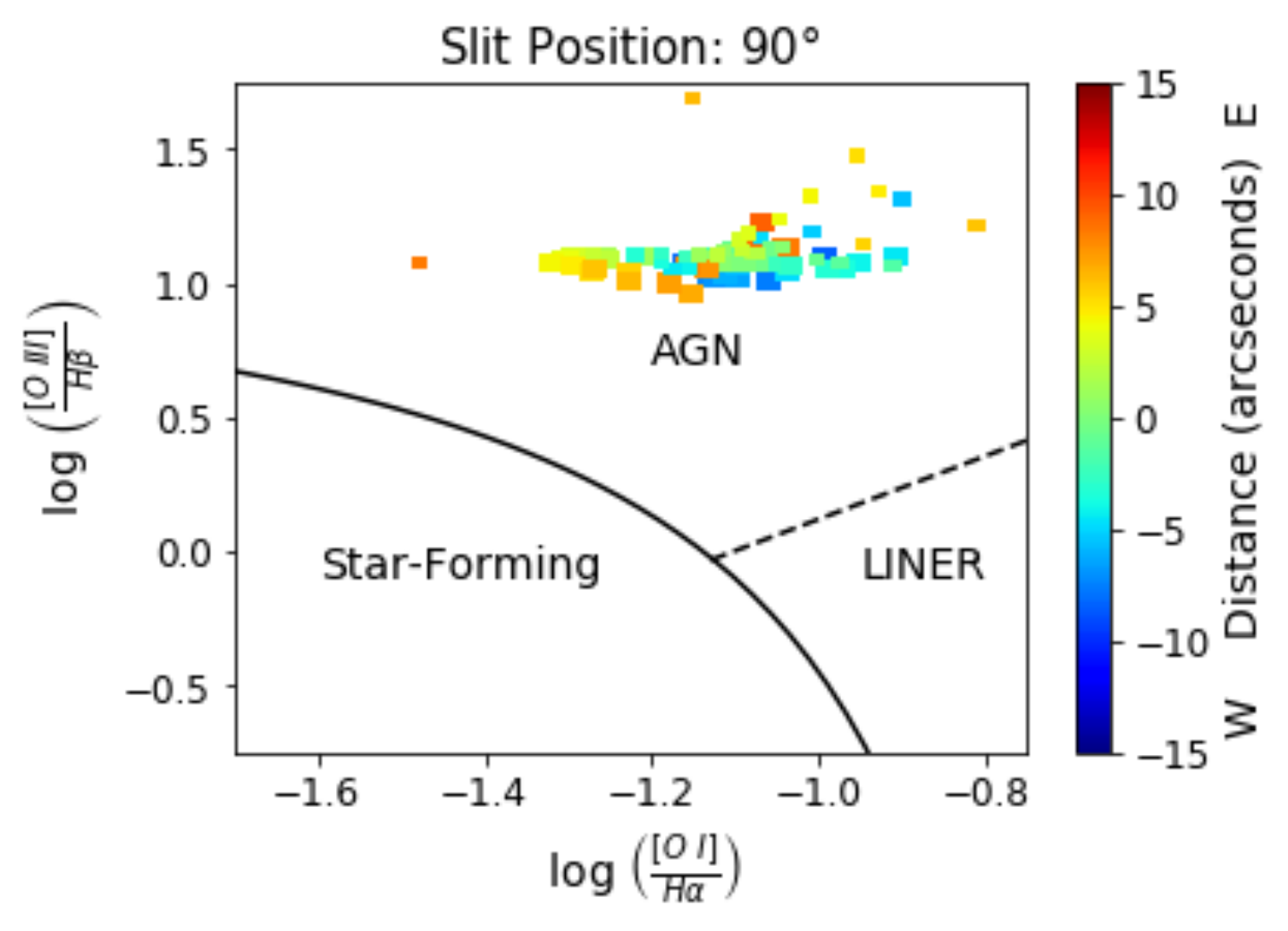}}
\subfigure{

\includegraphics[scale=0.4]{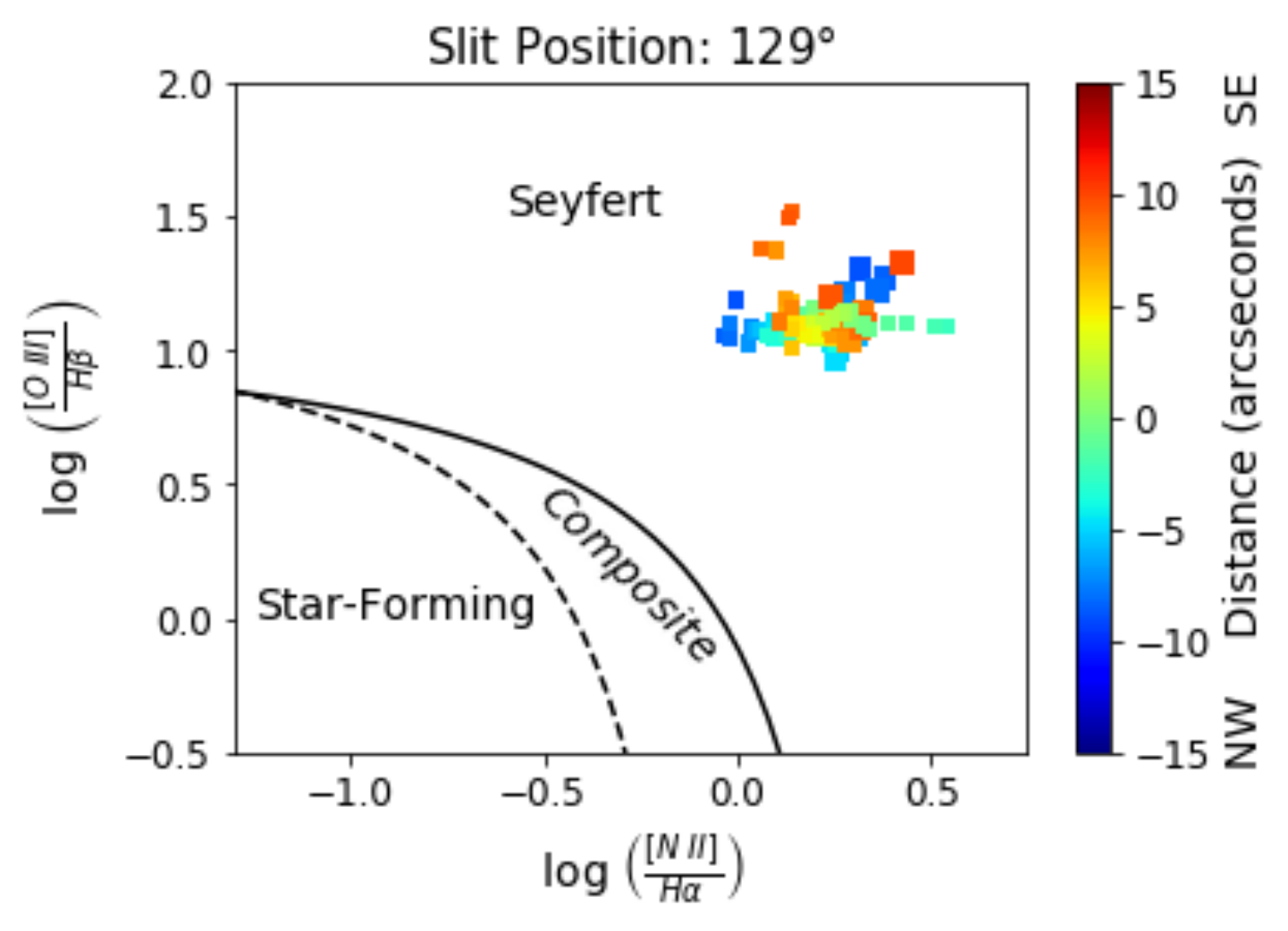}}%\hspace{-5ex}
\subfigure{
\includegraphics[scale=0.4]{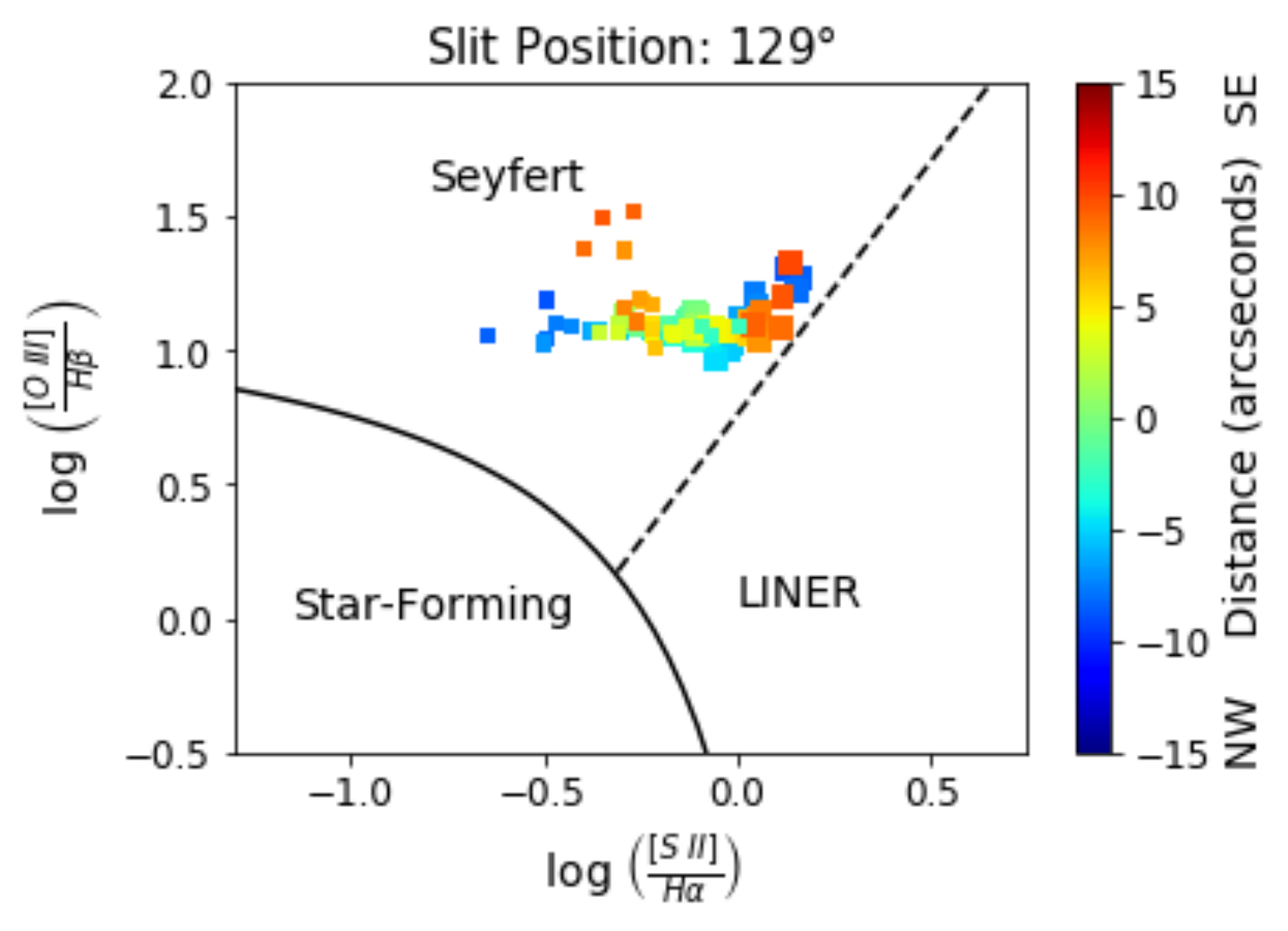}}%\hspace{-5ex}
\subfigure{
\includegraphics[scale=0.4]{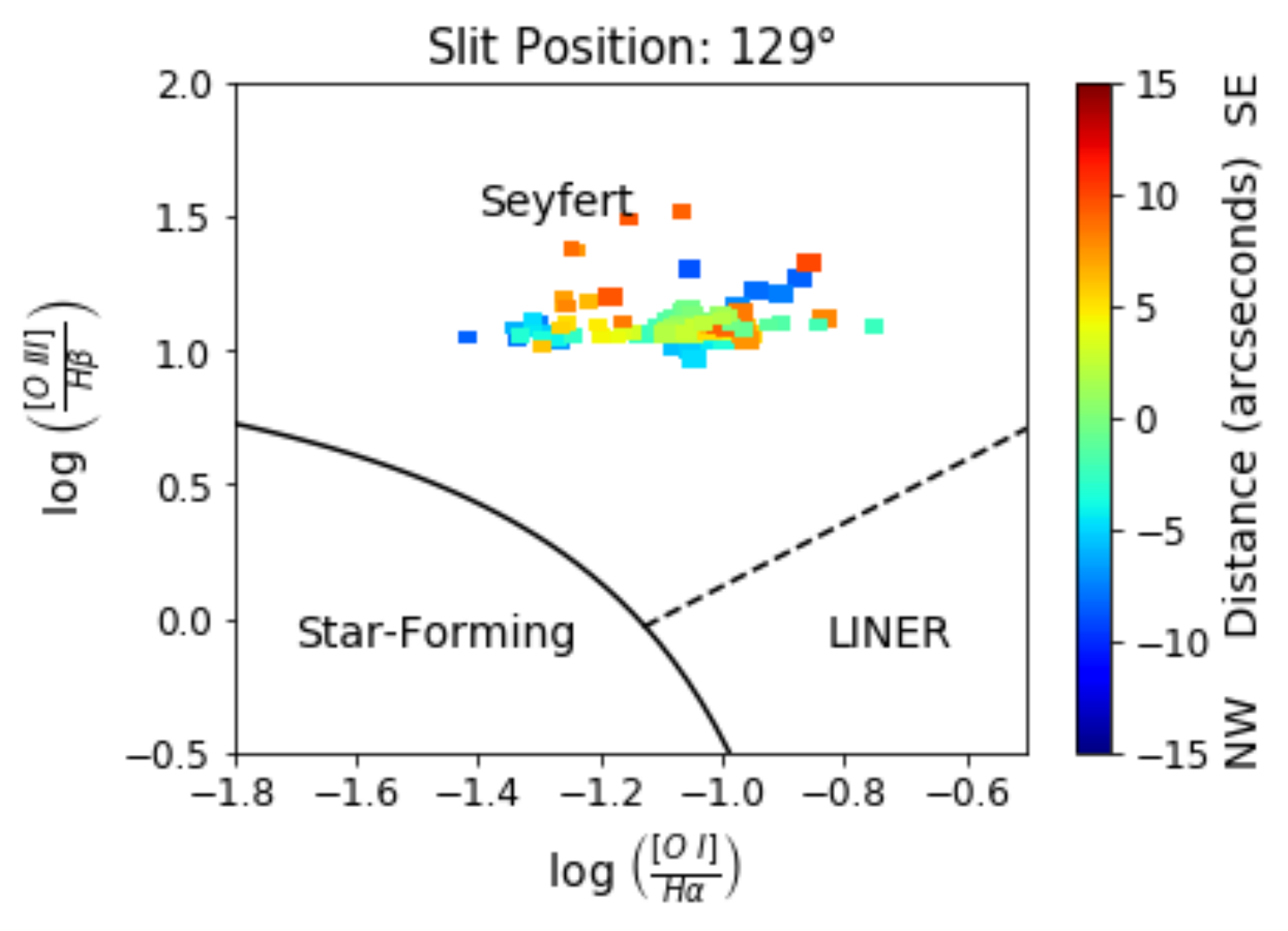}}%\hspace{-5ex}

\caption{BPT diagrams for the high flux (large squares) and low flux (small squares) components as a function of disance from the nucleus along the {\it APO} DIS long-slit at PA $=$ 30\arcdeg\ (top), 90\arcdeg\ (middle) and 129\arcdeg\ (bottom).
\label{fig:apo_bpt}}
\end{figure}

To investigate the source of ionization of the NLR and ENLR gas, we used our spectral fits to a number of different emission lines in the {\it APO} spectra as described in Section \ref{subsec:enlr_gas_apo}. We generated Baldwin-Phillips-Terlevich (BPT) diagrams \citep{Baldwin et al.(1981), Veilleux and Osterbrock(1987)} using the prescription of \citet{Kewley et al.(2001), Kewley et al.(2006)}. In Figure \ref{fig:apo_bpt}, we show these diagrams at each of the 3 {\it APO} slit positions for both flux components as a function of projected distance from the central nucleus, with the end points trimmed as in Section \ref{subsec:nlr_enlr_ionization}. Horizontal scatter is seen primarily in the low-flux components at large distances, and is likely due to large uncertainties in these weak-flux lines. Nevertheless, all of the data points lie in the AGN-ionized sections of the diagrams, far from the star-forming or even composite regions \citep{Kewley et al.(2006)}, with high enough ionization to place them firmly in the ``Seyfert'' category. We can trace the AGN-ionized gas that defines the ENLR to a distance of at least 12\arcsec\ (3.2 kpc) along the major axis of the gas/dust disk at PA $=$ 129\arcdeg. Given the apparent absence of significant star formation, the more distant, scattered points in Figure \ref{fig:apo_kinematics} are also likely ionized by the AGN, and the ionized disk therefore extends out to $\sim$20\arcsec\ ($\sim$5.4 kpc) in this direction.

Overall, our kinematic and ionization studies of the ENLR are consistent with the presence of a large-scale gas/dust disk in Mrk~3 at PA $\approx$ 129\arcdeg, as indicated by our earlier study of the geometries of the NLR and ENLR \citep{Crenshaw et al.(2010)}. Based on our photometry and kinematics of the host galaxy, the gas/dust disk is offset in PA by $\sim$100\arcdeg\ from the stellar major axis. There is no evidence for ionization of the gas in the disk due to stars, and hence no evidence for recent star formation in the host galaxy of Mrk~3. For comparison, {\it Spitzer} IRS observations show no evidence for significant star formation in the inner kpc of Mrk~3 \citep{Deo et al.(2007), Melendez et al.(2008)}, with an upper limit on the star formation rate of $\sim$4 M$_\sun$ yr$^{-1}$ determined by \citet{Melendez et al.(2008)}.

\bigskip\bigskip\bigskip

\subsection{Galactic Environment} \label{subsec:environment}

\begin{figure}[ht!]
\plottwo{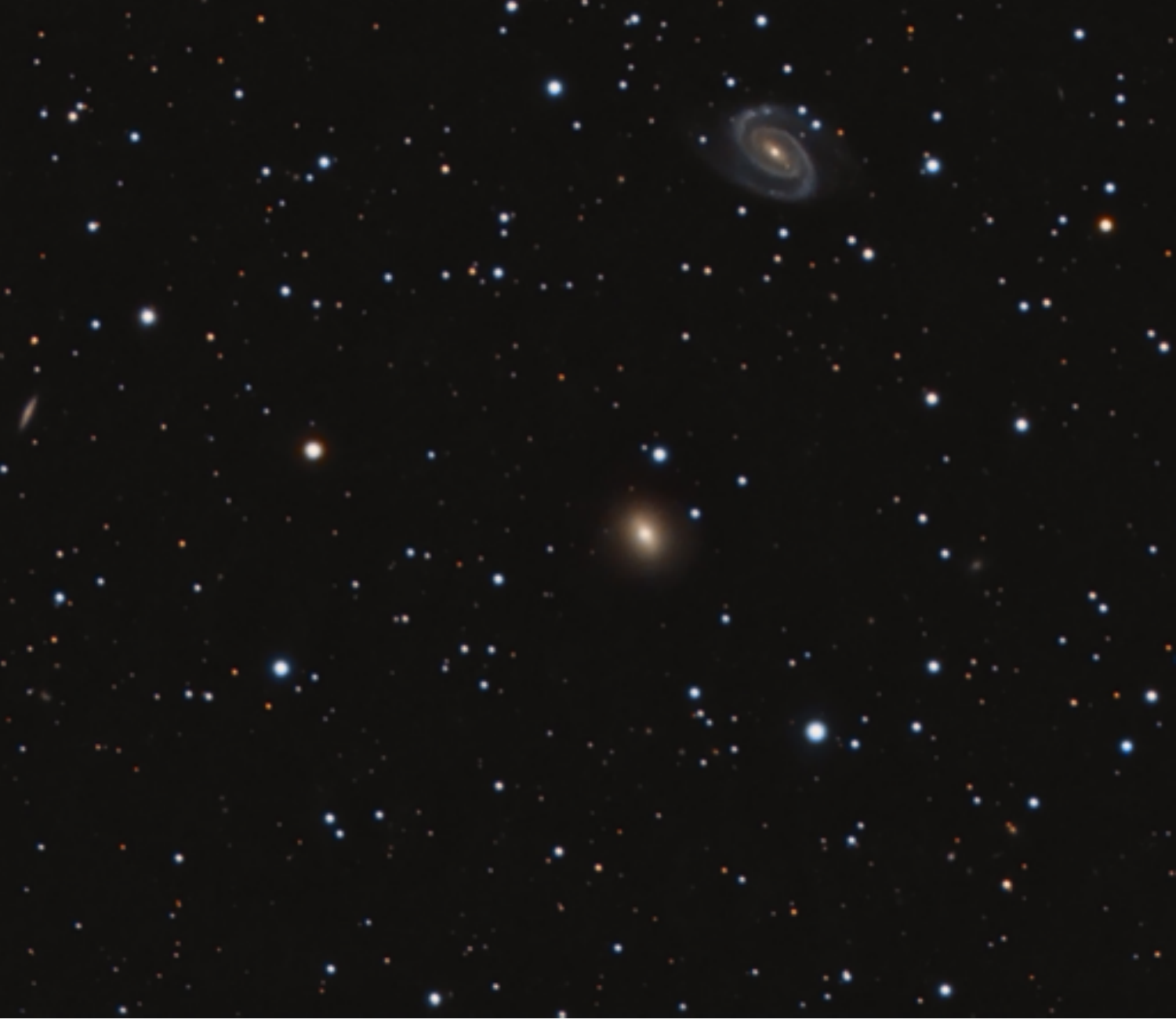}{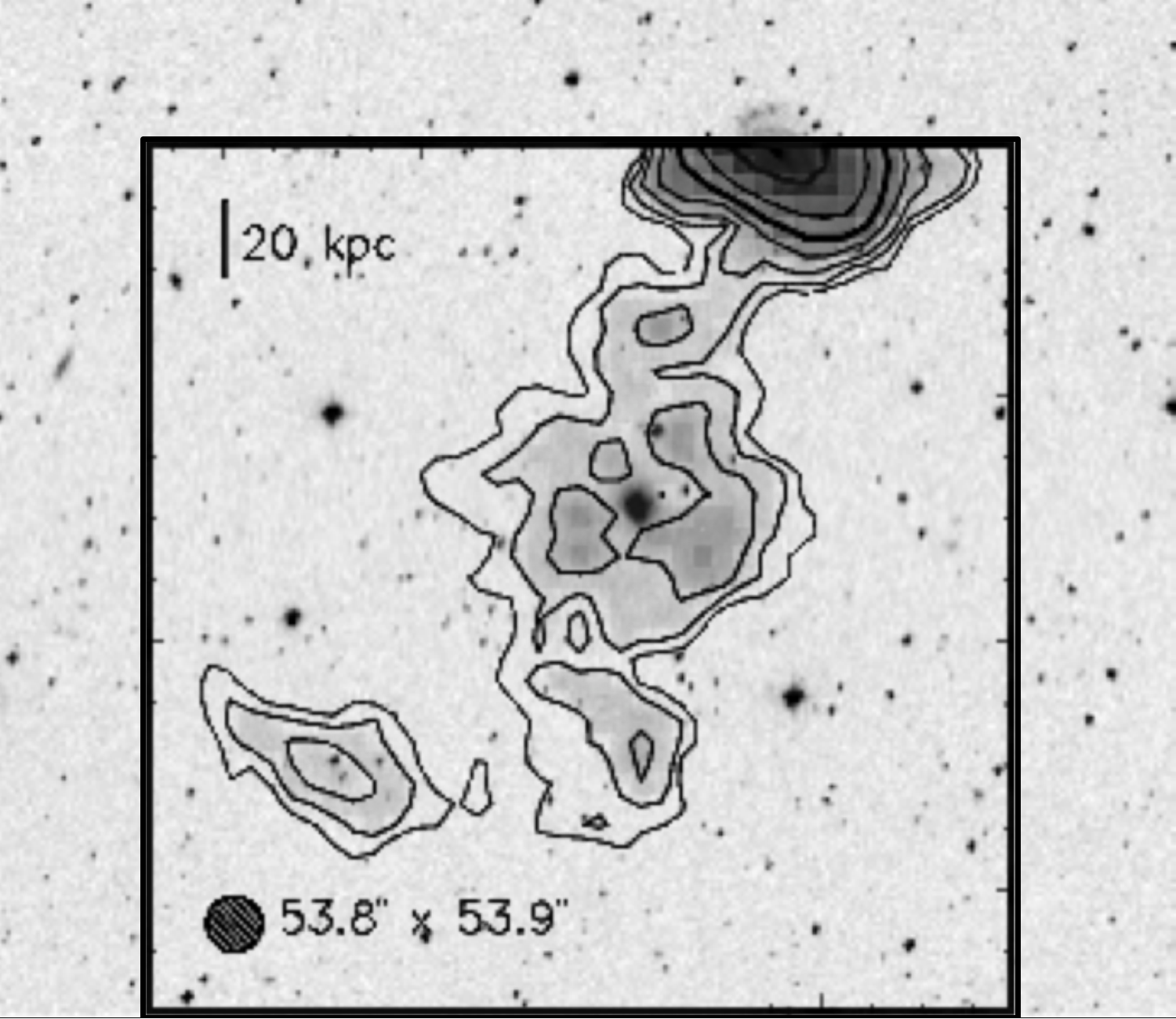}
\caption{Left: Color image centered near Mrk~3 (UGC~3426) and its gas-rich spiral companion UGC~3422 ~100 kpc to the NW. Image credit: This image was obtained by Rick Johnson through ManTrapSkies.com, and has been cropped from the original version (license located at https://creativecommons.org/licenses/by-nd/4.0/legalcode). Right: Contour map of H~I 21cm emission from \citet{Noordermeer et al.(2005)} on the same scale superimposed on a DSS red image, showing a tidal tail of H~I gas extending from UGC~3422 to Mrk~3 and beyond.
\label{fig:mrk3_HI}}
\end{figure}

Mrk~3 (UGC~3426) has a companion galaxy (UGC~3422) that is a gas-rich barred spiral galaxy that is $\sim$100 kpc to the NW, as shown in Figure \ref{fig:mrk3_HI}. \citet{Noordermeer et al.(2005)} show that there is a bridge of H~I emitting gas extending from UGC~3422 to Mrk~3 and beyond, which is likely due to a tidal encounter.
Furthermore, \citet{Noordermeer et al.(2005)} find no well-defined H~I radius for the host galaxy of Mrk~3.
We show their H~I surface density map overlayed on a DSS image in Figure \ref{fig:mrk3_HI}. The tidal stream is at the same general orientation as the gas/dust disk and is almost certainly its source. Thus, Mrk~3 is clear local example of external AGN fueling in an otherwise quiescent galaxy due to a tidal interaction with a gas-rich neighbor.

\citet{Noordermeer et al.(2005)} find that that the projected surface density of H~I is very low in the host galaxy of Mrk~3 (0.2 -- 0.3 M$_\sun$ pc$^{-2}$) compared to most of the early-type galaxies in their sample. These values are far below that those needed to instigate star formation at any location in this type of galaxy ($>$ 2 M$_\sun$ pc$^{-2}$) \citep{Kennicutt(1989)}, and it is therefore not surprising that we find no evidence for star formation in Mrk~3. However, it would be interesting to obtain an H~I 21 cm map at higher angular resolution than that of \citet{Noordermeer et al.(2005)} ($\sim$13\arcsec\ for their unsmoothed maps) to see if there are higher density clumps, as might be expected from the arcs and other structures in the dust lanes and ENLR ionized gas seen in Figure \ref{fig:mrk3_galaxy}.

\section{SUMMARY AND DISCUSSION} \label{sec:discussion}

Our photometric and kinematic analyses of the inner and outer stellar component of Mrk~3 show that it is a typical S0 galaxy with a large bulge and disk and no evidence for stellar spiral arms. The unusual features for an S0 galaxy are the dust lanes in the NE portion that are roughly perpendicular (offset by $\sim$100\arcdeg) to the stellar photometric and kinematic axes at PA $=$ 28\arcdeg, and regions of ionized gas (the NLR and ENLR) that appear to be the continuation of the dust lanes/spirals into the AGN ionizing bicone. The AGN fueled by the dusty gas is powerful enough to drive outflows to a distance of at least $\sim$320 pc, and further ionize and kinematically disturb the gas to a distance of at least 3 kpc. Based on our BPT diagrams, there is no evidence for significant star formation in either the large-scale galaxy or the nuclear region.

\subsection{Feeding on Extragalactic, Galactic, and Nuclear Scales} \label{subsec:feeding}

Our analysis of the kinematics of the ENLR from {\it APO} DIS long-slit spectra confirm the presence of a large-scale gas/dust disk at a PA $\approx$ 129\arcdeg\ that we previously identified based on the geometry of the NLR, ENLR, and dust lanes \citep{Crenshaw et al.(2010)}. This disk presumably feeds the AGN through nuclear dust spirals within $\sim$1 kpc of the nucleus as seen in {\it HST} images (Figure \ref{fig:mrk3_slits}). On very large scales, the disk is fed by a tidal stream of H~I gas from a companion gas-rich spiral galaxy (UGC~3422) 100 kpc away and in the same general direction as the gas/dust disk. Although this H~I gas flow is sufficient to fuel a luminous AGN, it's surface density is apparently too low to prompt star formation, and there is no evidence for AGN-induced star formation (``positive feedback'') from the outflows.

Recent work summarized by \citet{Storchi-Bergmann and Schnorr-Muller(2019)} identify several ways to fuel AGN on galactic or extragalactic scales (in addition to chaotic cold accretion onto AGN in the bright center galaxies of rich clusters). 1) Major mergers, where the mass ratio of the two galaxies is $\leq$ 4, are more likely to occur for luminous quasars at high (z $\geq$ 2) redshifts. 2) Minor mergers with mass ratios $>$ 4, have been identified as fueling mechanisms for a number of local AGN \citep{Martini et al.(2013), Fischer et al.(2015), Riffel et al.(2015b), Raimundo et al.(2017)}. 3) Tidal interactions between two galaxies that are not as severe as mergers can result in an exchange of gas from one galaxy to the other to fuel the AGN \citep{Davies et al.(2017)}.
4) Secular processes within a galaxy, particularly inflows along a large-scale stellar bar, can drive gas to within a few hundred to a thousand pc of the SMBH \citep{Shlosman et al.(1989), Regan et al.(1999)}. However, as noted by \citet{Storchi-Bergmann and Schnorr-Muller(2019)}, fueling along a bar often results in piling up of the gas at the Inner Lindblad Resonance (and often a starburst ring), and an additional mechanism such as a nuclear spiral or bar is needed to drive the gas further into the nucleus.
%We note that some of the above mechanisms are not mutually exclusive, because, for example, a merger or tidal interaction can instigate a bar in a galaxy that can subsequently drive fuel to the nucleus.

Mrk~3 is a clear case of the third example above, a tidal interaction that results in a broad flow of neutral gas that engulfs the S0 galaxy to form a galactic-scale disk, which, in this case, does not require a stellar bar to fuel the central SMBH. \citet{Davies et al.(2017)} provide evidence that, in general,  AGN in S0 galaxies are more likely to be fueled by external accretion due to galaxy interactions than AGN in spiral galaxies. They find that while the fraction of galaxies that are S0s increases with galaxy group size up to rich clusters (as found by others), the fraction of AGN in S0 galaxies decreases with increasing group size. They suggest that the intracluster medium in rich clusters is too hot for efficient accretion, driving down the number of AGN in S0 galaxies, whereas in small groups the gas from companions due to minor mergers or tidal interactions is cold and more easily accreted. 
Our results for Mrk~3 provide support for this scenario, because we can actually see the tidal stream of cold gas coming from a nearby gas-rich spiral to the host S0 galaxy, which settles into a large-scale disk to presumably fuel the central SMBH.

Additional evidence that AGN in S0 galaxies are externally fueled by minor mergers or tidal streams comes from the finding that nuclear gas disks in S0s are much more likely to be counter-rotating or offset in position angle with respect to their stellar disks compared to spiral galaxies \citep{Hicks et al.(2013), Davies et al.(2014), Raimundo et al.(2017)}, which favors external accretion for S0 galaxies over secular inflow. Thus, the nuclear disk of warm molecular and ionized gas in Mrk~3 that is counter-rotating with respect to the stellar velocity field at distances within 1.\arcsec5 ($\sim$400 pc) of the SMBH is not unusual in this respect.

One of the more interesting aspects of the fueling of Mrk~3 is the change in position of the gas disk(s) over different scales. The tidal stream from its gas-rich companion galaxy to the NW establishes a galactic-scale disk at PA $=$ 129\arcdeg\ (inclination $\approx$ 64\arcdeg), the circumnuclear gas disk is detected at PA $\approx$ 45\arcdeg, close to that of the stellar component but counter-rotating (inclination $\approx$ 65\arcdeg), and the putative molecular torus that confines the ionizing bicone should be at a PA $\approx$ 0\arcdeg\ (inclination $\approx$ 85\arcdeg). If these gas disks indeed represent the sources of the fueling flows, they indicate significant warping at different scales to presumably match the gravitational potentials of their environments. Interestingly, the H$_2$ emission outside of the nomimal bicone at NL and SL appears to connects to the nucleus in the N-S direction (Figure \ref{fig:nifs_fluxes}) at the approximate PA of the putative torus, suggesting that this connection may be the current fueling flow to the torus. However, we are not able to isolate this component of possible inflow, because the region around the nucleus is dominated by outflows in H$_2$. Velocity mapping of cold  gas at high resolution around the nucleus would therefore be very interesting.

\subsection{Feedback on Nuclear and Galactic Scales} \label{subsec:feedback}

\begin{figure}[ht!]
%\vspace{-8pt}
\centering
\subfigure{
\includegraphics[scale=0.4]{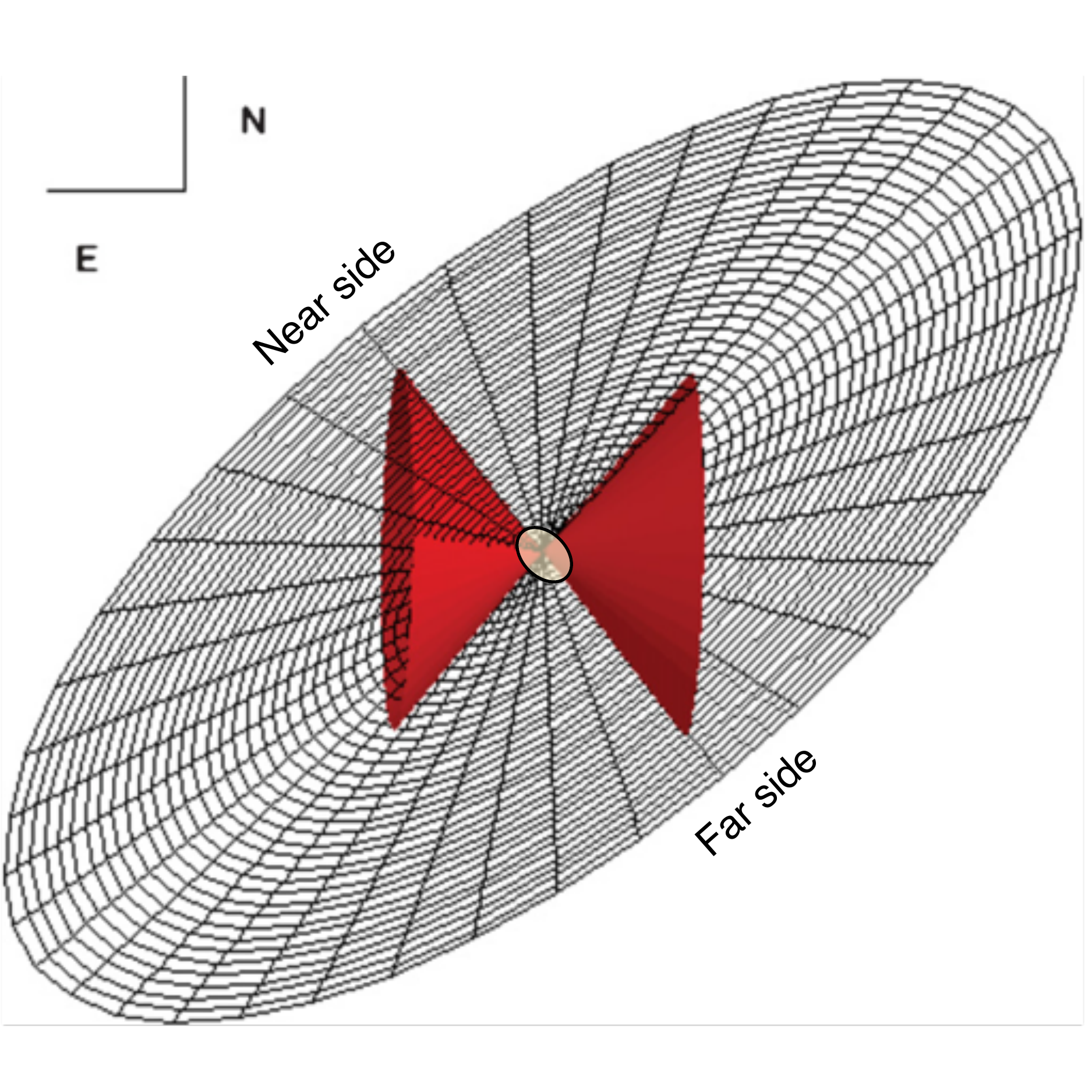}}\hspace{5ex}
\subfigure{
\includegraphics[scale=0.4]{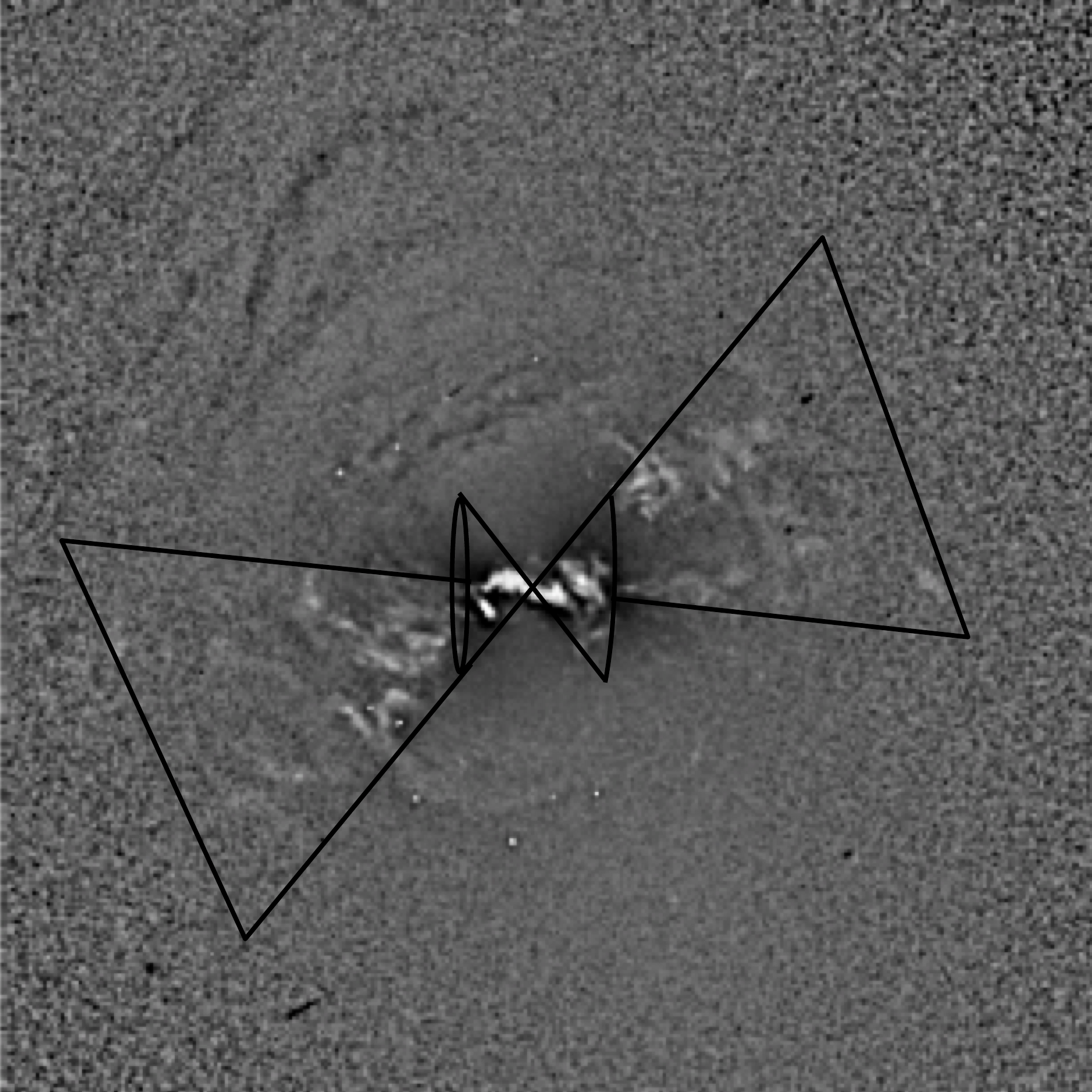}}
\caption{Left: Geometric model of the NLR bicone and large-scale gas/dust disk in Mrk 3 as viewed from the Earth, adopted from \citet{Crenshaw et al.(2010)}. The NLR bicone axis is nearly in the plane of the sky (the eastern axis lies 5\arcdeg\ out of the plane). The NE side of the large-scale gas/dust disk at PA $=$ 129\arcdeg\ is closer than the SW side and lies in front of most of the eastern cone. The orientation of the counter-rotating nuclear gas disk at PA$\approx$ 45\arcdeg\ is indicated by the small ellipse. Right: Structure map of Mrk 3 with the geometric model superimposed. The inner bicone of ionizing radiation encompasses the backward S shape of the NLR. The outer triangular regions show the intersection between the bicone of ionizing radiation and the gas/dust disk in the ENLR.
\label{fig:mrk3_geometry}}
\end{figure}

Our {\it Gemini} NIFS observations of [S~III] emission in the NLR of Mrk~3 confirm the dominance of outflowing ionized gas within at least 1.\arcsec2 ($\sim$320 pc) of the central continuum peak, which is the apex of the ionizing bicone and the likely location of the SMBH. Compared to the original detection of outflows from {\it HST} STIS long-slit spectra of [O~III] along the linear portion of the backwards ``S''-shaped NLR \citep{Ruiz et al.(2001), Crenshaw et al.(2010)}, the NIFS observations allow us to map the entire velocity field of the ionized gas around the SMBH and reveal even higher blueshifts than detected with STIS (up to $-$1500 km$^{-1}$), particularly in and around the previously unmapped eastern lobe (EL) and western lobe (WL) of [S~III] emission at the endpoints of the backwards ``S''. Nevertheless, the presence of redshifted knots of emission on each side of the AGN with radial velocities up to $+$800 km s$^{-1}$, often co-located or adjacent to blueshifted knots, confirms the overall picture of biconical outflow with the bicone axis close to the plane of the sky. The excess of blueshifted emission, particularly in the east, is consistent with the reddening trend determined by \citet{Collins et al.(2005)} and our specific geometric model of the NLR, ENLR, and large-scale gas/dust disk in Mrk~3 \citet{Crenshaw et al.(2010)}, which is reproduced in Figure \ref{fig:mrk3_geometry}.
Overall, we confirm our previous geometric and kinematic models of Mrk~3 \citep{Ruiz et al.(2001), Crenshaw et al.(2010)}, with the understanding that the latter could be tweaked to include somewhat higher maximum velocities in the NLR.

Our {\it Gemini} NIFS observations in the K-band provide the first high-resolution image and velocity maps of the warm (T $\approx$ 1000 K) molecular gas in Mrk~3 via the H$_2$ emission. We find that the bright H$_2$ emission matches the overall extent of the bright [S~III] gas but the structure is different, with H$_2$ emission knots that are often offset from the [S~III] knots, including the offset between the brightest H$_2$ knot at the nucleus (N) and the brightest [S~III] knot (EP) that is 0.\arcsec25 NE of the nucleus. There is also significant H$_2$ emission outside of the nominal ionizing bicone, similar to that in Mrk~573 \citep{Fischer et al.(2017)}. There is some general correspondence of this emission with dust lanes, particularly at the NL and SL positions in Figure \ref{fig:sm_nifs}, and the latter two lanes appear to connect to the nucleus outside of the ionizing bicone.

Despite the different morphologies, the kinematics of [S~III] and H$_2$ are similar in the sense that most H$_2$ emission knots are moving in the same directions as [S~III] knots at the same or adjacent locations but with smaller amplitudes. In particular, the H$_2$ north lane (NL) and south lane (SL) are outside of the nominal bicone, but share similar directions of motion with nearby [S~III] knots of emission.
%In particular, the H$_2$ emission in the SL is moving at the maxiumum H$_2$ radial velocity of $-$600 km s$^{-1}$, and is adjacent to the [S~III] EL, which contains very little H$_2$ but is moving at $-$1500 km s$^{-1}$.
This correspondence suggests that the H$_2$ knots both inside and outside of the nominal bicone originate from cold molecular gas that has been heated and accelerated away from a cold molecular reservoir (likely associated with the dust lanes) and are further ionized and accelerated in the ionization bicone to become the [S~III] knots. This picture is consistent with evidence for in situ ionization and acceleration of ambient gas as the primary source for the ionized gas outflows in nearby AGN \citep{Crenshaw et al.(2015), Revalski et al.(2018a), Revalski et al.(2018b)}, including Mrk~3 (Revalski et al. 2020, in preparation). We note, however, that some nearby AGN show more distinct kinematic differences between the ionized and molecular gas in their circumnuclear regions \citep{Riffel et al.(2013), Diniz et al.(2019), Schonell et al.(2019)}, and these must be taken into account in a fully developed model of the feeding and feedback processes.

 The NIFS observations may help to solve a puzzle from our previous paper \citep{Crenshaw et al.(2010)}. If the arcs in the ENLR and lobes in the NLR can be explained by the intersection of dust spirals in the gas disk with the ionizing bicone, why does the inner, linear portion of the NLR lie outside of this intersection as shown in Figure \ref{fig:mrk3_geometry}? The answer is that the latter may lie in the nuclear disk of warm and ionized gas detected in our NIFS observations, which is in the same general direction as the linear portion of the NLR. The backwards ``S'', which resembles the grand-design nuclear dust spirals seen in a number of other AGN \citep{Martini et al.(2003), Deo et al.(2006)}, would follow the warp of the gas disk and could represent the original fueling flow to the central SMBH as suggested by \citet{Crenshaw et al.(2010)}. After the AGN turned on, the radiation began to eat away at this feature, accelerating the warm molecular gas and ionizing it to produce the in-situ NLR outflows that we observe. In this scenario, the lack of H$_2$ emission from the EL could be due to the removal of the gas reservoir at this location, leaving only the outflowing ionized gas. Overall, the mass of the warm H$_2$ gas in the NIFS FOV is only $\sim$47 M$_\sun$ \citep{Riffel et al.(2018)}, on the low end compared to Mrk~573 \citep{Fischer et al.(2017)} and other nearby AGN \citep{Riffel et al.(2018)}, suggesting that Mrk~3 may be well on its way to exhausting its nuclear fuel (although this low mass may be due to other factors such as geometry or local excitation conditions).

The ionized gas in Mrk~3 is dominated by outflows within $\sim$320 pc of the SMBH and transitions to rotation within $\sim$1.1 kpc, which is only a fraction of the size of the large bulge and galaxy. However, the rotating, kinematically disturbed gas indicated by high FWHM extends to at least $\sim$3.2 kpc and, in general, the ionized gas extends to a distance of at least $\sim$5.4 kpc, which is about half of the extent host galaxy. Thus, although the current galaxy and bulge are well established from previous mergers and outflows, it appears that Mrk~3 is now in maintenance mode with continuous extragalactic, galactic, and nuclear-scale feeding and feedback from the AGN and little or no star formation.

\section{CONCLUSIONS} \label{sec:conclusions}

As discussed in Section \ref{subsec:feeding}, there are numerous ways that an AGN can be fueled on extragalactic, galactic, and nuclear scales \citep{Storchi-Bergmann and Schnorr-Muller(2019)}.
Our NIFS observations, for example, show that Mrk~509 is fueled by a minor merger with a gas-rich dwarf galaxy \citep{Fischer et al.(2015)}, and Mrk~573 is likely a case of secular fueling along a large-scale stellar bar followed by inflow along nuclear dust spirals \citep{Fischer et al.(2017)}.
Mrk~3 is clearly being fueled by a tidal interaction with a gas-rich companion galaxy, which has formed a galactic-scale gas disk that is offset from the stellar major axis of this S0 galaxy. Within $\sim$400 pc of the nucleus, the disk has apparently realigned close to the stellar axis, indicating a possible warp, although the gas is counter-rotating with respect to the stars. Offset or counter-rotating disks appear to be common in S0 galaxies with AGN \citep{Hicks et al.(2013), Davies et al.(2014)}, indicating an external origin is common in these sources. The final paths to the AGN in Mrk~3 may be nuclear dust lanes/spirals that feed into the observed north and south lanes (NL and SL) of H$_2$ emission and then along a N-S connection to the nucleus that feeds the putative torus and accretion disk. Observations of H~I and/or cold molecular gas in Mrk~3 at high spatial ($\leq$ 0.\arcsec1) and spectral ($\leq$ 10 km s$^{-1}$) resolutions would be helpful for mapping the inner fueling flow and detecting warps in the gas disk. Unfortunately, Mrk~3 cannot be observed by {\it ALMA} due to its high declination.

We have speculated that the backwards ``S'' shape of the NLR may be due to the original fueling flow that is now being ionized and accelerated away after the AGN turned on, leaving only the above N-S pathway for fueling. The lack of significant H$_2$ emission in the western lobe (WL) of [S~III] emission may indicate that the gas reservoir for outflows has been depleted and perhaps removed. Outflows dominate the kinematics of the ionized and warm molecular gas in Mrk~3 within $\sim$320 pc of the SMBH and the transition to pure rotation occurs within $\sim$1.1 kpc, which is only a small portion of the bulge and disk of Mrk~3, as has been found in other AGN at z $\lesssim$ 0.1 \citep{Fischer et al.(2017), Fischer et al.(2018)}. However, the gas continues to be kinematically disturbed by the AGN to distances of at least 3.2 kpc and ionized to at least 5.4 kpc. Although mergers and powerful outflows could have shaped the bulge and overall host galaxy of Mrk~3 at earlier epochs, it is currently in AGN maintenance mode with little or no star formation due to both the low surface density of the gas disk and disturbance/heating of the gas by the AGN. Nevertheless, this mode could continue as long as the gas disk is feed by a tidal tail from its companion galaxy, unless the flow is interrupted close to the nucleus.

\acknowledgements

This study has been supported by the National Science Foundation under grant No. 1211651 to DMC. RAR acknowledges financial support from CNPq and FAPERGS. This work is based in part on observations obtained at the Gemini Observatory (processed using the Gemini IRAF package), which is operated by the Association of Universities for Research in Astronomy, Inc., under a cooperative agreement with the NSF on behalf of the Gemini partnership: the National Science Foundation (United States), National Research Council (Canada), CONICYT (Chile), Ministerio de Ciencia, Tecnología e Innovación Productiva (Argentina), Ministério da Ciência, Tecnologia e Inovação (Brazil), and Korea Astronomy and Space Science Institute (Republic of Korea). This research has made use of the NASA/IPAC Extragalactic Database (NED), which is operated by the Jet Propulsion Laboratory, California Institute of Technology, under contract with the National Aeronautics and Space Administration. This research has made use of NASA’s Astrophysics Data System.


%\vspace{5mm}
\facilities{Gemini (NIFS), APO 3.5 m (DIS), HST (STIS, WFPC2)}

\end{document}